\documentstyle[amssymb,11pt]{article}

\newlength{\minitwocolumn}
\font\teneufm=eufm10
\font\seveneufm=eufm7
\font\fiveeufm=eufm5
\newfam\eufmfam
\textfont\eufmfam=\teneufm
\scriptfont\eufmfam=\seveneufm
\scriptscriptfont\eufmfam=\fiveeufm

\makeatletter
\@addtoreset{equation}{section}
\makeatother

\newtheorem{thm}{Theorem}[section]
\newtheorem{prop}[thm]{Proposition}

\newtheorem{dfn}[thm]{Definition}
\title{
\Large{\bf
Diagonalization of
infinite transfer matrix of\\
boundary $U_{q,p}(A_{N-1}^{(1)})$ face model
}}
\begin{document}
\maketitle
\begin{center}

\begin{center}
{\it Dedicated to Professor Etsuro Date on the occasion
of the 60th birthday}
\end{center}

~\\
{TAKEO KOJIMA}
\\~\\
{\it
Department of Mathematics and Physics,
Graduate School of Science and Engineering,\\
Yamagata University, Jonan 4-3-16, Yonezawa 992-8510,
Japan\\
kojima@yz.yamagata-u.ac.jp}
\end{center}

~\\

\begin{abstract}
We study infinitely many commuting operators $T_B(z)$,
which we call infinite transfer matrix of boundary $U_{q,p}(A_{N-1}^{(1)})$ face model.
We diagonalize the infinite transfer matrix $T_B(z)$ by using free field realizations of the vertex operators
of the elliptic quantum group $U_{q,p}(A_{N-1}^{(1)})$.
\end{abstract}

~\\

\section{Introduction}

There have been many developments in the field of
exactly solvable models.
Various models were found to be exactly solvable 
and various methods were
invented to solve these models
\cite{McWu, Bax, Gaudin, IK, Sm, JM}.
The vertex operator approach 
\cite{JM, DFJMN}
provides a powerful method to study exactly solvable 
lattice models.
This paper is devoted to boundary problem of 
the vertex operator approach to
exactly solvable lattice model.

Solvability of lattice models
was understood by means of the method of
commuting transfer matrix.
Two-dimensional solvable lattice models can be
related to one-dimensional solvable Hamiltonians
of one-dimensional quantum mechanics.
The transfer matrix of two-dimensional
lattice model is a generating function of
these Hamiltonians.
The commuting transfer matrix methods are
classified to four ways of doing this :
(A)~Clifford algebras and fermion operators \cite{McWu},
(B)~Bethe ansatz and quantum inverse scatting methods
\cite{Gaudin, IK},
(C)~Finite-lattice matrix functional relations
(Baxter's $T$-$Q$ relations)
\cite{Bax}, and
(D)~Vertex operator approach \cite{JM}.
The first three methods (A), (B), (C) work
for finite systems.
Meanwhile,
the vertex operator approach (D)
works only for infinite systems.
In infinite systems,
the transfer matrix is written by using vertex operators,
this can be related to the corner transfer matrix method
\cite{Bax}.
The corner transfer matrix method suggested
a link between solvable lattice models and representation
theory of infinite algebra.
The vertex operator approach arose from
interplay between
exactly solvable lattice models
and representation theory of infinite algebra.
The vertex operator approach (D) is completely
different from the first three methods (A), (B), (C).
The first paper on boundary problem of 
the vertex operator approach, is devoted to
the XXZ spin chain \cite{JKKKM},
in which vertex operators act on the highest
weight representation of the quantum affine group
$U_q(A_1^{(1)})$.
In \cite{JKKKM} the authors diagonalize
infinitely many commuting transfer matrices $T_B(z)$, 
using free field realizations of the vertex operators.
There have been two generalizations of
this theory \cite{JKKKM}.
The $U_q(A_{N-1}^{(1)})$ analogue of XXZ spin chain,
which give higher rank generalization of \cite{JKKKM},
was studied in \cite{FK}.
The boundary ABF model,
which give elliptic deformation of \cite{JKKKM},
was studied in \cite{MW}.
The boundary ABF is related to the elliptic quantum group
$U_{q,p}(A_1^{(1)})$.
This paper is devoted to a generalization of
the above papers \cite{JKKKM, MW, FK}.

Exactly solvable lattice model with open boundary
is defined by both Boltzmann weight functions and
boundary Boltzmann weight functions,
which satisfy Yang-Baxter equation and
boundary Yang-Baxter equation \cite{Sklyanin} 
respectively.
In this paper we are going to study
the $U_{q,p}(\widehat{sl_N})$ face model
defined by elliptic solution of Yang-Baxter equation 
\cite{JMO2}
and
boundary Yang-Baxter equation
\cite{BFKZ}, 
associated with
the elliptic quantum group 
$U_{q,p}(\widehat{sl_N})$ \cite{JKOS, KK}.
We are going to study
the infinite transfer matrix $T_B(z)$
associated with the boundary $U_{q,p}(A_{N-1}^{(1)})$ 
face model.
The boundary $U_{q,p}(A_{N-1}^{(1)})$ face model gives a
generalization of both 
the $U_q(A_{N-1}^{(1)})$ analogue of XXZ spin chain \cite{FK},
and the boundary ABF model \cite{MW}.
In this paper
we diagonalize
infinitely many commuting transfer matrices $T_B(z)$
acting on the bosonic Fock space,
in which the free field realization of 
the elliptic quantum group $U_{q,p}(A_{N-1}^{(1)})$ is constructed.
We construct free field realization of the boundary state
$|k\rangle_B$, which satisfies
$T_B(z)|k \rangle_B=|k \rangle_B$, by using
free field realizations of the vertex operators
\cite{AJMP}.
Multiplying the type-II vertex operators \cite{FKQ},
we diagonalize
infinitely many commuting transfer matrices $T_B(z)$.
We give complete proof of the 
eigenvalue problem $T_B(z)|k\rangle_B=|k\rangle_B$, 
this was only conjectured
for $U_{q,p}(A_2^{(1)})$ case \cite{Kojima1}, by using
identities of the elliptic theta functions.
We calculate the norm $~_B\langle k|k \rangle_B$
by using the coherent state.
We discuss physical and graphical interpretation
of the boundary state
$|k\rangle_B$ for 
the boundary $U_{q,p}(A_{N-1}^{(1)})$ face model.

The plan of the paper is as follows.
In section 2 we prepare the Boltzmann weight function,
the boundary Boltzmann weight function, and
the vertex operators.
We introduce infinite transfer matrix
of the boundary $U_{q,p}(A_{N-1}^{(1)})$ face model,
and formulate the problem.
In section 3 we prepare
the free field realization of
the vertex operators.
Section 4 is main part of this paper.
In section 4, we construct 
the free field realization of
the boundary state $|k\rangle_B$.
Multiplying the type-II vertex operators
we diagonalize the infinite transfer matrix
$T_B(z)$ of the boundary $U_{q,p}(A_{N-1}^{(1)})$ face model.
In section 5, we calculate the norm
$~_B\langle k|k\rangle_B$ by using coherent state.
In appendix we consider physical meaning of our problem.
We discuss the relation between our diagonalization
problem and
the boundary $U_{q,p}(A_{N-1}^{(1)})$ face model.

\section{Infinite transfer matrix}

In this section we introduce
infinitely many commuting operators $T_B(z)$,
which we call infinite transfer matrix of
boundary $U_{q,p}(A_{N-1}^{(1)})$ face model.

\subsection{Notation}

We prepare some notations.
Let us set integer $N=2,3,\cdots$.
We assume that $0<x<1$ and $r \geq N+2~(r \in {\mathbb Z})$.
We use parameterizations $z, u$ and $\tau$.
\begin{eqnarray}
z=x^{2u},~~~x=e^{-\pi i/r \tau}.
\nonumber
\end{eqnarray}
We set the elliptic theta function 
$[u]$ by
\begin{eqnarray}
~[u]=x^{\frac{u^2}{r}-u}\Theta_{x^{2r}}(x^{2u}),~~~
\Theta_q(z)&=&(q,q)_\infty (z;q)_\infty (q/z;q)_\infty.
\end{eqnarray}
Here we have used
\begin{eqnarray}
(z;q_1,q_2,\cdots,q_m)_\infty&=&\prod_{j_1,j_2,\cdots,j_m=0}^\infty
(1-q_1^{j_1}q_2^{j_2}\cdots q_m^{j_m}z).
\end{eqnarray}
The elliptic theta function $[u]$ satisfies the following quasi-periodicities.
\begin{eqnarray}
~[u+r]=-[u],~~~[u+r\tau]=-e^{-\pi i \tau -\frac{2\pi i u}{r}}[u].
\nonumber
\end{eqnarray}

Let $\epsilon_\mu (1\leq \mu \leq N)$ be the orthonormal
basis of ${\mathbb R}^N$ with the inner
product $(\epsilon_\mu |\epsilon_\nu)=\delta_{\mu,\nu}$.
Let us set $\bar{\epsilon}_\mu=\epsilon_\mu-\epsilon$
where $\epsilon=\frac{1}{N}\sum_{\nu=1}^N \epsilon_\nu$.
Note that $\sum_{\mu=1}^N \bar{\epsilon}_\mu=0$.
Let $\alpha_\mu~(1\leq \mu \leq N-1)$ the simple root :
\begin{eqnarray}
\alpha_\mu=\bar{\epsilon}_\mu-\bar{\epsilon}_{\mu+1}.
\end{eqnarray}
Let $\omega_\mu~(1\leq \mu \leq N-1)$ be 
the fundamental weights, which satisfy
\begin{eqnarray}
(\alpha_\mu|\omega_\nu)=\delta_{\mu,\nu},
~~(1\leq \mu,\nu\leq N-1).
\end{eqnarray}
Explicitly we set
$\omega_\mu=\sum_{\nu=1}^\mu \bar{\epsilon}_\nu.$
The type $A_{N-1}$ weight lattice is the
linear span of $\bar{\epsilon}_\mu$ or $\omega_\mu$.
\begin{eqnarray}
P=\sum_{\mu=1}^{N-1} {\mathbb Z}\bar{\epsilon}_\mu
=\sum_{\mu=1}^{N-1} {\mathbb Z}\omega_\mu.
\end{eqnarray}
For $a \in P$ we set $a_\mu$ and $a_{\mu,\nu}$ by
\begin{eqnarray}
a_{\mu,\nu}=a_{\mu}-a_{\nu},~~~
a_{\mu}=(a+\rho|\bar{\epsilon}_\mu),~~~(\mu, \nu \in P).
\end{eqnarray}
Here we set
$\rho=\sum_{\mu=1}^{N-1}\omega_\mu$.
Let us set the restricted path $P_{r-N}^+$ by
\begin{eqnarray}
P_{r-N}^+=\{a=\sum_{\mu=1}^{N-1}c_\mu \omega_\mu \in P|
c_\mu \in {\mathbb Z}, c_\mu \geq 0, 
\sum_{\mu=1}^{N-1}c_\mu \leq r-N \}.
\end{eqnarray}
For $a \in P_{r-N}^+$, condition 
$0<a_{\mu,\nu}<r,~(1\leq \mu<\nu\leq N-1)$ holds.

\subsection{Yang-Baxter equation}

We recall elliptic solutions of
the Yang-Baxter equation of face type.
An ordered pair $(b,a)\in P^2$ is called
admissible if and only if there exists
$\mu~(1\leq \mu \leq N)$ such that
\begin{eqnarray}
b-a=\bar{\epsilon}_\mu.\nonumber
\end{eqnarray}
An ordered set of four weights $(a,b,c,d)\in P^4$
is called an admissible configuration
around a face if and only if
the ordered pairs $(b,a)$, $(c,b)$, $(d,a)$ and $(c,d)$
are admissible.
Let us set
the Boltzmann weight functions of 
the face model \cite{JMO2},
$$
W\left(\left.\begin{array}{cc}
c&d\\
b&a
\end{array}\right|u\right)
$$
associated with admissible configuration
$(a,b,c,d)\in P^4$.
For $a \in P_{r-N}^+$, we set
\begin{eqnarray}
&&W\left(\left.
\begin{array}{cc}
a+2\bar{\epsilon}_\mu & a+\bar{\epsilon}_\mu\\
a+\bar{\epsilon}_\mu & a
\end{array}\right|u\right)=R(u),\\
&&W\left(\left.
\begin{array}{cc}
a+\bar{\epsilon}_\mu+\bar{\epsilon}_\nu & 
a+\bar{\epsilon}_\mu\\
a+\bar{\epsilon}_\nu & a
\end{array}\right|u\right)=R(u)\frac{[u][a_{\mu,\nu}-1]}
{[u-1][a_{\mu,\nu}]},\\
&&W\left(\left.
\begin{array}{cc}
a+\bar{\epsilon}_\mu+\bar{\epsilon}_\nu 
& a+\bar{\epsilon}_\nu\\
a+\bar{\epsilon}_\nu & a
\end{array}\right|u\right)=R(u)\frac{[u-a_{\mu,\nu}][1]}{
[u-1][a_{\mu,\nu}]}.
\end{eqnarray}
The normalizing function $R(u)$ is 
given by 
\begin{eqnarray}
R(u)&=&z^{\frac{r-1}{r}\frac{N-1}{N}}
\frac{\varphi(z^{-1})}{\varphi(z)},~~~
\varphi(z)=\frac{
(x^{2}z;x^{2r},x^{2N})_\infty 
(x^{2r+2N-2}z;x^{2r},x^{2N})_\infty}{
(x^{2r}z;x^{2r},x^{2N})_\infty 
(x^{2N}z;x^{2r},x^{2N})_\infty}.
\label{def:norBoltzmann}
\end{eqnarray}
Because $0<a_{\mu,\nu}<r~(1\leq \mu<\nu \leq N-1)$
holds for $a \in P_{r-N}^+$,
the Boltzmann weight functions
are well defined.

The Boltzmann weight functions 
satisfy the following relations 
(1), (2), (3) and (4).
\cite{JMO2, JMO3}
\\
\\
{\bf (1)}~~Yang-Baxter equation :
\begin{eqnarray}
&&\sum_{g}
W\left(\left.\begin{array}{cc}
d&e\\
c&g
\end{array}
\right|u_1\right)
W\left(\left.\begin{array}{cc}
c&g\\
b&a
\end{array}
\right|u_2\right)
W\left(\left.\begin{array}{cc}
e&f\\
g&a
\end{array}
\right|u_1-u_2\right)
\nonumber\\
&=&
\sum_{g}
W\left(\left.\begin{array}{cc}
g&f\\
b&a
\end{array}
\right|u_1\right)
W\left(\left.\begin{array}{cc}
d&e\\
g&f
\end{array}
\right|u_2\right)
W\left(\left.\begin{array}{cc}
d&g\\
c&b
\end{array}
\right|u_1-u_2\right).
\label{eqn:Boltzmann1}
\end{eqnarray}
\\
{\bf (2)}~~The first inversion relation :
\begin{eqnarray}
\sum_{g}
W\left(\left.\begin{array}{cc}
c&g\\
b&a
\end{array}
\right|-u\right)
W\left(\left.\begin{array}{cc}
c&d\\
g&a
\end{array}
\right|u\right)
=\delta_{b,d}.
\label{eqn:Boltzmann2}
\end{eqnarray}
\\
{\bf (3)}~~The second inversion relation :
\begin{eqnarray}
\sum_{g}G_g
W\left(\left.\begin{array}{cc}
g&b\\
d&c
\end{array}
\right|N-u\right)
W\left(\left.\begin{array}{cc}
g&d\\
b&a
\end{array}
\right|u\right)
=\delta_{a,c}\frac{G_{b}G_d}{G_a},
\label{eqn:Boltzmann3}
\end{eqnarray}
where we have set $G_a=\prod_{1\leq j<k \leq N}[a_{j,k}]$.\\
{\bf (4)}~~Initial conditions :
\begin{eqnarray}
W\left(\left.\begin{array}{cc}
g&b\\
d&c
\end{array}
\right|0\right)=\delta_{b,d}.
\label{eqn:Boltzmann4}
\end{eqnarray}
The normalization function $R(u)$ (\ref{def:norBoltzmann}) 
is chosen to satisfy
conditions (\ref{eqn:Boltzmann2}),
(\ref{eqn:Boltzmann3}) and (\ref{eqn:Boltzmann4}).
Upon this normalization $R(u)$,
the corner transfer matrix method works well,
and the maximum eigenvalue of the 
corner transfer matrix becomes $1$.

\subsection{Boundary Yang-Baxter equation}

An order set of three weights $(a,b,g)
\in P^3$
is called 
an admissible configuration at a boundary
if and only if
the ordered pairs $(g,a)$ and $(g,b)$ are admissible. 
Let us set the boundary Boltzmann weight functions
$
K\left(
\left.\begin{array}{cc}
&a\\
g&\\
&b
\end{array}
\right|u\right)
$ for admissible weights $(a,b,g)$ by
\begin{eqnarray}
K\left(
\left.\begin{array}{cc}
&a\\
a+\bar{\epsilon}_\mu&\\
&b
\end{array}
\right|u\right)=
z^{\frac{r-1}{r}\frac{N-1}{N}-\frac{2}{r}
a_1}\frac{h(z)}{h(z^{-1})}
\frac{[c-u][a_{1,\mu}+c+u]}
{[c+u][a_{1,\mu}+c-u]}
\delta_{a,b}.
\end{eqnarray}
In this paper, we consider
the case that continuous parameter $0<c<1$.
In what follows we need the case for $a \in P_{r-N}^+$.
The normalizing function
$h(z)$ is given by
\begin{eqnarray}
h(z)&=&
\frac{
(x^{2r+2N-2}/z^2;x^{2r},x^{4N})_\infty 
(x^{2N+2}/z^2;x^{2r},x^{4N})_\infty}{
(x^{2r}/z^2;x^{2r},x^{4N})_\infty 
(x^{4N}/z^2;x^{2r},x^{4N})_\infty}\nonumber\\
&\times&
\frac{
(x^{2N+2c}/z;x^{2r},x^{2N})_\infty
(x^{2r-2c}/z;x^{2r},x^{2N})_\infty}{
(x^{2N+2r-2c-2}/z;x^{2r},x^{2N})_\infty
(x^{2c+2}/z;x^{2r},x^{2N})_\infty}\nonumber\\
&\times&
\prod_{j=2}^N 
\frac{
(x^{2r+2N-2c-2a_{1,j}}/z;x^{2r},x^{2N})_\infty
(x^{2c+2a_{1,j}}/z;x^{2r},x^{2N})_\infty}{
(x^{2r+2N-2c-2a_{1,j}-2}/z;x^{2r},x^{2N})_\infty
(x^{2c+2+2a_{1,j}}/z;x^{2r},x^{2N})_\infty}.
\label{def:h}
\end{eqnarray}

The boundary Boltzmann weight functions 
and the Boltzmann weight functions
satisfy the following relations.
\\
(1)~Boundary Yang-Baxter equation :
\begin{eqnarray}
&&\sum_{f,g}
W\left(\left.\begin{array}{cc}
c&f\\
b&a
\end{array}\right|u-v\right)
W\left(\left.\begin{array}{cc}
c&d\\
f&g
\end{array}
\right|u+v\right)
K\left(\left.\begin{array}{cc}
~&g\\
f&\\
~&a
\end{array}
\right|u\right)
K\left(\left.\begin{array}{cc}
~&e\\
d&\\
~&g
\end{array}
\right|v\right)
\nonumber\\
&=&
\sum_{f,g}
W\left(\left.\begin{array}{cc}
c&d\\
f&e
\end{array}\right|u-v\right)
W\left(\left.\begin{array}{cc}
c&f\\
b&g
\end{array}
\right|u+v\right)
K\left(\left.\begin{array}{cc}
~&e\\
f&\\
~&g
\end{array}
\right|u\right)
K\left(\left.\begin{array}{cc}
~&g\\
b&\\
~&a
\end{array}
\right|v\right).\nonumber\\
\label{eqn:bBoltzmann1}
\end{eqnarray}
(2) Boundary unitary condition :
\begin{eqnarray}
K\left(\left.\begin{array}{cc}
~&a\\
b&\\
~&a
\end{array}
\right|u\right)
K\left(\left.\begin{array}{cc}
~&a\\
b&\\
~&a
\end{array}
\right|-u\right)=1.
\label{eqn:bBoltzmann2}
\end{eqnarray}
(3) Initial conditions :
\begin{eqnarray}
K\left(\left.\begin{array}{cc}
~&a\\
c&\\
~&b
\end{array}
\right|0\right)=\delta_{a,b}.
\label{eqn:bBoltzmann3}
\end{eqnarray}
Here we comment on the normalization $h(z)$.
The boundary Boltzmann weight functions associated with
$A_{1}^{(1)}$, $B_{N}^{(1)}$, 
$C_N^{(1)}$, $D_N^{(1)}$, $A_N^{(2)}$
satisfy the boundary crossing relations 
(except $A_{N \geq 2}^{(1)}$) like following
\cite{BFKZ}.
\begin{eqnarray}
\sum_g \left(\frac{G_g}{G_b}\right)^{\frac{1}{2}}
W\left(\left.\begin{array}{cc}
a&g\\
b&c\end{array}
\right|2u+\lambda\right)
K\left(\left.\begin{array}{cc}
~&a\\
g&\\
~&c
\end{array}
\right|u+\lambda\right)
=
K\left(\left.\begin{array}{cc}
~&a\\
b&\\
~&c
\end{array}
\right|-u\right).\label{eqn:bBoltzmann4}
\end{eqnarray}
In paper on $U_{q,p}(A_1^{(1)})$-face model \cite{MW},
the authors choose the normalization $h(z)$
of boundary Boltzmann weights,
from the equations
(\ref{eqn:bBoltzmann2}), (\ref{eqn:bBoltzmann3}) 
and (\ref{eqn:bBoltzmann4}).
This normalization makes
the maximum eigenvalue of the boundary state 
$|k\rangle_B$ becomes $1$ \cite{MW}.
However there does not exist
boundary crossing symmetry for higher-rank
case $A_{N \geq 2}^{(1)}$.
It is nontrivial to find
the normalization function $h(z)$ for
higher-rank case $A_{N\geq2}^{(1)}$.
In this paper the author choose the normalization function
$h(z)$ such that
there exists the boundary state $|k \rangle_B$
whose eigenvalue is $1$, without using
the boundary crossing relations
(\ref{eqn:bBoltzmann4}).

\subsection{Vertex operator}

In this section we introduce the vertex operator
$\Phi^{(b,a)}(z)$
for the $U_{q,p}(A_{N-1}^{(1)})$ face model
in the Regime III.
In appendix, we explain 
the physical interpretation of the vertex operators
$\Phi^{(b,a)}(z)$
for the elliptic quantum group $U_{q,p}(A_{N-1}^{(1)})$.
In this section we treat them symbolically.
In what follows we consider the so-called Regime III :
$0<u<1$ \cite{ABF, JMO2, LP}.
The vertex operator $\Phi^{(b,a)}(z)$ 
and the dual vertex operator $\Phi^{*(a,b)}(z)$
of admissible pair $(b,a)\in P^2$, for
the $U_{q,p}(A_{N-1}^{(1)})$ face model,
are the operators
which satisfy the following functional relations
(1) and (2), (resp. (1) and (2')).
\\
{\bf (1)}~Commutation relation :
\begin{eqnarray}
\Phi^{(a,b)}(z_1)
\Phi^{(b,c)}(z_2)
&=&\sum_{g}
W\left(\left.\begin{array}{cc}
a&g\\
b&c
\end{array}
\right|u_2-u_1\right)
\Phi^{(a,g)}(z_2)
\Phi^{(g,c)}(z_1),\label{eqn:VO1}\\
%%%%%%%%%%%%%%%%%%%%%%%%%%%%%%%%%%%%
\Phi^{*(a,b)}(z_1)
\Phi^{*(b,c)}(z_2)
&=&
\sum_{g}
W\left(\left.\begin{array}{cc}
c&b\\
g&a
\end{array}
\right|u_2-u_1\right)
\Phi^{*(a,g)}(z_2)
\Phi^{*(g,c)}(z_1),\label{eqn:VO2}\\
%%%%%%%%%%%%%%%%%%%%%%%%%%%%%%%%%%%%
\Phi^{(a,b)}(z_1)
\Phi^{*(b,c)}(z_2)
&=&
\sum_{g}
W\left(\left.\begin{array}{cc}
g&c\\
a&b
\end{array}
\right|u_1-u_2\right)
\Phi^{*(a,g)}(z_2)
\Phi^{(g,c)}(z_1).\label{eqn:VO3}
\end{eqnarray}
{\bf (2)}~Inversion relation :
\begin{eqnarray}
\Phi^{(a,g)}(z)\Phi^{*(g,b)}(z)=\delta_{a,b}.
\label{eqn:inversion1}
\end{eqnarray}
{\bf (2')}~Inversion relation :
\begin{eqnarray}
\sum_{g}
\Phi^{*(a,g)}(z)\Phi^{(g,b)}(z)=\delta_{a,b}.
\label{eqn:inversion2}
\end{eqnarray}
We have used parameterization $z=x^{2u}$.
The relations (1) and (2) is equivalent with the relations
(1) and (2').
In this section we don't mention the space which
these operators act.
There exist the operators which satisfy the
functional relations (1) and (2) (resp. (1) and (2')).
The free field realization of the vertex operators
$\Phi^{(b,a)}(z)$ and $\Phi^{*(a,b)}(z)$
are given in \cite{AJMP}.
In this paper we treat the problem in the Fock space 
${\cal F}_{l,k}$, which are introduced in \cite{AJMP}.
We review free field realization of vertex 
operators in the next section.

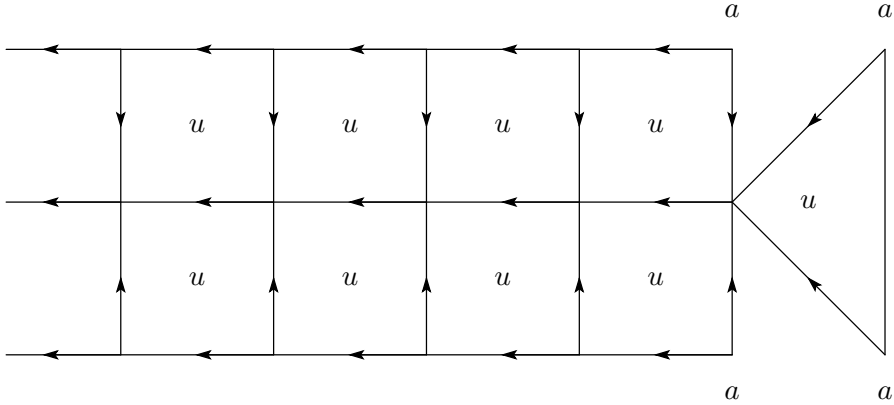
\subsection{Boundary transfer matrix}

We define the infinite transfer matrix $T_B(z)$
of the boundary $U_{q,p}(A_{N-1}^{(1)})$ face model,
by using the vertex operators.
\begin{eqnarray}
T_B(z)=\sum_{\mu=1}^N
\Phi^{*(a,a+\bar{\epsilon}_\mu)}(z^{-1})
K\left(\left.
\begin{array}{cc}
~& a\\
a+\bar{\epsilon}_\mu &\\
~& a
\end{array}
\right|u\right)
\Phi^{(a+\bar{\epsilon}_\mu,a)}(z).
\label{def:boundary-transfer}
\end{eqnarray}
Later, in appendix, we explain physical and
graphical
interpretation of
the infinite transfer matrix $T_B(z)$.

\begin{prop}~~~
The infinite transfer matrix $T_B(z)$ commute with each other.
\begin{eqnarray}
~[T_B(z_1),T_B(z_2)]=0.
\end{eqnarray}
\end{prop}
{\it Proof.}~~
Let's start from the product
$T_B(z_1)T_B(z_2)$,
\begin{eqnarray}
%T_B(z_1)T_B(z_2)&=&
\sum_{g_1,g_2}
\Phi^{*(a,g_1)}(z_1^{-1})\Phi^{(g_1,a)}(z_1)
\Phi^{*(a,g_2)}(z_2^{-1})\Phi^{(g_2,a)}(z_2)
%\nonumber\\
%&\times&
K\left(\left.\begin{array}{cc}
~&a\\
g_1&\\
~&a
\end{array}
\right|u_1\right)
K\left(\left.\begin{array}{cc}
~&a\\
g_2&\\
~&a
\end{array}
\right|u_2\right).
\nonumber
\end{eqnarray}
Exchange the ordering of
$
\Phi^{(g_1,a)}(z_1)
\Phi^{*(a,g_2)}(z_2^{-1})\Phi^{(g_2,a)}(z_2)
$
by using the commutation relations of vertex operators
(\ref{eqn:VO1}), (\ref{eqn:VO2}) , 
and use the boundary Yang-Baxter equation 
(\ref{eqn:bBoltzmann1}).
We have
\begin{eqnarray}
&&\sum_{g_1,g_2,g_3,g_4}
\Phi^{*(a,g_1)}(z_1^{-1})
\Phi^{*(g_1,g_3)}(z_2^{-1})
\Phi^{(g_3,g_4)}(z_2)
\Phi^{(g_4,a)}(z_1)
\nonumber\\
&\times&
W\left(\left.\begin{array}{cc}
g_3&g_4\\
g_2&a
\end{array}
\right|u_2+u_1\right)
W\left(\left.\begin{array}{cc}
g_3&g_2\\
g_1&a
\end{array}
\right|u_2-u_1\right)
K\left(\left.\begin{array}{cc}
~&a\\
g_4&\\
~&a
\end{array}
\right|u_1\right)
K\left(\left.\begin{array}{cc}
~&a\\
g_2&\\
~&a
\end{array}
\right|u_2\right).\nonumber
\end{eqnarray}
Exchanging the ordering of the vertex operators 
$
\Phi^{*(a,g_1)}(z_1^{-1})
\Phi^{*(g_1,g_3)}(z_2^{-1})
\Phi^{(g_3,g_4)}(z_2)
$
by using (\ref{eqn:VO2}), (\ref{eqn:VO3}), we get
$T_B(z_2)T_B(z_1)$.
\\
Q.E.D.

\subsection{Boundary state}

The commutativity
of the transfer matrix
$[T_B(z_1),T_B(z_2)]=0$
ensues that,
if the transfer matrices $T_B(z)$ are diagonalizable,
the transfer matrices $T_B(z)$
are diagonalized by the basis which is independent
of the spectral parameter $z$.

\begin{dfn}~~~We call eigenvector $|k \rangle_B$
with eigenvalue $1$ the boundary state.
\begin{eqnarray}
T_B(z)|k\rangle_B=|k\rangle_B. \label{def:boundary-state}
\end{eqnarray}
\end{dfn}

In this paper we construct
the free field realization
of the boundary state $|k\rangle_B$.
The construction of the boundary state $|k\rangle_B$
is main result of this paper.

\subsection{Type-II vertex operator}

In this section we introduce 
the type-II vertex operator
$\Psi^{*(b,a)}(z)$
for the $U_{q,p}(A_{N-1}^{(1)})$ face model,
which we use for diagonalization of
the transfer matrix $T_B(z)$.
For this purpose we prepare some functions.
Let us set $r^*=r-1$.
Let us set the elliptic theta function $[u]^*$ by
\begin{eqnarray}
~[u]^*=x^{\frac{u^2}{r^*}-u}\Theta_{x^{2r^*}}(x^{2u}).
\end{eqnarray}
Let us set
the Boltzmann weight functions 
$$
W^*\left(\left.\begin{array}{cc}
a&b\\
c&d
\end{array}\right|u\right)
$$
for admissible configuration $(a,b,c,d)\in P^4$.
For $a \in P_{r-1-N}^+$,
we set
\begin{eqnarray}
&&W^*\left(\left.
\begin{array}{cc}
a+2\bar{\epsilon}_\mu & a+\bar{\epsilon}_\mu\\
a+\bar{\epsilon}_\mu & a
\end{array}\right|u\right)=R^*(u),\\
&&W^*\left(\left.
\begin{array}{cc}
a+\bar{\epsilon}_\mu+\bar{\epsilon}_\nu & 
a+\bar{\epsilon}_\mu\\
a+\bar{\epsilon}_\nu & a
\end{array}\right|u\right)=R^*(u)
\frac{[u]^*[a_{\mu,\nu}-1]^*}
{[u-1]^*[a_{\mu,\nu}]^*},\\
&&W^*\left(\left.
\begin{array}{cc}
a+\bar{\epsilon}_\mu+\bar{\epsilon}_\nu 
& a+\bar{\epsilon}_\nu\\
a+\bar{\epsilon}_\nu & a
\end{array}\right|u\right)=R^*(u)
\frac{[u-a_{\mu,\nu}]^*[1]^*}{
[u-1]^*[a_{\mu,\nu}]^*}.
\end{eqnarray}
The function $R^*(u)$ is given by 
\begin{eqnarray}
R^*(u)&=&z^{-\frac{r}{r^*}\frac{N-1}{N}}
\frac{\varphi^*(z^{-1})}{\varphi^*(z)},~~~
\varphi^*(z)=\frac{
(z;x^{2r^*},x^{2N})_\infty 
(x^{2N+2r-2}z;x^{2r^*},x^{2N})_\infty}{
(x^{2r}z;x^{2r^*},x^{2N})_\infty 
(x^{2N-2}z;x^{2r^*},x^{2N})_\infty}.
\end{eqnarray}
The Boltzmann weight function 
$
W^*\left(\left.\begin{array}{cc}
a&b\\
c&d
\end{array}\right|u\right)
$ satisfies (1) the Yang-Baxter equation,
(2) the first inversion relation,
(3) the second inversion relation,
and
(4) initial condition, like
(\ref{eqn:Boltzmann1}), 
(\ref{eqn:Boltzmann2}), 
(\ref{eqn:Boltzmann3}) and 
(\ref{eqn:Boltzmann4}).

The type-II vertex operator $\Psi^{*(b,a)}(z)$
of admissible pair $(b,a)\in P^2$, for
the $U_{q,p}(A_{N-1}^{(1)})$ face model,
are the operators
which satisfy the following functional relations
(1) and (2).
\\
{\bf (1)}~Commutation relation :
\begin{eqnarray}
\Psi^{*(a,b)}(z_1)
\Psi^{*(b,c)}(z_2)
&=&-\sum_{g}
W^*\left(\left.\begin{array}{cc}
a&g\\
b&c
\end{array}
\right|u_1-u_2\right)
\Psi^{*(a,g)}(z_2)
\Psi^{*(g,c)}(z_1),\label{eqn:IIVO1}.
\end{eqnarray}
{\bf (2)}~Commutation relation with vertex operator:
\begin{eqnarray}
\Phi^{(d,c)}(z_1)\Psi^{*(b,a)}(z_2)=
\chi(z_2/z_1)
\Psi^{* (b,a)}(z_2)\Phi^{(d,c)}(z_1),
\label{eqn:IIVO2}\\
\Phi^{*(c,d)}(z_1)\Psi^{*(b,a)}(z_2)=
\chi(z_1/z_2)
\Psi^{* (b,a)}(z_2)\Phi^{*(c,d)}(z_1).
\label{eqn:IIVO3}
\end{eqnarray}
where we have set
\begin{eqnarray}
\chi(z)=z^{-\frac{N-1}{N}}
\frac{
(-x^{2N-1}z^{-1};x^{2N})_\infty 
(-xz;x^{2N})_\infty}{
(-xz^{-1};x^{2N})_\infty 
(-x^{2N-1}z;x^{2N})_\infty},
\label{def:chi}.
\end{eqnarray}
The free field realization of the type-II
vertex operators
$\Psi^{*(b,a)}(z)$
are given in \cite{FKQ}.

\begin{dfn}~~
We set the vectors $|\xi_1,\xi_2,\cdots,\xi_M
\rangle_{\mu_1,\mu_2,\cdots,\mu_M}$
$(1\leq \mu_1,\mu_2,\cdots,\mu_M  N)$. 
\begin{eqnarray}
&&
|\xi_1,\xi_2,\cdots,\xi_M
\rangle_{\mu_1,\mu_2,\cdots,\mu_M}
\label{def:excitations}
\\
&=&
\Psi^{*(b+\bar{\epsilon}_{\mu_1}
+\bar{\epsilon}_{\mu_2}+\cdots
+\bar{\epsilon}_{\mu_M},
b+\bar{\epsilon}_{\mu_2}+\cdots
+\bar{\epsilon}_{\mu_M})}(\xi_1)
\cdots
\Psi^{*(b+\bar{\epsilon}_{\mu_{M-1}}
+\bar{\epsilon}_{\mu_M},b+\bar{\epsilon}_{\mu_M})}(\xi_{M-1})
\Psi^{*(b+\bar{\epsilon}_{\mu_M},b)}(\xi_M)
|k \rangle_B.
\nonumber
\end{eqnarray}
We call
the vectors
$
|\xi_1,\xi_2,\cdots,\xi_M
\rangle_{\mu_1,\mu_2,\cdots,\mu_M}$
the excitation of the boundary state $|k \rangle_B$.
\end{dfn}

\begin{prop}~~~
The excitations
$
|\xi_1,\xi_2,\cdots,\xi_M
\rangle_{\mu_1,\mu_2,\cdots,\mu_M}$
are eigenvectors of the transfer matrix
$T_B(z)$.
\begin{eqnarray}
T_B(z)|\xi_1,\xi_2,\cdots,\xi_M
\rangle_{\mu_1,\mu_2,\cdots,\mu_M}
=
\prod_{j=1}^M \chi(\xi_j/z)\chi(1/\xi_j z)~
|\xi_1,\xi_2,\cdots,\xi_M
\rangle_{\mu_1,\mu_2,\cdots,\mu_M}.
\end{eqnarray}
\label{thm:main2}
\end{prop}

\section{Free field realization}

In this section we give free field realization
of the vertex operators $\Phi^{(b,a)}(z)$ and
the type-II vertex operators $\Psi^{*(d,c)}(z)$.

\subsection{Boson}

We set the bosonic oscillators $\beta_m^i, (i=1,2,\cdots,N-1;
m \in {\mathbb Z})$ by
\begin{eqnarray}
~[\beta_m^j,\beta_n^k]=
\left\{
\begin{array}{cc}
\displaystyle
m \frac{[(r-1)m]_x }{[rm]_x}
\frac{[(N-1)m]_x}{[Nm]_x}\delta_{m+n,0}
&(j=k)\\
\displaystyle
-m x^{Nm~sgn(j-k)}
\frac{[(r-1)m]_x}{[rm]_x}
\frac{[m]_x}{[Nm]_x}\delta_{m+n,0}
&(j \neq k).
\end{array}
\right.
\end{eqnarray}
Here the symbol $[a]_x=\frac{x^a-x^{-a}}{x-x^{-1}}$.
Let us set $\beta_m^N$ by
$\sum_{j=1}^Nx^{-2jm}\beta_m^j=0$.
The above commutation relations
are valid for all $1\leq j,k \leq N$.
We also introduce
the zero-mode operators $P_\alpha, Q_\alpha$, $(\alpha \in P)$
by
\begin{eqnarray}
~[iP_\alpha,Q_\beta]=(\alpha|\beta),~~(\alpha,\beta \in P).
\end{eqnarray}
In what follows we 
deal with the bosonic Fock space 
${\cal F}_{l,k}$, generated by 
$\beta_{-m}^j (m>0)$ over the vacuum vector 
$|l,k \rangle$, where
\begin{eqnarray}
l=b+\rho,~~~k=a+\rho,
\end{eqnarray}
for $a \in P_{r-N}^+$ and $b \in P_{r-1-N}^+$.
\begin{eqnarray}
{\cal F}_{l,k}={\mathbb C}[\{\beta_{-1}^j,
\beta_{-2}^j,\cdots\}_{j=1,\cdots,N-1}]|l,k\rangle,
\end{eqnarray}
where
\begin{eqnarray}
&&\beta_m^j |l,k\rangle=0,~~(m>0),\nonumber
\\
&&P_\alpha |l,k\rangle=\left(\alpha\left|
\sqrt{\frac{r}{r-1}}l-\sqrt{\frac{r-1}{r}}k
\right.\right)|l,k\rangle,\\
&&|l,k\rangle=e^{i\sqrt{\frac{r}{r-1}}Q_l-
\sqrt{\frac{r-1}{r}}Q_k}|0,0\rangle.\nonumber
\end{eqnarray}

\subsection{Vertex operator}

We give a free field realization of
the vertex operators $\Phi^{(b,a)}(z)$ \cite{AJMP},
associated with the elliptic algebra
$U_{q,p}(A_{N-1}^{(1)})$
\cite{JKOS, KK}.
Let us set basic operators 
$P_-(z),Q_-(z)$,
%$P_-^*(z),Q_-^*(z)$,
$R_-^j(z),S_-^j(z)$, 
$(1\leq j \leq N-1)$ 
by
\begin{eqnarray}
P_-(z)&=&\sum_{m>0}\frac{1}{m}\beta_{-m}^1z^{m},\\
Q_-(z)&=&-\sum_{m>0}\frac{1}{m}\beta_m^1z^{-m},\\
%P_-^*(z)&=&-\sum_{m>0}\frac{x^{Nm}}{m}\beta_{-m}^N z^m,\\
%Q_-^*(z)&=&\sum_{m>0}\frac{x^{-Nm}}{m}\beta_m^N z^{-m},\\
R_-^j(z)&=&
-\sum_{m>0}\frac{1}{m}(\beta_{-m}^j-\beta_{-m}^{j+1})x^{jm}z^m,\\
S_-^j(z)&=&
\sum_{m>0}\frac{1}{m}(\beta_m^j-\beta_m^{j+1})x^{-jm}z^{-m}.
\end{eqnarray}
Let us set the basic operators $U(z)$,$F_{\alpha_j}(z)$,
$(1\leq j \leq N-1)$
on the Fock space ${\cal F}_{l,k}$.
\begin{eqnarray}
U(z)&=&
z^{\frac{r-1}{2r}\frac{N-1}{N}}
e^{-i\sqrt{\frac{r-1}{r}}Q_{\bar{\epsilon}_1}}
z^{-\sqrt{\frac{r-1}{r}}P_{\bar{\epsilon}_1}}
e^{P_-(z)}e^{Q_-(z)},\\
%U^*(z)&=&
%z^{\frac{r-1}{2r}\frac{N-1}{N}}
%e^{i\sqrt{\frac{r-1}{r}}Q_{\bar{\epsilon}_N}}
%z^{\sqrt{\frac{r-1}{r}}P_{\bar{\epsilon}_N}}
%e^{P_-^*(z)}e^{Q_-^*(z)},\\
F_{\alpha_j}(z)&=&
z^{\frac{r-1}{r}}
e^{i\sqrt{\frac{r-1}{r}}Q_{\alpha_j}}
z^{\sqrt{\frac{r-1}{r}}P_{\alpha_j}}
e^{R_-^j(z)}e^{S_-^j(z)}.
\end{eqnarray}
In what follows we set
\begin{eqnarray}
\pi_\mu=\sqrt{r(r-1)}P_{\bar{\epsilon}_\mu},~~
\pi_{\mu, \nu}=\pi_\mu-\pi_\nu.
\end{eqnarray}
Then $\pi_{\mu \nu}$ acts on ${\cal F}_{l,k}$
as an integer $(\epsilon_\mu-\epsilon_\nu|rl-(r-1)k)$.
We give the free field realization of
the vertex operators
$\Phi^{(k+\bar{\epsilon}_{\mu},k)}(z)$, 
$(1\leq \mu \leq N-1)$ by
\begin{eqnarray}
\Phi^{(k+\bar{\epsilon}_1,k)}
(z_0^{-1})&=&
U(z_0),\nonumber
\\
\Phi^{(k+\bar{\epsilon}_\mu,k)}(z_0^{-1})
&=&
\oint
\cdots \oint 
\prod_{j=1}^{\mu-1}
\frac{dz_j}{2\pi i z_j} U(z_0)
F_{\alpha_1}(z_1)F_{\alpha_2}(z_2)
\cdots
F_{\alpha_{\mu-1}}(z_{\mu-1})\nonumber\\
&\times&
\prod_{j=1}^{\mu-1}
\frac{[u_j-u_{j-1}+\frac{1}{2}-\pi_{j,\mu}]}
{[u_j-u_{j-1}-\frac{1}{2}]}.
\end{eqnarray}
Here we set $z_j=x^{2u_j}$.
We take the integration contour to be simple closed 
curve that encircles 
$z_j=0, x^{1+2rs}z_{j-1}, (s \in {\mathbb N})$
but not
$z_j=x^{-1-2rs}z_{j-1}, (s \in {\mathbb N})$
for $1\leq j \leq \mu-1$.
The $\Phi^{(k+\bar{\epsilon}_\mu,k)}(z)$ 
is an operator such that
$\Phi^{(k+\bar{\epsilon}_\mu,k)}(z):
{\cal F}_{l,k}\to {\cal F}_{l,k+\bar{\epsilon}_\mu}$.
The vertex operators $\Phi^{(b,a)}(z)$
satisfy the commutation relation
(\ref{eqn:VO1}).
The free field realization of the dual vertex operator
$\Phi^{*(a,b)}(z)$
is given by similar way \cite{AJMP}.
In this paper we omit the free field realization of
the dual vertex operator $\Phi^{*(a,b)}(z)$,
because we don't use the explicit formulae of the dual 
vertex operators in this paper.
We only use the inversion relation
$\Phi^{(a,g)}(z)\Phi^{*(g,b)}(z)=\delta_{a,b}$,
(\ref{eqn:inversion1}).

\subsection{Type-II Vertex operator}

In this section we give the free field realization of
the type-II vertex operators
$\Psi^{*(b,a)}(z)$
\cite{FKQ}, associated with the elliptic algebra
$U_{q,p}(A_{N-1}^{(1)})$.
Let us set $P^*_+(z),Q^*_+(z)$,
$R_+^j(z),S_+^j(z)$,
$(1\leq j \leq N-1)$ by
\begin{eqnarray}
P^*_+(z)&=&
-\sum_{m>0}\frac{[rm]_x}{m[r^*m]_x}\beta_{-m}^1z^{m},\\
Q^*_+(z)&=&
\sum_{m>0}\frac{[rm]_x}{m[r^*m]_x}\beta_m^1z^{-m},\\
%P_+(z)&=&-\sum_{m>0}\frac{x^{Nm}}{m}\frac{[rm]_x}{[r^*m]_x}
%\beta_{-m}^N z^m,\\
%Q_+(z)&=&\sum_{m>0}\frac{x^{-Nm}}{m}
%\frac{[rm]_x}{[r^*m]_x}
%\beta_m^N z^{-m},\\
R_+^j(z)&=&
\sum_{m>0}\frac{[rm]_x}{m[r^*m]_x}
(\beta_{-m}^j-\beta_{-m}^{j+1})x^{jm}z^m,\\
S_+^j(z)&=&
-\sum_{m>0}\frac{[rm]_x}{m[r^*m]_x}
(\beta_m^j-\beta_m^{j+1})x^{-jm}z^{-m}.
\end{eqnarray}
Let us set the basic operators $V(z), E_{\alpha_j}(z)$
acting on the Fock space ${\cal F}_{l,k}$.
\begin{eqnarray}
%V(z)&=&
%z^{\frac{r}{2r^*}\frac{N-1}{N}}
%e^{-i\sqrt{\frac{r}{r^*}}Q_{\bar{\epsilon}_N}}
%z^{-\sqrt{\frac{r}{r^*}}P_{\bar{\epsilon}_N}}
%e^{P_+(z)}e^{Q_+(z)},\\
V^*(z)&=&
z^{\frac{r}{2r^*}\frac{N-1}{N}}
e^{i\sqrt{\frac{r}{r^*}}Q_{\bar{\epsilon}_1}}
z^{\sqrt{\frac{r}{r^*}}P_{\bar{\epsilon}_1}}
e^{P^*_+(z)}e^{Q^*_+(z)},\\
E_{\alpha_j}(z)&=&
z^{\frac{r}{r^*}}
e^{-i\sqrt{\frac{r}{r^*}}Q_{\alpha_j}}
z^{-\sqrt{\frac{r}{r^*}}P_{\alpha_j}}
e^{R_+^j(z)}e^{S_+^j(z)}.
\end{eqnarray}
We give a free field realization of
the type-II vertex operators
$\Psi^{*(l+\bar{\epsilon}_\mu,l)}(z)$ 
for $1\leq \mu \leq N-1$.
\begin{eqnarray}
\Psi^{*(l+\bar{\epsilon}_1,l)}(-z_0^{-1})&=&V^*(z_0),\nonumber
\\
\Psi^{*(l+\bar{\epsilon}_\mu,l)}(-z_0^{-1})&=&
\oint \cdots \oint 
\prod_{j=1}^{\mu-1} 
\frac{dz_j}{z_j} 
V^*(z_0)
E_{\alpha_1}(z_1)E_{\alpha_2}(z_2)
\cdots E_{\alpha_{\mu-1}}(z_{\mu-1})\nonumber\\
&\times&
\prod_{j=1}^{\mu-1}
\frac{[u_j-u_{j-1}-\frac{1}{2}+\pi_{j,\mu}]^*}
{[u_j-u_{j-1}+\frac{1}{2}]^*}.
\end{eqnarray}
The integration should be carried out in
the order $z_{\mu-1},\cdots,z_2,z_1$
along the contour for $z_j$-integration which encircles
$z_j=x^{-1+2sr^*}z_{j-1}$,$(s=0,1,2,\cdots)$,
but $z_j=x^{1-2sr^*}z_{j-1}$,
$(s=0,1,2,\cdots)$.
The operator $\Psi^{*(l+\bar{\epsilon}_\mu,l)}(z)$ 
is an operator such that
$\Psi^{*(l+\bar{\epsilon}_\mu,l)}(z):
{\cal F}_{l,k} \to {\cal F}_{l+\bar{\epsilon}_\mu,k}$.

\section{Boundary state}

In this section we give a free field realization of the
boundary state $|k\rangle_B$ 
in the bosonic Fock space ${\cal F}_{k,k}$,
\begin{eqnarray}
T_B(z)|k\rangle_B=|k\rangle_B.\nonumber
\end{eqnarray}
In this paper, we consider the case $l=k$,
which represents the ground state of the boundary
$U_{q,p}(A_{N-1}^{(1)})$ face model.
In the appendix, we explain
the physical meaning of $(l,k)$.
The label $k$ represents the condition at the origin
and the label $l$ represents the asymptotic 
boundary condition at $\infty$.
The condition $l=k \in P_{r-1-N}^+$
represents the ground state for $0<c<1$ in the Regime III.
i.e., the case $0<u<1$.

~\\
Using the inversion relation of the vertex operator
$\Phi^{(a,g)}(z)\Phi^{*(g,b)}(z)=\delta_{a,b}$,
we have the following proposition.

\begin{prop}~~The relation (\ref{def:boundary-state})
is equivalent to
the following relation.
\begin{eqnarray}
K\left(\left.\begin{array}{cc}
~&k\\
k+\bar{\epsilon}_j&\\
~&k
\end{array}
\right|u\right)
\Phi^{(k+\bar{\epsilon}_j,k)}(z)|k\rangle_B
=
\Phi^{(k+\bar{\epsilon}_j,k)}(z^{-1})|k\rangle_B.
\label{def:boundary-state2}
\end{eqnarray}
\end{prop}
In what follows we will show
the relation (\ref{def:boundary-state2}).

\subsection{Main result}

In this section we explain
the main result of this paper.
The commutation relation of bosons $\beta_m^j$
is not symmetric.
It is convenient to introduce new generators
of bosons $\alpha_m^j$,
whose commutation relation is symmetric.
Let us set $\alpha_m^j~(m \in {\mathbb Z}_{\neq 0};
1\leq j \leq N-1)$ by
\begin{eqnarray}
\alpha_m^j=x^{-jm}(\beta_m^j-\beta_m^{j+1}).
\end{eqnarray}
They satisfy the following commutation relations.
\begin{eqnarray}
~[\alpha_m^j,\alpha_n^k]=m\frac{[(r-1)m]_x}{[rm]_x}
\frac{[A_{j,k}m]_x}{[m]_x}\delta_{m+n,0},
\end{eqnarray}
where $A_{j,k}$ is a matrix element of the Cartan matrix of
$A_{N-1}$ type.
\begin{eqnarray}
(A_{j,k})_{1\leq j,k \leq N-1}=\left(
\begin{array}{cccccc}
2&-1&0&\cdots&\cdots&0\\
-1&2&-1&0&\cdots&0\\
0&-1&2&-1&0&\cdots\\
\cdots&\cdots&\cdots&\cdots&\cdots&\cdots\\
0&\cdots&0&-1&2&-1\\
0&\cdots&\cdots&0&-1&2
\end{array}\right).
\end{eqnarray}
Let us set $I_{j,k}(m)~(m\in {\mathbb Z}_{\neq 0};1\leq
j,k \leq N-1)$ by
\begin{eqnarray}
I_{j,k}(m)=\frac{[jm]_x[(N-k)m]_x}{[m]_x[Nm]_x}
=I_{k,j}(m)~~(1\leq j\leq k \leq N-1).
\end{eqnarray}
The matrix $(I_{j,k}(m))_{1\leq j,k \leq N-1}$
gives the inverse matrix of
$\left(\frac{[A_{j,k}m]_x}{[m]_x}\right)_{1\leq j,k \leq N-1}$.
\\
Let us set
\begin{eqnarray}
~[a]_x^+=x^a+x^{-a},~~\theta_m(x)=
\left\{\begin{array}{cc}
x& m~even,\\
0& m~odd.
\end{array}\right.
\end{eqnarray}

\begin{thm}~~~
The free field realization of 
the boundary state $|k\rangle_B$
is given by
\begin{eqnarray}
|k\rangle_B=e^F|k,k\rangle.
\end{eqnarray}
This bosonic vector satisfies
\begin{eqnarray}
T_B(z)|k \rangle_B=|k \rangle_B.
\end{eqnarray}
Here we have set
\begin{eqnarray}
F&=&
-\frac{1}{2}\sum_{m>0}
\sum_{j=1}^{N-1}\sum_{k=1}^{N-1}
\frac{1}{m}\frac{[rm]_x}{[(r-1)m]_x}
I_{j,k}(m)\alpha_{-m}^j \alpha_{-m}^k
+\sum_{m>0}\sum_{j=1}^{N-1}\frac{1}{m}
D_j(m)\beta_{-m}^j,
\end{eqnarray}
where we have set
\begin{eqnarray}
D_j(m)&=&
-\theta_m\left(\frac{[(N-j)m/2]_x[rm/2]_x^+
x^{\frac{(3j-N-1)m}{2}}}
{[(r-1)m/2]_x}\right)\nonumber\\
&&+\frac{x^{(j-1)m}[(-r+2\pi_{1,j}+2c-j+2)m]_x}{[(r-1)m]_x}
\nonumber\\
&&+\frac{[m]_x 
x^{(r-2c+2j-2)m}}{[(r-1)m]_x}
\left(\sum_{k=j+1}^{N-1}x^{-2m \pi_{1,k}}\right)\nonumber\\
&&
+\frac{x^{(2j-N)m}[(r-2\pi_{1,N}-2c+N-1)m]_x}{[(r-1)m]_x}.
\end{eqnarray}
\label{thm:main}
\end{thm}

\begin{thm}~~~
The free field realization of 
the boundary state $~_B\langle k|$
is given by
\begin{eqnarray}
~_B\langle k|=\langle k,k |e^G.
\end{eqnarray}
This bosonic vector satisfies
\begin{eqnarray}
~_B\langle k|T_B(z)=~_B\langle k|.
\end{eqnarray}
Here we have set the bosonic operator $G$ by
\begin{eqnarray}
G&=&
-\frac{1}{2}\sum_{m>0}
\sum_{j=1}^{N-1}\sum_{k=1}^{N-1}
\frac{x^{2Nm}}{m}\frac{[rm]_x}{[(r-1)m]_x}
I_{j,k}(m)\alpha_{m}^j \alpha_{m}^k
+\sum_{m>0}\sum_{j=1}^{N-1}\frac{1}{m}
E_j(m)\beta_{m}^j,
\end{eqnarray}
where we have set
\begin{eqnarray}
E_j(m)&=&
\theta_m\left(\frac{[(N-j)m/2]_x[rm/2]_x^+
x^{\frac{(3N+1-3j)m}{2}}}
{[(r-1)m/2]_x}\right)\nonumber\\
&&-\frac{x^{(N+1-j)m}[(r-2c+N-j-2\pi_{1,j})m]_x}{[(r-1)m]_x}
\nonumber\\
&&-\frac{[m]_x 
x^{(r-2c+2N-2j)m}}{[(r-1)m]_x}
\left(\sum_{k=j+1}^{N-1}x^{-2m \pi_{1,k}}\right)\nonumber\\
&&
-\frac{x^{2(N-j)m}[(-r+2c+1+2\pi_{1,N})m]_x}{[(r-1)m]_x}.
\end{eqnarray}
\end{thm}

We construct diagonalization of
the transfer matrix $T_B(z)$ on the subspace of
${\cal F}_{k,k}$, which is spanned by
the excitations $|\xi_1 \cdots \xi_M\rangle_{\mu_1
\cdots \mu_M}$.
See proposition \ref{thm:main2}.
This subspace is expected to become the physical space
of the boundary 
$U_{q,p}(A_{N-1}^{(1)})$ face model explained in appendix.

\subsection{Proof}

In this section we give a proof of theorem \ref{thm:main}.
At first we prepare some propositions,
which are derived by direct calculations.

\begin{prop}~~~The adjoint action of $e^F$
has the effect of a Bogoliubov transformation.
\begin{eqnarray}
&&e^{-F}\alpha_m^je^F=\alpha_m^j-\alpha_{-m}^j\\
&&+\frac{[(r-1)m]_x}{[rm]_x}(x^{(-j+1)m}D_j(m)
-x^{(-j-1)m}D_{j+1}(m)),~(m>0; 1\leq j \leq N-2),
\nonumber\\
&&e^{-F}\alpha_m^{N-1}e^F=\alpha_m^{N-1}-\alpha_m^{N-1}
+\frac{[(r-1)m]_x}{[rm]_x}x^{(-N+2)m}D_{N-1}(m),~
(m>0),\\
&&e^{-F}\beta_m^1e^F=\beta_m^1-\beta_{-m}^1\nonumber\\
&&+\frac{[(r-1)m]_x}{[rm]_x [Nm]_x}
([(N-1)m]_xD_1(m)-[m]_x x^{-Nm}\sum_{j=2}^{N-2}D_j(m)).
\end{eqnarray}
\end{prop}

~\\

\begin{prop}~~~
The adjoint action of $e^G$
has the effect of a Bogoliubov transformation.
\begin{eqnarray}
&&e^{G}\alpha_{-m}^je^{-G}=
\alpha_{-m}^j-x^{2Nm}\alpha_{m}^j\\
&&+\frac{[(r-1)m]_x}{[rm]_x}(x^{(j-1)m}E_j(m)
-x^{(j+1)m}E_{j+1}(m)),~(m>0; 1\leq j \leq N-2),
\nonumber\\
&&e^{G}\alpha_{-m}^{N-1}e^{-G}=\alpha_{-m}^{N-1}
-x^{2Nm}\alpha_{m}^{N-1}
+\frac{[(r-1)m]_x}{[rm]_x}x^{(N-2)m}E_{N-1}(m),~
(m>0),\\
&&e^G \beta_{-m}^N e^{-G}
=\beta_{-m}^N-\beta_m^N-\frac{[(r-1)m]_x [m]_x x^{-Nm}}{[rm]_x[Nm]_x}
\sum_{j=1}^{N-1}E_j(m).
\end{eqnarray}
\end{prop}

~\\

\begin{prop}~~The actions of the basic operators on the
boundary state are given by
\begin{eqnarray}
e^{Q_-(z)}|k\rangle_B&=&h(z)e^{P_-(1/z)}|k\rangle_B,\\
e^{S_-^j(w)}|k\rangle_B&=&g_j(w)e^{R_-^j(1/w)}|k\rangle_B,
~~(1\leq j \leq N-1).
\end{eqnarray}
Here we have set
\begin{eqnarray}
h(z)&=&
\frac{
(x^{2r+2N-2}/z^2;x^{2r},x^{4N})_\infty 
(x^{2N+2}/z^2;x^{2r},x^{4N})_\infty}{
(x^{2r}/z^2;x^{2r},x^{4N})_\infty 
(x^{4N}/z^2;x^{2r},x^{4N})_\infty}\nonumber\\
&\times&
\frac{
(x^{2N+2c}/z;x^{2r},x^{2N})_\infty
(x^{2r-2c}/z;x^{2r},x^{2N})_\infty}{
(x^{2N+2r-2c-2}/z;x^{2r},x^{2N})_\infty
(x^{2c+2}/z;x^{2r},x^{2N})_\infty}\nonumber\\
&\times&
\prod_{j=2}^N 
\frac{
(x^{2r+2N-2c-2\pi_{1,j}}/z;x^{2r},x^{2N})_\infty
(x^{2c+2\pi_{1,j}}/z;x^{2r},x^{2N})_\infty}{
(x^{2r+2N-2c-2\pi_{1,j}-2}/z;x^{2r},x^{2N})_\infty
(x^{2c+2+2\pi_{1,j}}/z;x^{2r},x^{2N})_\infty},
\label{def:h}
\\
g_j(w)&=&(1-w^{-2})
\frac{
(x^{2\pi_{1,j}+2c+2-j}/w;x^{2r})_\infty
(x^{-2\pi_{1,j+1}+2r-2c+j}/w;x^{2r})_\infty
}{
(x^{2r-2\pi_{1,j}-2c-2+j}/w;x^{2r})_\infty
(x^{2c+2\pi_{1,j+1}-j}/w;x^{2r})_\infty}.
\end{eqnarray}
\end{prop}

~\\

\begin{prop}~
The actions of the basic operators on the dual boundary state
are given by
\begin{eqnarray}
~_B\langle k|e^{P^*_-(z/x^N)}&=&
h^*(z)~_B\langle k|e^{Q^*_-(1/x^{N}z)},\\
~_B\langle k|e^{R_-^j(w/x^N)}&=&
g_j^*(w)~_B\langle k|e^{S_-^j(1/x^{N}w)},~(1\leq j \leq N-1).
\end{eqnarray}
Here we have set
\begin{eqnarray}
h^*(z)&=&
\frac{
(x^{2r+2N-2}z^2;x^{2r},x^{4N})_\infty 
(x^{2N+2}z^2;x^{2r},x^{4N})_\infty}{
(x^{2r}z^2;x^{2r},x^{4N})_\infty 
(x^{4N}z^2;x^{2r},x^{4N})_\infty}\nonumber\\
&\times&
\frac{
(x^{2N+2c+2\pi_{1,N}}z;x^{2r},x^{2N})_\infty
(x^{2r-2c-2\pi_{1,N}}z;x^{2r},x^{2N})_\infty}{
(x^{2N+2r-2c-2-2\pi_{1,N}}z;x^{2r},x^{2N})_\infty
(x^{2c+2+2\pi_{1,N}}z;x^{2r},x^{2N})_\infty}\nonumber\\
&\times&
\prod_{j=1}^{N-1} 
\frac{
(x^{2r+2N-2c-2\pi_{1,j}}z;x^{2r},x^{2N})_\infty
(x^{2c+2\pi_{1,j}}z;x^{2r},x^{2N})_\infty}{
(x^{2r+2N-2c-2\pi_{1,j}-2}z;x^{2r},x^{2N})_\infty
(x^{2c+2+2\pi_{1,j}}z;x^{2r},x^{2N})_\infty},\\
g_j^*(w)&=&(1-w^2)
\frac{
(x^{2c+2-N+j+2\pi_{1,j+1}}w;x^{2r})_\infty
(x^{2r-2c+N-j-2\pi_{1,j}}w;x^{2r})_\infty
}{
(x^{2r-2c-2+N-j-2\pi_{1,j+1}}w;x^{2r})_\infty
(x^{2c-N+j+2\pi_{1,j}}w;x^{2r})_\infty}.
\end{eqnarray}
\end{prop}

~\\
\\
In order to write proof compactly,
we introduce "weak equality" in the following sense.

\begin{dfn}~~~
When functions $F(z_1,z_2,\cdots,z_L)$ and
$G(z_1,z_2,\cdots,z_L)$ satisfy
\begin{eqnarray}
\sum_{\epsilon_1=\pm 1}
\cdots
\sum_{\epsilon_L=\pm 1}
F(z_1^{\epsilon_1},z_2^{\epsilon_2},\cdots, z_L^{\epsilon_L})
=
\sum_{\epsilon_1=\pm 1}
\cdots
\sum_{\epsilon_L=\pm 1}
G(z_1^{\epsilon_1},z_2^{\epsilon_2},\cdots, 
z_L^{\epsilon_L}),
\end{eqnarray}
we write 
\begin{eqnarray}
F(z_1,z_2,\cdots,z_L) \sim_{\atop{(z_1 \cdots z_L)}} 
G(z_1,z_2,\cdots,z_L)
\end{eqnarray}
showing the weak equality.
\end{dfn}
We note that the meanings of weak equality,
$$F(z_1,z_2,\cdots,z_L) \sim_{\atop{(z_1 \cdots z_L)}} 
G(z_1,z_2,\cdots,z_L)$$
and weak equality,
$$F(z_1,z_2,\cdots,z_L) \sim_{\atop{(z_1 \cdots z_{L-1})}} 
G(z_1,z_2,\cdots,z_L)$$
are different.
We prepare propositions in terms of weakly equality.

\begin{prop}~~~The function $g_j(w)$, $(1\leq j \leq N-1)$
satisfy the following.
\begin{eqnarray}
g_j(w)w
\Theta_{x^{2r}}(x^{2c+2\pi_{1,j}-j+2}w)
\Theta_{x^{2r}}(x^{2c+2\pi_{1,j+1}-j}/w)\sim_{\atop{w}}0.
\label{eqn:theta1}
\end{eqnarray}
\end{prop}

\begin{prop}~~~The following theta identity holds.
\begin{eqnarray}
&&\Theta_{x^{2r}}(x^{2c+2k-1}w)
\Theta_{x^{2r}}(x^{2c+1}/w)\nonumber\\
&\times&
\left(z
\Theta_{x^{2r}}(x^{2c}/z)
\Theta_{x^{2r}}(x^{2k+2c}z)
\Theta_{x^{2r}}(x^{2k-1}/wz)
\Theta_{x^{2r}}(xw/z)\right.\nonumber
\\
&-&\left.
z^{-1}
\Theta_{x^{2r}}(x^{2c}z)
\Theta_{x^{2r}}(x^{2k+2c}/z)
\Theta_{x^{2r}}(x^{2k-1}z/w)
\Theta_{x^{2r}}(xwz)\right)\sim_{\atop{w}}0.
\label{eqn:theta2}
\end{eqnarray}
\end{prop}

~\\
{\it Proof of main theorem}~~~
Now let us start a proof of main theorem \ref{thm:main}.
We would like to show
\begin{eqnarray}
T_B(z)|k\rangle_B=|k\rangle_B.
\end{eqnarray}
Multiplying the vertex operators 
$\Phi^{(k+\bar{\epsilon}_\mu,k)}(z)$ from
the left, and using the inversion relation of vertex
operators (\ref{eqn:inversion1}), we get a necessary and sufficient 
condition for $\mu=1,2,\cdots,N-1$,
\begin{eqnarray}
&&
\Phi^{(k+\bar{\epsilon}_\mu,k)}(z)
z^{\frac{r-1}{2r}\frac{N-1}{N}-\frac{1}{r}\pi_1}
h(z)[c-u][\pi_{1,\mu}+c+u]|k\rangle_B\nonumber\\
&&=
\Phi^{(k+\bar{\epsilon}_\mu,k)}(z^{-1})
z^{-\frac{r-1}{2r}\frac{N-1}{N}+\frac{1}{r}\pi_1}h(z^{-1})
[c+u][\pi_{1,\mu}+c-u]
|k\rangle_B.
\label{eqn:proof}
\end{eqnarray}
When we change variable $z \to z^{-1}$ in LHS,
we get RHS.
We will show (\ref{eqn:proof}),
which has good symmetry.\\
$\bullet$~The case of $\mu=1$.~
Using the free field realization of the vertex operators,
we get LHS of (\ref{eqn:proof}) as following.
\begin{eqnarray}
e^{-i\sqrt{\frac{r-1}{r}}Q_{\bar{\epsilon}_1}}
h(z)h(z^{-1})[c-u][c+u]
e^{P_-(z)+P_-(1/z)}|k\rangle_B.
\end{eqnarray}
This is invariant under $z \to z^{-1}$.
Hence LHS and RHS coincide.
\\
%%%%%%%%%%%%%%%%%%%%%%%%%%%%%%%%%%
%%%%%%%%%%%%%%%%%%%%%%%%%%%%%%%%%%
$\bullet$~The case of $\mu=2$.~
In this case the two theta relations 
(\ref{eqn:theta1}) and (\ref{eqn:theta2})
play important roles.
Using the free field realization of the vertex operators,
we get LHS of (\ref{eqn:proof}) for
$\mu=2$ as following.
\begin{eqnarray}
&&
e^{-i\sqrt{\frac{r-1}{r}}Q_{\bar{\epsilon}_2}}c(\pi_{1,2})
zh(z)h(z^{-1})
\Theta_{x^{2r}}(x^{2c}/z)
\Theta_{x^{2r}}(x^{2\pi_{1,2}+2c}z)
\nonumber\\
&\times&
\oint \frac{dw}{w}g_1(w)w
\frac{
\Theta_{x^{2r}}(xw/z)
\Theta_{x^{2r}}(x^{2\pi_{1,2}-1}/zw)}{
D(z,w)}
e^{P_-(z)+P_-(1/z)+R_-^1(w)+R_-^1(1/w)}|k\rangle_B.
\end{eqnarray}
where $c(\pi_{1,2})$ is independent of $w,z$ 
and we have set
\begin{eqnarray}
D(z,w)=
(xzw;x^{2r})_\infty 
(xz/w;x^{2r})_\infty 
(xw/z;x^{2r})_\infty 
(x/wz;x^{2r})\infty.
\end{eqnarray}
The integration contour encircles 
$w=0, x^{1+2rs}z^{\pm 1}, (s \in {\mathbb N})$ 
but not $x^{-1-2rs}z^{\pm 1}, (s \in {\mathbb N})$.
Note that the operator part,
\begin{eqnarray}
e^{P_-(z)+P_-(1/z)+R_-^1(w)+R_-^1(1/w)}|k\rangle_B,\nonumber
\end{eqnarray}
and the function
$D(z,w)$ are invariant under 
$(z,w)\to (1/z,w), (z,1/w), (1/z,1/w)$.
We have LHS-RHS of (\ref{eqn:proof}) for
$\mu=2$ as following.
\begin{eqnarray}
&&
e^{-i\sqrt{\frac{r-1}{r}}Q_{\bar{\epsilon}_2}}c(\pi_{1,2})
h(z)h(z^{-1})\oint \frac{dw}{w}\frac{g_1(w)w}{D(z,w)}
\\
&\times&
\left(
z
\Theta_{x^{2r}}(x^{2c}/z)
\Theta_{x^{2r}}(x^{2\pi_{1,2}+2c}z)
\Theta_{x^{2r}}(xw/z)
\Theta_{x^{2r}}(x^{2\pi_{1,2}-1}/zw)\right.\nonumber\\
&&\left.
-
z^{-1}
\Theta_{x^{2r}}(x^{2c}z)
\Theta_{x^{2r}}(x^{2\pi_{1,2}+2c}/z)
\Theta_{x^{2r}}(xwz)
\Theta_{x^{2r}}(x^{2\pi_{1,2}-1}z/w)
\right)
e^{P_-(z)+P_-(1/z)+R_-^1(w)+R_-^1(1/w)}|k\rangle_B,\nonumber
\end{eqnarray}
where
the integration contour encircles 
$w=0, x^{1+2rs}z^{\pm 1}, (s \in {\mathbb N})$ 
but not $w=x^{-1-2rs}z^{\pm 1}, (s \in {\mathbb N})$.
The contour of integral is invariant under $w \to w^{-1}$.
Hence sufficient condition of
LHS$-$RHS$=0$ becomes
the following weakly sense relation.
\begin{eqnarray}
g_1(w)w&\times&(z 
\Theta_{x^{2r}}(x^{2c}/z)\Theta_{x^{2r}}(x^{2\pi_{1,2}+2c}z)
\Theta_{x^{2r}}(x^{2\pi_{1,2}-1}/zw)\Theta_{x^{2r}}(xw/z)\nonumber\\
&&-
z^{-1}
\Theta_{x^{2r}}(x^{2c}z)
\Theta_{x^{2r}}(x^{2\pi_{1,2}+2c}/z)
\Theta_{x^{2r}}(x^{2\pi_{1,2}-1}z/w)\Theta_{x^{2r}}(xzw))
\sim_{\atop{w}}0.
\end{eqnarray}
Using theta identity (\ref{eqn:theta1}) for $g_1(w)$, we have
the exactly the same relation as (\ref{eqn:theta2})
for $k=\pi_{1,2}$.
We have shown the relation (\ref{eqn:proof}) for $\mu=2$.
\\
%%%%%%%%%%%%%%%%%%%%%%%%%%%%%%%%%
%%%%%%%%%%%%%%%%%%%%%%%%%%%%%%%%%
$\bullet$~~
The case of $\mu=3$.~~
Using the free field realization of the vertex operators,
we get LHS of (\ref{eqn:proof}) for $\mu=3$ as following.
\begin{eqnarray}
&&e^{-i\sqrt{\frac{r-1}{r}}Q_{\bar{\epsilon}_3}}
c(\pi_{1,3},\pi_{2,3})z h(z)h(z^{-1})
\Theta_{x^{2r}}(x^{2c}/z)
\Theta_{x^{2r}}(x^{2\pi_{1,3}+2c}z)
\nonumber\\
&\times&
\oint \frac{dw_1}{w_1}g_1(w_1)
\oint \frac{dw_2}{w_2}g_2(w_2)w_2
\frac{
\Theta_{x^{2r}}(xw_1/z)
\Theta_{x^{2r}}(x^{2\pi_{1,3}-1}/zw_1)
\Theta_{x^{2r}}(xw_1w_2)
\Theta_{x^{2r}}(x^{2\pi_{2,3}-1}w_1/w_2)
}
{
D(z,w_1)D(w_1,w_2)}\nonumber\\
&\times&
e^{P_-(z)+P_-(1/z)+R_-^1(w_1)+R_-^1(1/w_1)+R_-^2(w_2)+R_-^2(1/w_2)}
|k\rangle_B.
\end{eqnarray}
where $c(\pi_{1,3},\pi_{2,3})$ is independent of $w_1,w_2,z$.
The integration contour
encircles 
$w_1=0, x^{1+2rs}z^{\pm 1}, (s \in {\mathbb N})$
but not $w_1=x^{-1-2rs}z^{\pm 1}, (s \in {\mathbb N})$.
The integration contour encircles
$w_2=0, x^{1+2rs}w_1^{\pm 1}, (s \in {\mathbb N})$ 
but not $w_2=x^{-1-2rs}w_1^{\pm 1}, (s \in {\mathbb N})$.
Note that the operator part,
\begin{eqnarray}
e^{P_-(z)+P_-(1/z)+R_-^1(w_1)+R_-^1(1/w_1)+R_-^2(w_2)+
R_-^2(1/w_2)}|k\rangle_B,\nonumber
\end{eqnarray}
and the function
$D(z,w_1)D(w_1,w_2)$ are invariant under 
$$(z,w_1,w_2) \to 
(z^{\pm 1},w_1^{\pm 1},w_2^{\pm 1}),
(z^{\pm 1},w_1^{\pm 1},w_2^{\mp 1}),
(z^{\pm 1},w_1^{\mp 1},w_2^{\pm 1}),
(z^{\pm 1},w_1^{\mp 1},w_2^{\mp 1}).$$
We have LHS-RHS of (\ref{eqn:proof}) for $\mu=3$
as following.
\begin{eqnarray}
&&e^{-i\sqrt{\frac{r-1}{r}}Q_{\bar{\epsilon}_3}}
c(\pi_{1,3},\pi_{2,3})h(z)h(z^{-1})
\oint \oint \frac{dw_1}{w_1}g_1(w_1)
\frac{dw_2}{w_2}g_2(w_2)w_2\nonumber\\
&\times&
\left(
z
\Theta_{x^{2r}}(x^{2c}/z)
\Theta_{x^{2r}}(x^{2\pi_{1,3}+2c}z)
\Theta_{x^{2r}}(xw_1/z)
\Theta_{x^{2r}}(x^{2\pi_{1,3}-1}/zw_1)
\right.\nonumber\\
&&\left.-
z^{-1}
\Theta_{x^{2r}}(x^{2c}z)
\Theta_{x^{2r}}(x^{2\pi_{1,3}+2c}/z)
\Theta_{x^{2r}}(xw_1z)
\Theta_{x^{2r}}(x^{2\pi_{1,3}-1}z/w_1)
\right)\nonumber\\
&\times&
\frac{\Theta_{x^{2r}}(xw_1w_2)
\Theta_{x^{2r}}(x^{2\pi_{2,3}-1}w_1/w_2)}
{D(z,w_1)D(w_1,w_2)}
\nonumber\\
&\times&
e^{P_-(z)+P_-(1/z)+R_-^1(w_1)+R_-^1(1/w_1)+R_-^2(w_2)+R_-^2(1/w_2)}
|k\rangle_B.
\end{eqnarray}
Here the integration contour
encircles 
$w_1=0, x^{1+2rs}z^{\pm 1}, (s \in {\mathbb N})$
but not $w_1=x^{-1-2rs}z^{\pm 1}, (s \in {\mathbb N})$.
The integration contour encircles
$w_2=0, x^{1+2rs}w_1^{\pm 1}, (s \in {\mathbb N})$ 
but not $w_2=x^{-1-2rs}w_1^{\pm 1}, (s \in {\mathbb N})$.
The integration contour is invariant under
$(w_1,w_2)\to (w_1,1/w_2), 
(1/w_1, w_2), (1/w_1,1/w_2)$.
Hence sufficient condition of
LHS$-$RHS$=0$ becomes
the following weakly sense relation.
\begin{eqnarray}
&&g_1(w_1)g_2(w_2)w_2
\Theta_{x^{2r}}(xw_1w_2)
\Theta_{x^{2r}}(x^{2\pi_{2,3}-1}w_1/w_2)\nonumber
\\
&\times&(z 
\Theta_{x^{2r}}(x^{2c}/z)
\Theta_{x^{2r}}(x^{2\pi_{1,3}+2c}z)
\Theta_{x^{2r}}(x^{2\pi_{1,3}-1}/zw_1)\Theta_{x^{2r}}(xw_1/z)\nonumber\\
&&-
z^{-1}
\Theta_{x^{2r}}(x^{2c}z)
\Theta_{x^{2r}}(x^{2\pi_{1,3}+2c}/z)
\Theta_{x^{2r}}(x^{2\pi_{1,3}-1}z/w_1)
\Theta_{x^{2r}}(xzw_1))
\sim_{\atop{(w_1,w_2)}}0.
\end{eqnarray}
Using theta identity (\ref{eqn:theta2}) for $k=\pi_{1,3}$,
the sufficient condition is reduced to
the following.
\begin{eqnarray}
g_1(w_1)g_2(w_2)w_2 
\Theta_{x^{2r}}(x^{2\pi_{1,3}+2c-1}/w_1)
\Theta_{x^{2r}}(x^{2c+1}w_1)
\Theta_{x^{2r}}(xw_1w_2)
\Theta_{x^{2r}}(x^{2\pi_{2,3}-1}w_1/w_2)
\sim_{\atop{(w_1,w_2)}} 0.\nonumber
\end{eqnarray}
Using theta identity (\ref{eqn:theta1}) for $g_1(w)$
the sufficient condition is reduced to
the following.
\begin{eqnarray}
&&g_2(w_2)w_2(
w_1^{-1}
\Theta_{x^{2r}}(x^{2c+2\pi_{1,2}-1}w_1)
\Theta_{x^{2r}}(x^{2\pi_{1,3}+2c-1}/w_1)
\Theta_{x^{2r}}(x^{2\pi_{2,3}-1}w_1/w_2)
\Theta_{x^{2r}}(xw_1w_2)\nonumber\\
&&-
w_1
\Theta_{x^{2r}}(x^{2c+2\pi_{1,2}-1}/w_1)
\Theta_{x^{2r}}(x^{2\pi_{1,3}+2c-1}w_1)
\Theta_{x^{2r}}(x^{2\pi_{2,3}-1}/w_1w_2)
\Theta_{x^{2r}}(xw_2/w_1))\sim_{\atop{w_2}}0.
\end{eqnarray}
Using relation (\ref{eqn:theta1}) for $j=2$,
we get exactly the same relation (\ref{eqn:theta2}) with
substitutions $z \to w_1^{-1}, w\to w_2$,
$c\to c+\pi_{1,2}-\frac{1}{2}$
and $k \to \pi_{2,3}$.
We have shown the relation (\ref{eqn:proof}) for $\mu=3$.
\\
$\bullet$~The case of general $\mu$.~~
Using the free field realization of the vertex operators,
we get LHS of (\ref{eqn:proof}) for $\mu$ as following.
\begin{eqnarray}
&&e^{-i\sqrt{\frac{r-1}{r}}Q_{\bar{\epsilon}_\mu}}
c(\pi_{1,\mu},\pi_{2,\mu},\cdots,\pi_{\mu-1,\mu})
\nonumber\\
&\times&z h(z)h(z^{-1})
\Theta_{x^{2r}}(x^{2c}/z)
\Theta_{x^{2r}}(x^{2\pi_{1,\mu}+2c}z)
\oint \cdots \oint \prod_{j=1}^{\mu-1}
\frac{dw_j}{w_j}
\prod_{j=2}^{\mu-1}
g_j(w_j)w_{\mu-1}\nonumber\\
&\times&
\frac{
\displaystyle
\Theta_{x^{2r}}(xw_1/z)
\Theta_{x^{2r}}(x^{2\pi_{1,\mu}-1}/zw_1)
\prod_{j=2}^{\mu-1}
\Theta_{x^{2r}}(xw_{j-1}w_j)
\Theta_{x^{2r}}(x^{2\pi_{j,\mu}-1}w_{j-1}/w_j)
}
{
\displaystyle
D(z,w_1)\prod_{j=2}^{\mu-1}
D(w_{j-1},w_j)}\nonumber\\
&\times&
e^{P_-(z)+P_-(1/z)+\sum_{j=1}^{\mu-1}(
R_-^1(w_j)+R_-^1(1/w_j))}
|k\rangle_B.
\end{eqnarray}
where $c(\pi_{1,\mu},\cdots,\pi_{\mu-1,\mu})$ 
is independent of $w_1,w_2,\cdots,w_{\mu-1},z$.
Here the integration contour
encircles 
$w_1=0, x^{1+2rs}z^{\pm 1}, (s \in {\mathbb N})$
but not $w_1=x^{-1-2rs}z^{\pm 1}, (s \in {\mathbb N})$.
For $j=1,2,\cdots, \mu-1$, 
the integration contour encircles
$w_{j+1}=0, x^{1+2rs}w_{j}^{\pm 1}, (s \in {\mathbb N})$ 
but not $w_{j+1}=x^{-1-2rs}w_{j}^{\pm 1}, (s \in {\mathbb N})$.
Note that the operator part,
\begin{eqnarray}
e^{P_-(z)+P_-(1/z)+\sum_{j=1}^{\mu-1}
(R_-^j(w_j)+R_-^j(1/w_j))}|k\rangle_B,\nonumber
\end{eqnarray}
and the function
$D(z,w_1)\prod_{j=2}^{\mu-1}
D(w_{j-1},w_j)$ are invariant under 
$z\to z^{-1}$ or $w_j\to w_j^{-1}, (j=1,2,\cdots,\mu-1)$.
Hence we have
LHS-RHS of (\ref{eqn:proof}) for general $\mu$
as following.
\begin{eqnarray}
&&e^{-i\sqrt{\frac{r-1}{r}}Q_{\bar{\epsilon}_\mu}}
c(\pi_{1,\mu},\pi_{2,\mu},\cdots,\pi_{\mu-1,\mu})
h(z)h(z^{-1})
\oint \cdots \oint \prod_{j=1}^{\mu-1}
\frac{dw_j}{w_j}
\prod_{j=2}^{\mu-1}
g_j(w_j)w_{\mu-1}\nonumber\\
&\times&
\left(
z
\Theta_{x^{2r}}(x^{2c}/z)
\Theta_{x^{2r}}(x^{2\pi_{1,\mu}+2c}z)
\Theta_{x^{2r}}(xw_1/z)
\Theta_{x^{2r}}(x^{2\pi_{1,\mu}-1}/zw_1)
\right.\nonumber\\
&&\left.-
z^{-1}
\Theta_{x^{2r}}(x^{2c}z)
\Theta_{x^{2r}}(x^{2\pi_{1,\mu}+2c}/z)
\Theta_{x^{2r}}(xw_1z)
\Theta_{x^{2r}}(x^{2\pi_{1,\mu}-1}z/w_1)
\right)
\nonumber\\
&\times&\frac{
\displaystyle
\prod_{j=2}^{\mu-1}
\Theta_{x^{2r}}(xw_{j-1}w_j)
\Theta_{x^{2r}}(x^{2\pi_{j,\mu}-1}w_{j-1}/w_j)
}
{
\displaystyle
D(z,w_1)\prod_{j=2}^{\mu-1}
D(w_{j-1},w_j)}
~e^{P_-(z)+P_-(1/z)+\sum_{j=1}^{\mu-1}(
R_-^1(w_j)+R_-^1(1/w_j))}
|k\rangle_B.
\end{eqnarray}
Here the integration contour
encircles 
$w_1=0, x^{1+2rs}z^{\pm 1}, (s \in {\mathbb N})$
but not $w_1=x^{-1-2rs}z^{\pm 1}, (s \in {\mathbb N})$.
For $j=1,2,\cdots, \mu-1$, 
the contour of integral encircles
$w_{j+1}=0, x^{1+2rs}w_{j}^{\pm 1}, (s \in {\mathbb N})$ 
but not $w_{j+1}=x^{-1-2rs}w_{j}^{\pm 1}, (s \in {\mathbb N})$.
The contour of integral
is invariant under $w_j \to w_j^{-1}, (j=1,2,\cdots,\mu-1)$.
Hence sufficient condition of
LHS$-$RHS$=0$ becomes
the following weakly sense relation.
\begin{eqnarray}
&&\prod_{j=1}^{\mu-1}g_j(w_j)w_{\mu-1}
\prod_{j=2}^{\mu-1}
\Theta_{x^{2r}}(xw_{j-1}w_j)
\Theta_{x^{2r}}(x^{\pi_{j,\mu}-1}w_{j-1}/w_j)
\nonumber
\\
&\times&(z 
\Theta_{x^{2r}}(x^{2c}/z)
\Theta_{x^{2r}}(x^{2\pi_{1,\mu}+2c}z)
\Theta_{x^{2r}}(x^{2\pi_{1,\mu}-1}/zw_1)
\Theta_{x^{2r}}(xw_1/z)\label{eqn:proof1}
\\
&&-
z^{-1}
\Theta_{x^{2r}}(x^{2c}z)
\Theta_{x^{2r}}(x^{2\pi_{1,\mu}+2c}/z)
\Theta_{x^{2r}}(x^{2\pi_{1,\mu}-1}z/w_1)
\Theta_{x^{2r}}(xzw_1))
\sim_{\atop{(w_1,w_2,\cdots,w_{\mu-1})}}0.\nonumber
\end{eqnarray}
The next step is to show theta identity (\ref{eqn:proof1})
of $(\mu-1)$ variables $w_1,w_2,\cdots w_{\mu-1}$.
We show it by induction.
Using theta identity (\ref{eqn:theta1}) and
(\ref{eqn:theta2}),
the relation (\ref{eqn:proof1}) is reduced to
the weakly sense relation (\ref{eqn:proof3}) for $\mu=2$.
Using theta identity (\ref{eqn:theta1}) and
(\ref{eqn:theta2}) repeatedly,
the relation (\ref{eqn:proof3}) for $\mu=2$ 
is reduced to
the weakly sense relation (\ref{eqn:proof3}) 
for $2 \leq \nu \leq \mu-1$.
\begin{eqnarray}
&&\prod_{j=\nu}^{\mu-1}g_j(w_j)w_{\mu-1}
\prod_{j=\nu+1}^{\mu-1}
\Theta_{x^{2r}}(xw_{j-1}w_j)
\Theta_{x^{2r}}(x^{\pi_{j,\mu-1}-1}w_{j-1}/w_j)
\nonumber
\\
&\times&(w_{\nu-1}^{-1} 
\Theta_{x^{2r}}(x^{2\pi_{1,\nu}+2c-\nu+1}w_{\nu-1})
\Theta_{x^{2r}}(x^{2\pi_{1,\mu}+2c-\nu+1}/w_{\nu-1})
\Theta_{x^{2r}}(x^{2\pi_{\nu,\mu}-1}w_{\nu-1}/w_\nu)
\Theta_{x^{2r}}(xw_{\nu-1}w_{\nu})
\nonumber
\\
&&-
w_{\nu-1}
\Theta_{x^{2r}}(x^{2\pi_{1,\nu}+2c-\nu+1}/w_{\nu-1})
\Theta_{x^{2r}}(x^{2\pi_{1,\mu}+2c-\nu+1}w_{\nu-1})
\Theta_{x^{2r}}(x^{2\pi_{\nu,\mu}-1}/w_{\nu-1}w_\nu)
\Theta_{x^{2r}}(xw_{\nu}/w_{\nu-1})
)\nonumber\\
&&\sim_{\atop{(w_{\nu-1},w_\nu, \cdots,w_{\mu-1})}}0,
~~~~~(2\leq \nu \leq \mu-1).
\label{eqn:proof3}
\end{eqnarray}
The relation (\ref{eqn:proof3}) for $\nu$
is reduced to those for $\nu+1$.
The relation (\ref{eqn:proof3}) for $\nu=\mu-1$
is the same as (\ref{eqn:theta2}).
Now we have shown the theta identity
(\ref{eqn:proof1}).
Hence we have shown the identity 
(\ref{eqn:proof}) for general $\mu$.
\\
Q.E.D.
\\~\\~\\
We give a comment on a proof of the dual boundary state,
\begin{eqnarray}
~_B\langle k|T_B^{(k)}(z)=~_B\langle k|.
\end{eqnarray}
Multiplying the vertex operators 
$\Phi^{*(k,k+\bar{\epsilon}_\mu)}(z)$ from
the right, and using the inversion relation of vertex
operators (\ref{eqn:inversion1}), we get a necessary and sufficient 
condition for $1\leq \mu \leq N-1$,
\begin{eqnarray}
&&
~_B\langle k|
z^{\frac{r-1}{2r}\frac{N-1}{N}-\frac{1}{r}\pi_1}
h(z)[c-u][\pi_{1,\mu}+c+u]
\Phi_\mu^*(z^{-1})
\nonumber\\
&&=
~_B\langle k|
z^{-\frac{r-1}{2r}\frac{N-1}{N}+\frac{1}{r}\pi_1}h(z^{-1})
[c+u][\pi_{1,\mu}+c-u]
\Phi_\mu^{*}(z).
\end{eqnarray}
As the same manner the above we can give
a proof for the dual boundary state.
We omit details.
Here we note two useful relations of the dual boundary state.
\begin{prop}
\begin{eqnarray}
g_j^*(1/w)w 
\Theta_{x^{2r}}(x^{2c-N+j+2\pi_{1,j}}/w)
\Theta_{x^{2r}}(x^{2c+2-N+j+2\pi_{1,j+1}}w)
\sim_{\atop{w}}0,~~(1\leq j \leq N-1).
\end{eqnarray}
\end{prop}

\begin{prop}
\begin{eqnarray}
\frac{h(z)h^*(z)}{h(1/z)h^*(1/z)}=
\frac{\Theta_{x^{2r}}(x^{2c}z)
\Theta_{x^{2r}}(x^{2c+2\pi_{1,N}}/z)}{
\Theta_{x^{2r}}(x^{2c}/z)
\Theta_{x^{2r}}(x^{2c+2\pi_{1,N}}z)},
\end{eqnarray}
\end{prop}
The relation between the functions $h(z)$ and $h^*(z)$
gives self consistency between the boundary state and the
dual boundary state.

\section{Norm}

In this section we calculate the norm
of the boundary state,
\begin{eqnarray}
~_B\langle k | k\rangle_B.\nonumber
\end{eqnarray}
The free field realization of $e^F$ and $e^G$
are quadratic. 
Hence evaluation of the norm is reduced to 
the Gaussian integrals, which is possible to calculate.
Let us set
\begin{eqnarray}
&&\{z\}_\infty=(z;x^{4N},x^{2r})_\infty,~~
\{z\}_\infty^*=(z;x^{4N},x^{2r^*})_\infty,\nonumber\\
&&\left[z\right]_\infty=
(z;x^{2N},x^{2r})_\infty,~~
\left[z\right]_\infty^*
=(z;x^{2N},x^{2r^*})_\infty.\nonumber\\
\end{eqnarray}

\begin{thm}~~~The norm $~_B\langle k|k \rangle_B$
is evaluated by double infinite product.
\begin{eqnarray}
~_B\langle k|k \rangle_B
&=&\frac{1}
{(x^{2N};x^{2N})^{\frac{N(N-1)}{2}}}\times
\prod_{i=1}^{N-1}
\left(\frac{\sqrt{(x^{4N-2-2i};x^{4N})_\infty 
(x^{4N+2-2i};x^{4N})_\infty}}{
(x^{4N-2i};x^{4N})_\infty}\right)^{i(N-i)}\nonumber\\
&\times&
\prod_{i=1}^{N-1}
\left(
\frac{\{x^{4N-2-2i}\}_\infty^* 
\{x^{4N+2-2i}\}_\infty}{
\{x^{4N+2-2i}\}_\infty^* 
\{x^{4N-2-2i}\}_\infty}
\right)^{i(N-i)}
\prod_{i=1}^N
\prod_{j=1}^{N-i}
\frac{
\left[
x^{-2+2i+2j}
\right]^*_\infty}
{
\left[x^{-2+2i+2j}\right]_\infty}
\prod_{i=1}^N
\prod_{k=1}^{i-1}
\frac{
\left[x^{2N-2i+2j}\right]_\infty^*}{
\left[x^{2N-2i+2j}\right]_\infty}
\nonumber\\
&\times&
\prod_{1\leq i<j \leq N}
\frac{
\left[x^{4c+2\pi_{1,i}+2\pi_{1,j}}\right]_\infty^*
\left[x^{2r-4c+2N-2-2\pi_{1,i}-2\pi_{1,j}}\right]_\infty^*
\left[x^{2N+2\pi_{j,i}}\right]_\infty 
\left[x^{2N+2\pi_{i,j}}\right]_\infty
}{
\left[x^{4c+2+2\pi_{1,i}+2\pi_{1,j}}\right]_\infty
\left[x^{2r-4c+2N-2-2\pi_{1,i}-2\pi_{1,j}}\right]_\infty
\left[x^{2N+2\pi_{j,i}}\right]_\infty^*
\left[x^{2N+2\pi_{i,j}}\right]_\infty^*
}\nonumber\\
&\times&
\prod_{i=1}^N
\prod_{j=1}^{N-i}
\sqrt{\frac{
\left[x^{-2r+4c+2+2j+4\pi_{1,i}}\right]_\infty 
\left[x^{2r-4c+2N-2+2j-4\pi_{1,i}}\right]_\infty
}
{
\left[x^{-2r+4c+2+2j+4\pi_{1,i}}\right]_\infty^* 
\left[x^{2r-4c+2N-2+2j-4\pi_{1,i}}\right]_\infty^*}}
\nonumber\\
&\times&
\prod_{i=1}^N
\prod_{j=1}^{i-1}
\sqrt{
\frac{
\left[x^{-2r+4c+2+2j+4\pi_{1,i}}\right]_\infty
\left[x^{2r-4c+2N-2+2j-4\pi_{1,i}}\right]_\infty
}
{
\left[x^{-2r+4c+2+2j+4\pi_{1,i}}\right]_\infty^*
\left[x^{2r-4c+2N-2+2j-4\pi_{1,i}}\right]_\infty^*
}
}\nonumber
\\
&\times&
\prod_{i=1}^{N-1}
\prod_{j=1}^i
\prod_{k=1}^{N-i}
\sqrt{\frac{
\{x^{4N+2r-4c-2+2i-2j-2k-4\pi_{1,i}}\}_\infty 
\{x^{2N-2r+4c+2+2i-2j-2k+4\pi_{1,i}}\}_\infty
}{
\{x^{4N+2r-4c+2+2i-2j-2k-4\pi_{1,i+1}}\}_\infty 
\{x^{2N-2r+4c+6+2i-2j-2k+4\pi_{1,i}}\}_\infty
}}
\nonumber
\\
&&~~~~~~~~~~\times
\sqrt{\frac{
\{x^{4N-2r+4c+2-2i-2j-2k+4\pi_{1,i+1}}\}_\infty 
\{x^{6N+2r-4c-2-2i-2j-2k-4\pi_{1,i+1}}\}_\infty
}{
\{x^{4N-2r+4c+6-2i-2j-2k+4\pi_{1,i}}\}_\infty 
\{x^{6N+2r-4c+2-2i-2j-2k-4\pi_{1,i}}\}_\infty
}}\nonumber\\
&&~~~~~~~~~~\times
\sqrt{\frac{
\{x^{4N+2r-4c+2+2i-2j-2k-4\pi_{1,i+1}}\}_\infty^* 
\{x^{2N-2r+4c+6+2i-2j-2k+4\pi_{1,i}}\}_\infty^*}
{
\{x^{4N+2r-4c-2+2i-2j-2k-4\pi_{1,i}}\}_\infty^*
\{x^{2N-2r+4c+2+2i-2j-2k+4\pi_{1,i}}\}_\infty^*
}}
\nonumber
\\
&&~~~~~~~~~~\times
\sqrt{\frac{
\{x^{4N-2r+4c+6-2i-2j-2k+4\pi_{1,i}}\}_\infty^* 
\{x^{6N+2r-4c+2-2i-2j-2k-4\pi_{1,i}}\}_\infty^*
}
{
\{x^{4N-2r+4c+2-2i-2j-2k+4\pi_{1,i+1}}\}_\infty^*
\{x^{6N+2r-4c-2-2i-2j-2k-4\pi_{1,i+1}}\}_\infty^*
}}.\nonumber
\\
\end{eqnarray}
\end{thm}

In what follows 
we explain how to 
calculate the norm $~_B\langle k|k\rangle_B$.
We calculate it by using a decomposition of the identity
on ${\cal F}_{l,k}$ which employs coherent state. 
We define the coherent state
\begin{eqnarray}
&&
|\xi_1 \cdots \xi_{N-1}\rangle=
\exp\left(
\sum_{m>0}\sum_{j=1}^{N-1}
\frac{1}{m}\frac{[rm]_x}{[(r-1)m]_x}
\xi_j(m)\alpha_{-m}^j
\right)
|l,k\rangle,\\
&&
\langle \bar{\xi}_1 \cdots \bar{\xi}_{N-1}|
=\langle l,k |\exp\left(
\sum_{m>0}\sum_{j=1}^{N-1}
\frac{1}{m}\frac{[rm]_x}{[(r-1)m]_x}
\bar{\xi}_j(m)
\alpha_m^j \right).
\end{eqnarray}
The actions of the boson $\alpha_m^i$ are
given by
\begin{eqnarray}
&&\alpha_m^i |\xi_1 \cdots \xi_{N-1}\rangle=
\sum_{j=1}^{N^-1}\frac{[A_{i,j}m]_x}{[m]_x}\xi_j(m)
|\xi_1 \cdots \xi_{N-1}\rangle,\\
&&\langle \bar{\xi}_1 \cdots \bar{\xi}_{N-1}|
\alpha_{-m}^i=
\langle \bar{\xi}_1 \cdots \bar{\xi}_{N-1}|
\sum_{j=1}^{N-1}\frac{[A_{i,j}m]_x}{[m]_x}\bar{\xi}_j(m).
\end{eqnarray}
The action of the bosons $\beta_m^i$ on the coherent 
state are given by
\begin{eqnarray}
&&\beta_m^1 |\xi_1 \cdots \xi_{N-1}\rangle=
\xi_1(m)|\xi_1 \cdots \xi_{N-1}\rangle,\\
&&\beta_m^i |\xi_1 \cdots \xi_{N-1}\rangle=
(
-x^{im}\xi_{i-1}(m)
+x^{(i-1)m}\xi_i(m))|\xi_1 \cdots \xi_{N-1}\rangle,(2\leq i
\leq N-1),\\
&&\langle \bar{\xi}_1 \cdots \bar{\xi}_{N-1}|
\beta_{-m}^1=\bar{\xi}_1(m)
\langle \bar{\xi}_1 \cdots \bar{\xi}_{N-1}|,\\
&&\langle \bar{\xi}_1 \cdots \bar{\xi}_{N-1}|
\beta_{-m}^i=(x^{(-i+1)m}\bar{\xi}_i(m)
-x^{-im}\bar{\xi}_{i-1}(m))
\langle \bar{\xi}_1 \cdots \bar{\xi}_{N-1}|,(2\leq i \leq N-1).
\nonumber\\
\end{eqnarray}
Using the following Gaussian integral,
\begin{eqnarray}
&&\int_{-\infty}^\infty
\prod_{m>0} 
\mu_m d\xi_m d\bar{\xi}_m\exp\left(-\frac{1}{2}
\sum_{m>0}\mu_m
(\bar{\xi}_m, \xi_m){\cal A}_m \left(\begin{array}{c}
\bar{\xi}_m\\
\xi_m \end{array}\right)+\sum_{m>0}(\bar{\xi}_m,\xi_m)
{\cal B}_m\right)\nonumber\\
&=&\frac{1}{\prod_{m>0}\sqrt{-\det({\cal A}_m)}}
\exp\left(\frac{1}{2}\sum_{m>0}
\frac{1}{\mu_m}
~^t{\cal B}_m{\cal A}_m^{-1}{\cal B}_m
\right),\label{eqn:gaussian}
\end{eqnarray}
we conclude that 
the identity on ${\cal F}_{l,k}$ is decomposed by following.
\begin{eqnarray}
id&=&\int_{\infty}^\infty
 \prod_{m>0}\prod_{j=1}^{N-1}
\frac{1}{m}\frac{[(j+1)m]_x [rm]_x}{[m]_x [(r-1)m]_x}
d\xi_j(m)d\bar{\xi}_j(m)\nonumber\\
&\times&
\exp\left(-\sum_{m>0}\sum_{j,k=1}^{N-1}
\frac{1}{m}\frac{[A_{j,k}m]_x [rm]_x}{[m]_x [(r-1)m]_x}
\xi_j(m)\bar{\xi}_k(m)\right)
|\xi_1 \cdots \xi_{N-1}\rangle \langle \bar{\xi}_1
\cdots \bar{\xi}_{N-1}|.
\end{eqnarray}
Inserting this decomposition of the identity
between $e^G$ and $e^F$,
we find
\begin{eqnarray}
&&~_B\langle k|k\rangle_B
=\langle k,k|e^G e^F |k,k\rangle
\nonumber\\
&=&\int_{\infty}^\infty
 \prod_{m>0}\prod_{j=1}^{N-1}
\frac{1}{m}\frac{[(j+1)m]_x [rm]_x}{[m]_x [(r-1)m]_x}
d\xi_j(m)d\bar{\xi}_j(m)\nonumber\\
&\times&
\exp\left(-\frac{1}{2}\sum_{m>0}
\sum_{j,k=1}^{N-1}\frac{1}{m}
\frac{[rm]_x[A_{j,k}m]_x}{[(r-1)m]_x[m]_x}
(2\xi_j(m)\bar{\xi}_k(m)-x^{2Nm}\xi_j(m)\xi_k(m)-\bar{\xi}_j(m)\bar{\xi}_k(m)
)\right.\nonumber\\
&&~~~~~~+\left.
\sum_{m>0}
\sum_{j=1}^{N-1}\frac{1}{m}(Y_j(m)\xi_j(m)+X_j(m)\bar{\xi}_j(m))
\right),
\end{eqnarray}
where we have set
\begin{eqnarray}
Y_j(m)&=&x^{jm}
(x^{-m}E_j(m)-x^mE_{j+1}(m)))\nonumber\\
&=&
\frac{x^{Nm}}{[(r-1)m]_x}([(r-2c+N-j-2-2\pi_{1,j+1})m]_x
+[(-r+2c-N+j+2\pi_{1,j})m]_x)
\nonumber\\
&+&
\theta_m\left(\frac{x^{Nm}[m/2]_x[rm/2]_x^+}{[(r-1)m/2]}\right),
\label{def:Y}
\\
X_j(m)&=&x^{-jm}
(-x^{-m}D_{j+1}(m)+x^m D_j(m))\nonumber
\\
&=&
\frac{-1}{[(r-1)m]_x}
([(-r+2c-j+2\pi_{1,j+1})m]_x+[(r-2c+j-2-2\pi_{1,j})m]_x)
\nonumber\\
&-&\theta_m\left(\frac{[m/2]_x[rm/2]_x^+}{[(r-1)m/2]_x}\right).
\label{def:X}
\end{eqnarray}
Here we read $D_{N}(m)=E_{N}(m)=0$.\\
Let us repeat Gaussian integral
(\ref{eqn:gaussian}),
we get formulae without integrals in proposition.

\begin{prop}~~~
The norm of boundary states is written by
infinite product.
\begin{eqnarray}
&&~_B\langle k|k \rangle_B
\nonumber\\
&=&\frac{1}{(x^{2N};x^{2N})_\infty^{\frac{N-1}{2}}}
\exp\left(\sum_{m=1}^\infty
\frac{1}{m}\frac{1}{x^{2Nm}-1}\frac{[(r-1)m]_x}{[rm]_x}
\frac{1}{[Nm]_x[m]_x}
\right. \nonumber\\
&\times&
\sum_{j=1}^{N-1}[jm]_x[(N-j)m]_x
\left(\frac{x^{2Nm}}{2}X_j(m)^2+\frac{1}{2}Y_j(m)^2
-X_j(m)Y_j(m)\right)\\
&+&\left.
\sum_{1\leq j<l \leq N-1}
[jm]_x[(N-l)m]_x
(x^{2Nm}X_j(m)X_l(m)+Y_j(m)Y_l(m)-X_j(m)Y_l(m)-Y_j(m)X_l(m))
\right).\nonumber
\end{eqnarray}
Here $X_j(m)$ and $Y_j(m)$ are given by
(\ref{def:X}) and (\ref{def:Y}).
\end{prop}

At first glance, the norm $~_B\langle k| k\rangle_B$
is evaluated by
$$
(z;x^{2N},x^{2N},x^{2r},x^{2r^*})_\infty.
$$
After some calculations,
cancellations occur.
In order to write the norm $~_B\langle k |k \rangle_B$, 
we need only
$$
(z;x^{2N},x^{2r})_\infty,~~~(z;x^{2N},x^{2r^*})_\infty.
$$
Here we omit detailed calculations.

\subsection*{acknowledgments}~~
The author is supported by the Grant-in Aid
for Scientific Research {\bf C}(21540228) 
from Japan Society for
Promotion of Science.

\appendix

\section{Some formulae}

In this appendix we summarize some formulae.

\begin{eqnarray}
{\rm log}~h(z)&=&
-\sum_{m>0}\frac{1}{2m}\frac{
[(r-1)m]_x[(N-1)m]_x}{[rm]_x[Nm]_x}z^{-2m}\nonumber\\
&&-
\sum_{m>0}\frac{1}{m}
\frac{[(r-1)m]_x}{[rm]_x[N m]_x}
\left([(N-1)m]_x D_1(m)-[m]_x x^{-Nm}\sum_{k=2}^{N-1}
D_k(m)
\right)z^{-m},\nonumber\\
{\rm log}~g_j(z)&=&
-\sum_{m>0}
\frac{1}{2m}\frac{[(r-1)m]_x (x^m+x^{-m})}{[rm]_x}z^{-2m}
\nonumber\\
&&+\sum_{m>0}
\frac{1}{m}\frac{[(r-1)m]_x}{[rm]_x}
x^{-jm}(x^m D_j(m)-x^{-m}D_{j+1}(m))z^{-m},
~~~(j=1,2,\cdots,N-2),\nonumber\\
{\rm log}~g_{N-1}(z)&=&
-\sum_{m>0}
\frac{1}{2m}\frac{[(r-1)m]_x (x^m+x^{-m})}{[rm]_x}z^{-2m}
\nonumber\\
&&+\sum_{m>0}\frac{1}{m}
\frac{[(r-1)m]_x}{[rm]_x}
x^{(-N+2)m}D_{N-1}(m)z^{-m}.
\nonumber
\end{eqnarray}

\begin{eqnarray}
{\rm log}~h^*(z)&=&
-\sum_{m>0}\frac{1}{2m}\frac{
[(r-1)m]_x[(N-1)m]_x}{[rm]_x[Nm]_x}z^{-2m}\nonumber\\
&&+
\sum_{m>0}\frac{1}{m}
\frac{[(r-1)m]_x [m]_x x^{-Nm}}{[rm]_x[N m]_x}
\left(\sum_{k=1}^{N-1}
E_k(m)
\right)z^{-m},\nonumber\\
{\rm log}~g_j^*(z)&=&
-\sum_{m>0}
\frac{1}{2m}\frac{[(r-1)m]_x (x^m+x^{-m})}{[rm]_x}z^{-2m}
\nonumber\\
&&+\sum_{m>0}
\frac{1}{m}\frac{[(r-1)m]_x}{[rm]_x}
x^{(j-N)m}(x^m E_{j+1}(m)-x^{-m}E_{j}(m))z^{-m},
~~~(j=1,2,\cdots,N-2),\nonumber\\
{\rm log}~g_{N-1}^*(z)&=&
-\sum_{m>0}
\frac{1}{2m}\frac{[(r-1)m]_x (x^m+x^{-m})}{[rm]_x}z^{-2m}
\nonumber\\
&&-\sum_{m>0}\frac{1}{m}
\frac{[(r-1)m]_x}{[rm]_x}
x^{-2m}E_{N-1}(m)z^{-m}.
\nonumber
\end{eqnarray}

\section{Physical Interpretation}

In this appendix we explain
a physical interpretation of
our problem.
In main text of this paper we use no picture.
In this appendix we give
graphical definition of various quantities of our problem.
We present an ordered pair $(b,a)\in P^2$ as FIG.1.
%%%%%%%%%%%%%%%%%%%%%%%%%%%%%%%%%%%%
\begin{center}
\unitlength 0.1in
\begin{picture}( 16.0000,  0.9000)( 10.6500, -8.0500)
% VECTOR 2 0 3 0
% 2 2600 800 2000 800
% 
\special{pn 8}%
\special{pa 2600 800}%
\special{pa 2000 800}%
\special{fp}%
\special{sh 1}%
\special{pa 2000 800}%
\special{pa 2068 820}%
\special{pa 2054 800}%
\special{pa 2068 780}%
\special{pa 2000 800}%
\special{fp}%
% LINE 2 0 3 0
% 2 2000 800 1400 800
% 
\special{pn 8}%
\special{pa 2000 800}%
\special{pa 1400 800}%
\special{fp}%
% STR 2 0 3 0
% 3 2800 700 2800 800 5 0
% $a$
\put(28.0000,-8.0000){\makebox(0,0){$a$}}%
% STR 2 0 3 0
% 3 1200 700 1200 800 5 0
% $b$
\put(12.0000,-8.0000){\makebox(0,0){$b$}}%
\end{picture}%
~\\
~\\
~~~~~FIG.1.
Ordered pair\\~\\
\end{center}
%%%%%%%%%%%%%%%%%%%%%%%%%%%
For admissible pair $(a,g,h,f) \in P^4$
we present the Boltzmann weight functions
$W\left(\left.\begin{array}{cc}
h&f\\
g&a
\end{array}\right|u\right)$ by FIG.2.
%%%%%%%%%%%%%%%%%%%%%%%%%%%
\begin{center}
%WinTpicVersion3.08
\unitlength 0.1in
\begin{picture}( 16.0000, 16.0000)( 10.6500,-19.1500)
% VECTOR 2 0 3 0
% 8 2600 600 2000 600 1400 1800 1400 1200 2600 1800 2000 1800 2600 1800 2600 1200
% 
\special{pn 8}%
\special{pa 2600 600}%
\special{pa 2000 600}%
\special{fp}%
\special{sh 1}%
\special{pa 2000 600}%
\special{pa 2068 620}%
\special{pa 2054 600}%
\special{pa 2068 580}%
\special{pa 2000 600}%
\special{fp}%
\special{pa 1400 1800}%
\special{pa 1400 1200}%
\special{fp}%
\special{sh 1}%
\special{pa 1400 1200}%
\special{pa 1380 1268}%
\special{pa 1400 1254}%
\special{pa 1420 1268}%
\special{pa 1400 1200}%
\special{fp}%
\special{pa 2600 1800}%
\special{pa 2000 1800}%
\special{fp}%
\special{sh 1}%
\special{pa 2000 1800}%
\special{pa 2068 1820}%
\special{pa 2054 1800}%
\special{pa 2068 1780}%
\special{pa 2000 1800}%
\special{fp}%
\special{pa 2600 1800}%
\special{pa 2600 1200}%
\special{fp}%
\special{sh 1}%
\special{pa 2600 1200}%
\special{pa 2580 1268}%
\special{pa 2600 1254}%
\special{pa 2620 1268}%
\special{pa 2600 1200}%
\special{fp}%
% LINE 2 0 3 0
% 14 2600 1200 2600 600 2000 600 1400 600 1400 600 1400 600 1400 600 1400 600 1400 1200 1400 1200 1400 1800 2000 1800 1400 1200 1400 600
% 
\special{pn 8}%
\special{pa 2600 1200}%
\special{pa 2600 600}%
\special{fp}%
\special{pa 2000 600}%
\special{pa 1400 600}%
\special{fp}%
\special{pa 1400 600}%
\special{pa 1400 600}%
\special{fp}%
\special{pa 1400 600}%
\special{pa 1400 600}%
\special{fp}%
\special{pa 1400 1200}%
\special{pa 1400 1200}%
\special{fp}%
\special{pa 1400 1800}%
\special{pa 2000 1800}%
\special{fp}%
\special{pa 1400 1200}%
\special{pa 1400 600}%
\special{fp}%
% STR 2 0 3 0
% 3 2800 1900 2800 2000 5 0
% $a$
\put(28.0000,-20.0000){\makebox(0,0){$a$}}%
% STR 2 0 3 0
% 3 2800 300 2800 400 5 0
% $f$
\put(28.0000,-4.0000){\makebox(0,0){$f$}}%
% STR 2 0 3 0
% 3 1200 1900 1200 2000 5 0
% $g$
\put(12.0000,-20.0000){\makebox(0,0){$g$}}%
% STR 2 0 3 0
% 3 1200 300 1200 400 5 0
% $h$
\put(12.0000,-4.0000){\makebox(0,0){$h$}}%
% STR 2 0 3 0
% 3 2000 1110 2000 1210 5 0
% $u$
\put(20.0000,-12.1000){\makebox(0,0){$u$}}%
\end{picture}%
~\\
~\\
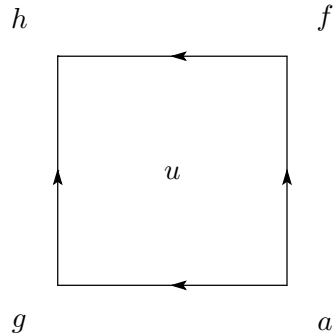
~~~~~FIG.2.~Boltzmann weight\\~\\
\end{center}
%%%%%%%%%%%%%%%%%%%%%%%%%%%%%%%%%%
In what follows we consider 
the restricted path $a\in P_{r-N}^+$. 
\\
For admissible pair
$(a,b,g)\in P^3$
we present boundary Boltzmann weight function 
$K\left(\left.\begin{array}{cc}
&a\\
g&\\
&b
\end{array}\right|u\right)$ by FIG.3.
\begin{center}
%WinTpicVersion3.08
\unitlength 0.1in
\begin{picture}( 12.0000, 16.8500)(  8.6500,-20.1000)
% VECTOR 2 0 3 0
% 2 2000 410 1600 810
% 
\special{pn 8}%
\special{pa 2000 410}%
\special{pa 1600 810}%
\special{fp}%
\special{sh 1}%
\special{pa 1600 810}%
\special{pa 1662 778}%
\special{pa 1638 772}%
\special{pa 1634 750}%
\special{pa 1600 810}%
\special{fp}%
% LINE 2 0 3 0
% 2 1600 810 1200 1210
% 
\special{pn 8}%
\special{pa 1600 810}%
\special{pa 1200 1210}%
\special{fp}%
% VECTOR 2 0 3 0
% 2 2000 2010 1600 1610
% 
\special{pn 8}%
\special{pa 2000 2010}%
\special{pa 1600 1610}%
\special{fp}%
\special{sh 1}%
\special{pa 1600 1610}%
\special{pa 1634 1672}%
\special{pa 1638 1648}%
\special{pa 1662 1644}%
\special{pa 1600 1610}%
\special{fp}%
% LINE 2 0 3 0
% 2 1600 1610 1200 1210
% 
\special{pn 8}%
\special{pa 1600 1610}%
\special{pa 1200 1210}%
\special{fp}%
% LINE 2 0 3 0
% 2 2000 2010 2000 410
% 
\special{pn 8}%
\special{pa 2000 2010}%
\special{pa 2000 410}%
\special{fp}%
% STR 2 0 3 0
% 3 1600 1110 1600 1210 5 0
% $u$
\put(16.0000,-12.1000){\makebox(0,0){$u$}}%
% STR 2 0 3 0
% 3 2200 310 2200 410 5 0
% $a$
\put(22.0000,-4.1000){\makebox(0,0){$a$}}%
% STR 2 0 3 0
% 3 2200 1910 2200 2010 5 0
% $b$
\put(22.0000,-20.1000){\makebox(0,0){$b$}}%
% STR 2 0 3 0
% 3 1000 1110 1000 1210 5 0
% $g$
\put(10.0000,-12.1000){\makebox(0,0){$g$}}%
\end{picture}%
~\\
~\\
FIG.3.~~Boundary Boltzmann weight\\~\\
\end{center}
%%%%%%%%%%%%%%%%%%%%%%%%%%%%%%%%%%%%%%%%%%%%%%%
In what follows 
we consider $a \in P_{r-N}^+$.
Using the Boltzmann weight functions
$W$ and the boundary Boltzmann weight functions $K$,
we introduce two dimensional solvable lattice model,
which we call the boundary $U_{q,p}(A_{N-1}^{(1)})$ face model.
We fix parameters $N=2,3,4,\cdots$,
$0<x<1$ and $r \geq N+2, (r \in {\mathbb N})$.
We fix a continuous parameter $c$ in the boundary Boltzmann weight
function as $0<c<1$.
In what follows we consider infinite product
of the Boltzmann weight functions $W$
and the boundary Boltzmann weight functions $K$
in the Regime III : $0<u<1$.
The boundary $U_{q,p}(A_{N-1}^{(1)})$ face model
defined by FIG.4.
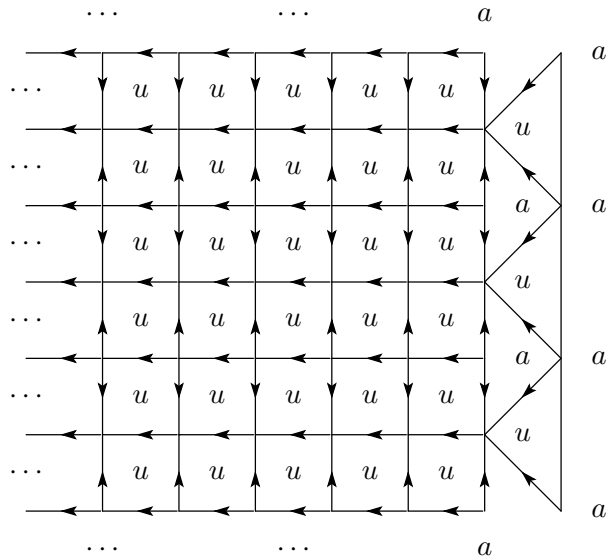
\begin{center}
~\\

%WinTpicVersion3.08
\unitlength 0.1in
%
~\\
~\\
FIG.4. Boundary face model

~\\
\end{center}
In order to consider the boundary $U_{q,p}(A_{N-1}^{(1)})$ 
face model, we divide it into some parts.
For this purpose we prepare the space ${\cal H}_{l,k}$,
on which the operators act.
Following the general scheme 
of algebraic approach
in solvable lattice models,
we consider the corner transfer matrix
$A^{(a)}(u)$ which represent north-west quadrant
as FIG.5.
\begin{center}
~\\

%WinTpicVersion3.08
\unitlength 0.1in
%
%
~\\
~\\
FIG.5.~Corner transfer matrix

~\\
\end{center}
In the large lattice limit $M \to \infty$, 
apart from a divergence scalar,
the corner transfer matrix $A^{(a)}(u)$,
$(a \in P_{r-N}^+)$ is of the form
\begin{eqnarray}
A^{(a)}(u) \sim x^{2u H},\nonumber
\end{eqnarray}
where the operator $H$ is independent of $u$.
The spectrum of $H$ is given in
\cite{JMO2}.
Let us set the space ${\cal H}_{l,k}$, where
\begin{eqnarray}
l=b+\rho,~~k=a+\rho,\nonumber
\end{eqnarray}
the space spanned by the eigenvectors of $A^{(a)}(u)$
with the asymptotic boundary condition
$(b,b+\omega_1,b+\omega_2,\cdots,b+\omega_{N-1})$
given by
the choice of $b \in P_{r-1-N}^+$.

Let us set the vertex operator $\Phi_N^{(a_1,a_2)}(-u)$
by FIG.6.

\begin{center}
~\\

%WinTpicVersion3.08
\unitlength 0.1in
\begin{picture}( 55.2500, 12.0000)( -9.2500,-15.1500)
% VECTOR 2 0 3 0
% 20 1200 600 1200 600 1200 600 1600 600 2000 600 2400 600 2800 600 3200 600 1200 1400 1200 1000 1200 1400 1600 1400 2000 1400 2400 1400 2000 1400 2000 1000 2800 1400 3200 1400 2800 1400 2800 1000
% 
\special{pn 8}%
\special{pa 1200 600}%
\special{pa 1200 600}%
\special{fp}%
\special{pa 1200 600}%
\special{pa 1600 600}%
\special{fp}%
\special{sh 1}%
\special{pa 1600 600}%
\special{pa 1534 580}%
\special{pa 1548 600}%
\special{pa 1534 620}%
\special{pa 1600 600}%
\special{fp}%
\special{pa 2000 600}%
\special{pa 2400 600}%
\special{fp}%
\special{sh 1}%
\special{pa 2400 600}%
\special{pa 2334 580}%
\special{pa 2348 600}%
\special{pa 2334 620}%
\special{pa 2400 600}%
\special{fp}%
\special{pa 2800 600}%
\special{pa 3200 600}%
\special{fp}%
\special{sh 1}%
\special{pa 3200 600}%
\special{pa 3134 580}%
\special{pa 3148 600}%
\special{pa 3134 620}%
\special{pa 3200 600}%
\special{fp}%
\special{pa 1200 1400}%
\special{pa 1200 1000}%
\special{fp}%
\special{sh 1}%
\special{pa 1200 1000}%
\special{pa 1180 1068}%
\special{pa 1200 1054}%
\special{pa 1220 1068}%
\special{pa 1200 1000}%
\special{fp}%
\special{pa 1200 1400}%
\special{pa 1600 1400}%
\special{fp}%
\special{sh 1}%
\special{pa 1600 1400}%
\special{pa 1534 1380}%
\special{pa 1548 1400}%
\special{pa 1534 1420}%
\special{pa 1600 1400}%
\special{fp}%
\special{pa 2000 1400}%
\special{pa 2400 1400}%
\special{fp}%
\special{sh 1}%
\special{pa 2400 1400}%
\special{pa 2334 1380}%
\special{pa 2348 1400}%
\special{pa 2334 1420}%
\special{pa 2400 1400}%
\special{fp}%
\special{pa 2000 1400}%
\special{pa 2000 1000}%
\special{fp}%
\special{sh 1}%
\special{pa 2000 1000}%
\special{pa 1980 1068}%
\special{pa 2000 1054}%
\special{pa 2020 1068}%
\special{pa 2000 1000}%
\special{fp}%
\special{pa 2800 1400}%
\special{pa 3200 1400}%
\special{fp}%
\special{sh 1}%
\special{pa 3200 1400}%
\special{pa 3134 1380}%
\special{pa 3148 1400}%
\special{pa 3134 1420}%
\special{pa 3200 1400}%
\special{fp}%
\special{pa 2800 1400}%
\special{pa 2800 1000}%
\special{fp}%
\special{sh 1}%
\special{pa 2800 1000}%
\special{pa 2780 1068}%
\special{pa 2800 1054}%
\special{pa 2820 1068}%
\special{pa 2800 1000}%
\special{fp}%
% LINE 2 0 3 0
% 20 1200 600 1200 1000 1600 1400 2000 1400 2400 1400 2800 1400 3200 1400 4000 1400 2000 1000 2000 600 1600 600 2000 600 2400 600 2800 600 2800 600 2800 1000 3200 600 4000 600 3600 1400 3600 600
% 
\special{pn 8}%
\special{pa 1200 600}%
\special{pa 1200 1000}%
\special{fp}%
\special{pa 1600 1400}%
\special{pa 2000 1400}%
\special{fp}%
\special{pa 2400 1400}%
\special{pa 2800 1400}%
\special{fp}%
\special{pa 3200 1400}%
\special{pa 4000 1400}%
\special{fp}%
\special{pa 2000 1000}%
\special{pa 2000 600}%
\special{fp}%
\special{pa 1600 600}%
\special{pa 2000 600}%
\special{fp}%
\special{pa 2400 600}%
\special{pa 2800 600}%
\special{fp}%
\special{pa 2800 600}%
\special{pa 2800 1000}%
\special{fp}%
\special{pa 3200 600}%
\special{pa 4000 600}%
\special{fp}%
\special{pa 3600 1400}%
\special{pa 3600 600}%
\special{fp}%
% VECTOR 2 0 3 0
% 2 3600 1400 3600 1000
% 
\special{pn 8}%
\special{pa 3600 1400}%
\special{pa 3600 1000}%
\special{fp}%
\special{sh 1}%
\special{pa 3600 1000}%
\special{pa 3580 1068}%
\special{pa 3600 1054}%
\special{pa 3620 1068}%
\special{pa 3600 1000}%
\special{fp}%
% STR 2 0 3 0
% 3 1600 900 1600 1000 5 0
% $u$
\put(16.0000,-10.0000){\makebox(0,0){$u$}}%
% STR 2 0 3 0
% 3 2400 900 2400 1000 5 0
% $u$
\put(24.0000,-10.0000){\makebox(0,0){$u$}}%
% STR 2 0 3 0
% 3 3200 900 3200 1000 5 0
% $u$
\put(32.0000,-10.0000){\makebox(0,0){$u$}}%
% STR 2 0 3 0
% 3 4000 900 4000 1000 5 0
% $\cdots$
\put(40.0000,-10.0000){\makebox(0,0){$\cdots$}}%
% LINE 2 0 3 0
% 4 4000 600 4600 600 4600 1400 4000 1400
% 
\special{pn 8}%
\special{pa 4000 600}%
\special{pa 4600 600}%
\special{fp}%
\special{pa 4600 1400}%
\special{pa 4000 1400}%
\special{fp}%
% STR 2 0 3 0
% 3 1200 300 1200 400 5 0
% $a_2$
\put(12.0000,-4.0000){\makebox(0,0){$a_2$}}%
% STR 2 0 3 0
% 3 1200 1500 1200 1600 5 0
% $a_1$
\put(12.0000,-16.0000){\makebox(0,0){$a_1$}}%
% STR 2 0 3 0
% 3 200 900 200 1000 5 0
% $\Phi_N^{(a_1,a_2)}(-u)=$
\put(2.0000,-10.0000){\makebox(0,0){$\Phi_N^{(a_1,a_2)}(-u)=$}}%
\end{picture}%
~\\
~\\
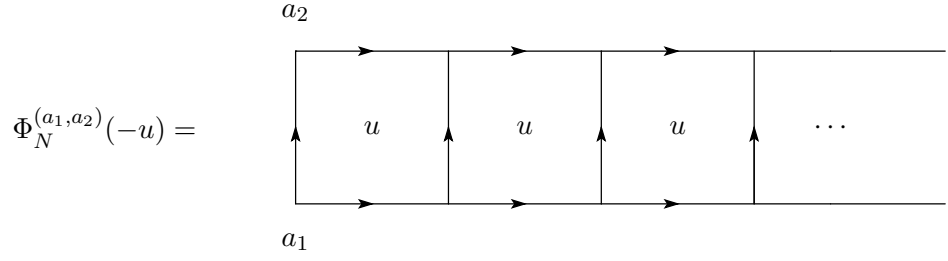
FIG.6.~Vertex operator

~\\
\end{center}
The vertex operator $\Phi_N^{(a,a+\bar{\epsilon}_\mu)}(u)$ 
is an operator
\begin{eqnarray}
\Phi_N^{(a,a+\bar{\epsilon}_\mu)}(u):
{\cal H}_{l,k+\bar{\epsilon}_\mu} \longrightarrow 
{\cal H}_{l,k}.\nonumber
\end{eqnarray}
Let us set the vertex operator $\Phi_W^{(a_1,a_2)}(u)$
by FIG.7.
\begin{center}
~\\

%WinTpicVersion3.08
\unitlength 0.1in
\begin{picture}( 52.8000, 12.0000)( -6.8000,-15.1500)
% VECTOR 2 0 3 0
% 20 1200 600 1600 600 2000 600 2400 600 2800 600 3200 600 3600 600 3600 1000 2800 600 2800 1000 2000 600 2000 1000 1200 600 1200 1000 1200 1400 1600 1400 2000 1400 2400 1400 2800 1400 3200 1400
% 
\special{pn 8}%
\special{pa 1200 600}%
\special{pa 1600 600}%
\special{fp}%
\special{sh 1}%
\special{pa 1600 600}%
\special{pa 1534 580}%
\special{pa 1548 600}%
\special{pa 1534 620}%
\special{pa 1600 600}%
\special{fp}%
\special{pa 2000 600}%
\special{pa 2400 600}%
\special{fp}%
\special{sh 1}%
\special{pa 2400 600}%
\special{pa 2334 580}%
\special{pa 2348 600}%
\special{pa 2334 620}%
\special{pa 2400 600}%
\special{fp}%
\special{pa 2800 600}%
\special{pa 3200 600}%
\special{fp}%
\special{sh 1}%
\special{pa 3200 600}%
\special{pa 3134 580}%
\special{pa 3148 600}%
\special{pa 3134 620}%
\special{pa 3200 600}%
\special{fp}%
\special{pa 3600 600}%
\special{pa 3600 1000}%
\special{fp}%
\special{sh 1}%
\special{pa 3600 1000}%
\special{pa 3620 934}%
\special{pa 3600 948}%
\special{pa 3580 934}%
\special{pa 3600 1000}%
\special{fp}%
\special{pa 2800 600}%
\special{pa 2800 1000}%
\special{fp}%
\special{sh 1}%
\special{pa 2800 1000}%
\special{pa 2820 934}%
\special{pa 2800 948}%
\special{pa 2780 934}%
\special{pa 2800 1000}%
\special{fp}%
\special{pa 2000 600}%
\special{pa 2000 1000}%
\special{fp}%
\special{sh 1}%
\special{pa 2000 1000}%
\special{pa 2020 934}%
\special{pa 2000 948}%
\special{pa 1980 934}%
\special{pa 2000 1000}%
\special{fp}%
\special{pa 1200 600}%
\special{pa 1200 1000}%
\special{fp}%
\special{sh 1}%
\special{pa 1200 1000}%
\special{pa 1220 934}%
\special{pa 1200 948}%
\special{pa 1180 934}%
\special{pa 1200 1000}%
\special{fp}%
\special{pa 1200 1400}%
\special{pa 1600 1400}%
\special{fp}%
\special{sh 1}%
\special{pa 1600 1400}%
\special{pa 1534 1380}%
\special{pa 1548 1400}%
\special{pa 1534 1420}%
\special{pa 1600 1400}%
\special{fp}%
\special{pa 2000 1400}%
\special{pa 2400 1400}%
\special{fp}%
\special{sh 1}%
\special{pa 2400 1400}%
\special{pa 2334 1380}%
\special{pa 2348 1400}%
\special{pa 2334 1420}%
\special{pa 2400 1400}%
\special{fp}%
\special{pa 2800 1400}%
\special{pa 3200 1400}%
\special{fp}%
\special{sh 1}%
\special{pa 3200 1400}%
\special{pa 3134 1380}%
\special{pa 3148 1400}%
\special{pa 3134 1420}%
\special{pa 3200 1400}%
\special{fp}%
% LINE 2 0 3 0
% 2 1600 600 4600 600
% 
\special{pn 8}%
\special{pa 1600 600}%
\special{pa 4600 600}%
\special{fp}%
% LINE 2 0 3 0
% 2 1600 1400 4600 1400
% 
\special{pn 8}%
\special{pa 1600 1400}%
\special{pa 4600 1400}%
\special{fp}%
% LINE 2 0 3 0
% 6 1200 1400 1200 1000 2000 1200 2000 1400 2000 1400 2000 1400
% 
\special{pn 8}%
\special{pa 1200 1400}%
\special{pa 1200 1000}%
\special{fp}%
\special{pa 2000 1200}%
\special{pa 2000 1400}%
\special{fp}%
\special{pa 2000 1400}%
\special{pa 2000 1400}%
\special{fp}%
% LINE 2 0 3 0
% 8 2800 1400 2800 1400 2800 1000 2800 1400 3600 1400 3600 1000 2000 1000 2000 1400
% 
\special{pn 8}%
\special{pa 2800 1400}%
\special{pa 2800 1400}%
\special{fp}%
\special{pa 2800 1000}%
\special{pa 2800 1400}%
\special{fp}%
\special{pa 3600 1400}%
\special{pa 3600 1000}%
\special{fp}%
\special{pa 2000 1000}%
\special{pa 2000 1400}%
\special{fp}%
% STR 2 0 3 0
% 3 1600 900 1600 1000 5 0
% $u$
\put(16.0000,-10.0000){\makebox(0,0){$u$}}%
% STR 2 0 3 0
% 3 2400 900 2400 1000 5 0
% $u$
\put(24.0000,-10.0000){\makebox(0,0){$u$}}%
% STR 2 0 3 0
% 3 3200 900 3200 1000 5 0
% $u$
\put(32.0000,-10.0000){\makebox(0,0){$u$}}%
% STR 2 0 3 0
% 3 4000 900 4000 1000 5 0
% $\cdots$
\put(40.0000,-10.0000){\makebox(0,0){$\cdots$}}%
% STR 2 0 3 0
% 3 1200 300 1200 400 5 0
% $a_2$
\put(12.0000,-4.0000){\makebox(0,0){$a_2$}}%
% STR 2 0 3 0
% 3 1200 1500 1200 1600 5 0
% $a_1$
\put(12.0000,-16.0000){\makebox(0,0){$a_1$}}%
% STR 2 0 3 0
% 3 400 900 400 1000 5 0
% $\Phi_W^{(a_1,a_2)}(u)=$
\put(4.0000,-10.0000){\makebox(0,0){$\Phi_W^{(a_1,a_2)}(u)=$}}%
\end{picture}%
~\\
~\\
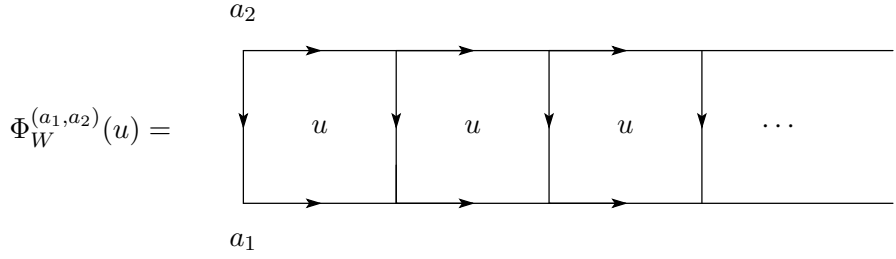
FIG.7.~Vertex operator

~\\
\end{center}
The vertex operator 
$\Phi_W^{(a+\bar{\epsilon}_\mu,a)}(u)$ 
is an operator
\begin{eqnarray}
\Phi_W^{(a+\bar{\epsilon}_\mu,a)}(u):
{\cal H}_{l,k} \longrightarrow 
{\cal H}_{l,k+\bar{\epsilon}_\mu}.\nonumber
\end{eqnarray}
The vertex operators 
$\Phi_N^{(a_1,a_2)}(u)$
and
$\Phi_W^{(a_1,a_2)}(u)$
satisfy
the following 
commutation relations (1), (2) and (2').
\\
{\bf (1)}~Commutation relation :
\begin{eqnarray}
\Phi_W^{(a,b)}(u_1)
\Phi^{(b,d)}(u_2)
&=&\sum_{g}
W\left(\left.\begin{array}{cc}
a&g\\
b&d
\end{array}
\right|u_2-u_1\right)
\Phi_W^{(a,g)}(u_2)
\Phi_W^{(g,d)}(u_1),\nonumber\\
%%%%%%%%%%%%%%%%%%%%%%%%%%%%%%%%%%%%
\Phi_N^{(a,b)}(u_1)
\Phi_N^{(b,d)}(u_2)
&=&
\sum_{g}
W\left(\left.\begin{array}{cc}
d&b\\
g&a
\end{array}
\right|u_2-u_1\right)
\Phi^{(a,g)}(u_2)
\Phi^{(g,d)}(u_1),\nonumber\\
%%%%%%%%%%%%%%%%%%%%%%%%%%%%%%%%%%%%
\Phi_W^{(a,b)}(u_1)
\Phi_N^{(b,d)}(u_2)
&=&
\sum_{g}
W\left(\left.\begin{array}{cc}
g&d\\
a&b
\end{array}
\right|u_1-u_2\right)
\Phi_N^{(a,g)}(u_2)
\Phi_W^{(g,d)}(u_1).\nonumber
\end{eqnarray}
{\bf (2)}~Inversion relation :
\begin{eqnarray}
\Phi_W^{(a,g)}(z)\Phi_N^{(g,b)}(z)=\delta_{a,b}.
\nonumber
\end{eqnarray}
{\bf (2')}~Inversion relation :
\begin{eqnarray}
\sum_{g}
\Phi_N^{(a,g)}(z)\Phi_W^{(g,b)}(z)=\delta_{a,b}.
\nonumber
\end{eqnarray}
Let us set the infinite transfer matrix $T_B^{(a)}(u)$
by
\begin{eqnarray}
T_B^{(a)}(u)=\sum_{\mu=1}^N
\Phi_N^{(a,a+\bar{\epsilon}_\mu)}(-u)
K\left(\left.
\begin{array}{cc}
~& a\\
a+\bar{\epsilon}_\mu &\\
~& a
\end{array}
\right|u\right)
\Phi_W^{(a+\bar{\epsilon}_\mu,a)}(u).\nonumber
\end{eqnarray}
The infinite transfer matrix $T_B^{(a)}(u)$ is an operator,
\begin{eqnarray}
T_B^{(a)}(u): {\cal H}_{l,k} \longrightarrow {\cal H}_{l,k}.
\nonumber
\end{eqnarray}
The infinite transfer matrix $T_B^{(a)}(u)$ is given 
by FIG.8 at the same time.

\begin{center}
~\\

%WinTpicVersion3.08
\unitlength 0.1in
\begin{picture}( 46.0000, 20.0000)(  2.0000,-21.1500)
% VECTOR 2 0 3 0
% 54 3200 400 2800 400 2400 400 2000 400 1600 400 1200 400 800 400 400 400 4000 400 3600 400 4000 400 4000 800 3200 400 3200 800 2400 400 2400 800 1600 400 1600 800 800 400 800 800 4000 1200 3600 1200 3200 1200 2800 1200 2400 1200 2000 1200 1600 1200 1200 1200 800 1200 400 1200 4000 2000 3600 2000 3200 2000 2800 2000 2400 2000 2000 2000 1600 2000 1200 2000 800 2000 400 2000 800 2000 800 1600 1600 2000 1600 1600 2400 2000 2400 1600 3200 2000 3200 1600 4000 2000 4000 1600 4800 400 4400 800 4800 2000 4400 1600
% 
\special{pn 8}%
\special{pa 3200 400}%
\special{pa 2800 400}%
\special{fp}%
\special{sh 1}%
\special{pa 2800 400}%
\special{pa 2868 420}%
\special{pa 2854 400}%
\special{pa 2868 380}%
\special{pa 2800 400}%
\special{fp}%
\special{pa 2400 400}%
\special{pa 2000 400}%
\special{fp}%
\special{sh 1}%
\special{pa 2000 400}%
\special{pa 2068 420}%
\special{pa 2054 400}%
\special{pa 2068 380}%
\special{pa 2000 400}%
\special{fp}%
\special{pa 1600 400}%
\special{pa 1200 400}%
\special{fp}%
\special{sh 1}%
\special{pa 1200 400}%
\special{pa 1268 420}%
\special{pa 1254 400}%
\special{pa 1268 380}%
\special{pa 1200 400}%
\special{fp}%
\special{pa 800 400}%
\special{pa 400 400}%
\special{fp}%
\special{sh 1}%
\special{pa 400 400}%
\special{pa 468 420}%
\special{pa 454 400}%
\special{pa 468 380}%
\special{pa 400 400}%
\special{fp}%
\special{pa 4000 400}%
\special{pa 3600 400}%
\special{fp}%
\special{sh 1}%
\special{pa 3600 400}%
\special{pa 3668 420}%
\special{pa 3654 400}%
\special{pa 3668 380}%
\special{pa 3600 400}%
\special{fp}%
\special{pa 4000 400}%
\special{pa 4000 800}%
\special{fp}%
\special{sh 1}%
\special{pa 4000 800}%
\special{pa 4020 734}%
\special{pa 4000 748}%
\special{pa 3980 734}%
\special{pa 4000 800}%
\special{fp}%
\special{pa 3200 400}%
\special{pa 3200 800}%
\special{fp}%
\special{sh 1}%
\special{pa 3200 800}%
\special{pa 3220 734}%
\special{pa 3200 748}%
\special{pa 3180 734}%
\special{pa 3200 800}%
\special{fp}%
\special{pa 2400 400}%
\special{pa 2400 800}%
\special{fp}%
\special{sh 1}%
\special{pa 2400 800}%
\special{pa 2420 734}%
\special{pa 2400 748}%
\special{pa 2380 734}%
\special{pa 2400 800}%
\special{fp}%
\special{pa 1600 400}%
\special{pa 1600 800}%
\special{fp}%
\special{sh 1}%
\special{pa 1600 800}%
\special{pa 1620 734}%
\special{pa 1600 748}%
\special{pa 1580 734}%
\special{pa 1600 800}%
\special{fp}%
\special{pa 800 400}%
\special{pa 800 800}%
\special{fp}%
\special{sh 1}%
\special{pa 800 800}%
\special{pa 820 734}%
\special{pa 800 748}%
\special{pa 780 734}%
\special{pa 800 800}%
\special{fp}%
\special{pa 4000 1200}%
\special{pa 3600 1200}%
\special{fp}%
\special{sh 1}%
\special{pa 3600 1200}%
\special{pa 3668 1220}%
\special{pa 3654 1200}%
\special{pa 3668 1180}%
\special{pa 3600 1200}%
\special{fp}%
\special{pa 3200 1200}%
\special{pa 2800 1200}%
\special{fp}%
\special{sh 1}%
\special{pa 2800 1200}%
\special{pa 2868 1220}%
\special{pa 2854 1200}%
\special{pa 2868 1180}%
\special{pa 2800 1200}%
\special{fp}%
\special{pa 2400 1200}%
\special{pa 2000 1200}%
\special{fp}%
\special{sh 1}%
\special{pa 2000 1200}%
\special{pa 2068 1220}%
\special{pa 2054 1200}%
\special{pa 2068 1180}%
\special{pa 2000 1200}%
\special{fp}%
\special{pa 1600 1200}%
\special{pa 1200 1200}%
\special{fp}%
\special{sh 1}%
\special{pa 1200 1200}%
\special{pa 1268 1220}%
\special{pa 1254 1200}%
\special{pa 1268 1180}%
\special{pa 1200 1200}%
\special{fp}%
\special{pa 800 1200}%
\special{pa 400 1200}%
\special{fp}%
\special{sh 1}%
\special{pa 400 1200}%
\special{pa 468 1220}%
\special{pa 454 1200}%
\special{pa 468 1180}%
\special{pa 400 1200}%
\special{fp}%
\special{pa 4000 2000}%
\special{pa 3600 2000}%
\special{fp}%
\special{sh 1}%
\special{pa 3600 2000}%
\special{pa 3668 2020}%
\special{pa 3654 2000}%
\special{pa 3668 1980}%
\special{pa 3600 2000}%
\special{fp}%
\special{pa 3200 2000}%
\special{pa 2800 2000}%
\special{fp}%
\special{sh 1}%
\special{pa 2800 2000}%
\special{pa 2868 2020}%
\special{pa 2854 2000}%
\special{pa 2868 1980}%
\special{pa 2800 2000}%
\special{fp}%
\special{pa 2400 2000}%
\special{pa 2000 2000}%
\special{fp}%
\special{sh 1}%
\special{pa 2000 2000}%
\special{pa 2068 2020}%
\special{pa 2054 2000}%
\special{pa 2068 1980}%
\special{pa 2000 2000}%
\special{fp}%
\special{pa 1600 2000}%
\special{pa 1200 2000}%
\special{fp}%
\special{sh 1}%
\special{pa 1200 2000}%
\special{pa 1268 2020}%
\special{pa 1254 2000}%
\special{pa 1268 1980}%
\special{pa 1200 2000}%
\special{fp}%
\special{pa 800 2000}%
\special{pa 400 2000}%
\special{fp}%
\special{sh 1}%
\special{pa 400 2000}%
\special{pa 468 2020}%
\special{pa 454 2000}%
\special{pa 468 1980}%
\special{pa 400 2000}%
\special{fp}%
\special{pa 800 2000}%
\special{pa 800 1600}%
\special{fp}%
\special{sh 1}%
\special{pa 800 1600}%
\special{pa 780 1668}%
\special{pa 800 1654}%
\special{pa 820 1668}%
\special{pa 800 1600}%
\special{fp}%
\special{pa 1600 2000}%
\special{pa 1600 1600}%
\special{fp}%
\special{sh 1}%
\special{pa 1600 1600}%
\special{pa 1580 1668}%
\special{pa 1600 1654}%
\special{pa 1620 1668}%
\special{pa 1600 1600}%
\special{fp}%
\special{pa 2400 2000}%
\special{pa 2400 1600}%
\special{fp}%
\special{sh 1}%
\special{pa 2400 1600}%
\special{pa 2380 1668}%
\special{pa 2400 1654}%
\special{pa 2420 1668}%
\special{pa 2400 1600}%
\special{fp}%
\special{pa 3200 2000}%
\special{pa 3200 1600}%
\special{fp}%
\special{sh 1}%
\special{pa 3200 1600}%
\special{pa 3180 1668}%
\special{pa 3200 1654}%
\special{pa 3220 1668}%
\special{pa 3200 1600}%
\special{fp}%
\special{pa 4000 2000}%
\special{pa 4000 1600}%
\special{fp}%
\special{sh 1}%
\special{pa 4000 1600}%
\special{pa 3980 1668}%
\special{pa 4000 1654}%
\special{pa 4020 1668}%
\special{pa 4000 1600}%
\special{fp}%
\special{pa 4800 400}%
\special{pa 4400 800}%
\special{fp}%
\special{sh 1}%
\special{pa 4400 800}%
\special{pa 4462 768}%
\special{pa 4438 762}%
\special{pa 4434 740}%
\special{pa 4400 800}%
\special{fp}%
\special{pa 4800 2000}%
\special{pa 4400 1600}%
\special{fp}%
\special{sh 1}%
\special{pa 4400 1600}%
\special{pa 4434 1662}%
\special{pa 4438 1638}%
\special{pa 4462 1634}%
\special{pa 4400 1600}%
\special{fp}%
% LINE 2 0 3 0
% 2 4000 1200 200 1200
% 
\special{pn 8}%
\special{pa 4000 1200}%
\special{pa 200 1200}%
\special{fp}%
% LINE 2 0 3 0
% 2 4000 2000 200 2000
% 
\special{pn 8}%
\special{pa 4000 2000}%
\special{pa 200 2000}%
\special{fp}%
% LINE 2 0 3 0
% 2 4000 400 200 400
% 
\special{pn 8}%
\special{pa 4000 400}%
\special{pa 200 400}%
\special{fp}%
% LINE 2 0 3 0
% 6 4400 800 4000 1200 4000 1200 4400 1600 4800 2000 4800 400
% 
\special{pn 8}%
\special{pa 4400 800}%
\special{pa 4000 1200}%
\special{fp}%
\special{pa 4000 1200}%
\special{pa 4400 1600}%
\special{fp}%
\special{pa 4800 2000}%
\special{pa 4800 400}%
\special{fp}%
% LINE 2 0 3 0
% 10 4000 800 4000 1600 3200 800 3200 1600 2400 800 2400 1600 1600 800 1600 1600 800 800 800 1600
% 
\special{pn 8}%
\special{pa 4000 800}%
\special{pa 4000 1600}%
\special{fp}%
\special{pa 3200 800}%
\special{pa 3200 1600}%
\special{fp}%
\special{pa 2400 800}%
\special{pa 2400 1600}%
\special{fp}%
\special{pa 1600 800}%
\special{pa 1600 1600}%
\special{fp}%
\special{pa 800 800}%
\special{pa 800 1600}%
\special{fp}%
% STR 2 0 3 0
% 3 4000 100 4000 200 5 0
% $a$
\put(40.0000,-2.0000){\makebox(0,0){$a$}}%
% STR 2 0 3 0
% 3 4800 100 4800 200 5 0
% $a$
\put(48.0000,-2.0000){\makebox(0,0){$a$}}%
% STR 2 0 3 0
% 3 4400 1100 4400 1200 5 0
% $u$
\put(44.0000,-12.0000){\makebox(0,0){$u$}}%
% STR 2 0 3 0
% 3 1200 700 1200 800 5 0
% $u$
\put(12.0000,-8.0000){\makebox(0,0){$u$}}%
% STR 2 0 3 0
% 3 2000 700 2000 800 5 0
% $u$
\put(20.0000,-8.0000){\makebox(0,0){$u$}}%
% STR 2 0 3 0
% 3 2800 700 2800 800 5 0
% $u$
\put(28.0000,-8.0000){\makebox(0,0){$u$}}%
% STR 2 0 3 0
% 3 3600 700 3600 800 5 0
% $u$
\put(36.0000,-8.0000){\makebox(0,0){$u$}}%
% STR 2 0 3 0
% 3 3600 1500 3600 1600 5 0
% $u$
\put(36.0000,-16.0000){\makebox(0,0){$u$}}%
% STR 2 0 3 0
% 3 2800 1500 2800 1600 5 0
% $u$
\put(28.0000,-16.0000){\makebox(0,0){$u$}}%
% STR 2 0 3 0
% 3 2000 1500 2000 1600 5 0
% $u$
\put(20.0000,-16.0000){\makebox(0,0){$u$}}%
% STR 2 0 3 0
% 3 1200 1500 1200 1600 5 0
% $u$
\put(12.0000,-16.0000){\makebox(0,0){$u$}}%
% STR 2 0 3 0
% 3 4000 2100 4000 2200 5 0
% $a$
\put(40.0000,-22.0000){\makebox(0,0){$a$}}%
% STR 2 0 3 0
% 3 4800 2100 4800 2200 5 0
% $a$
\put(48.0000,-22.0000){\makebox(0,0){$a$}}%
\end{picture}%
~\\
~\\
FIG.8.~~Boundary transfer matrix

~\\
\end{center}
Because of
the commutation relations of the vertex operators
and
the boundary Yang-Baxter equation,
we get the commutativity of the
infinite transfer matrix
$T_B^{(a)}(u)$.
\begin{eqnarray}
~[T_B^{(a)}(u),T_B^{(a)}(v)]=0.\nonumber
\end{eqnarray}
In this paper, we are interested in the ground state.
The boundary $U_{q,p}(A_{N-1}^{(1)})$ face model is given by
infinite product of the infinite transfer matrix $T_B^{(a)}(u)$.
We figure out the ground state of the problem
in the infinite transfer matrix $T_B^{(a)}(u)$.
Note that we consider the case $0<c<1, 0<a_{1,\mu}<r$.
In the limit $x \to 1$ we have

\begin{eqnarray}
\left|K\left(\left.\begin{array}{cc}
&a\\
a+\bar{\epsilon}_1&\\
&a
\end{array}\right|u\right)\right|
>
\left|K\left(\left.\begin{array}{cc}
&a\\
a+\bar{\epsilon}_\mu
&\\
&a
\end{array}\right|u\right)\right|~~~(\mu \neq 1).
\nonumber
\end{eqnarray}

~\\
Therefore,
the ground state configuration
is given by following figure.
Asymptotic boundary condition should be $b=a$.
The ground state is given by FIG.9.

\begin{center}
~\\
%WinTpicVersion3.08
\unitlength 0.1in
\begin{picture}( 46.0000, 20.0000)(  2.0000,-21.1500)
% VECTOR 2 0 3 0
% 54 3200 400 2800 400 2400 400 2000 400 1600 400 1200 400 800 400 400 400 4000 400 3600 400 4000 400 4000 800 3200 400 3200 800 2400 400 2400 800 1600 400 1600 800 800 400 800 800 4000 1200 3600 1200 3200 1200 2800 1200 2400 1200 2000 1200 1600 1200 1200 1200 800 1200 400 1200 4000 2000 3600 2000 3200 2000 2800 2000 2400 2000 2000 2000 1600 2000 1200 2000 800 2000 400 2000 800 2000 800 1600 1600 2000 1600 1600 2400 2000 2400 1600 3200 2000 3200 1600 4000 2000 4000 1600 4800 400 4400 800 4800 2000 4400 1600
% 
\special{pn 8}%
\special{pa 3200 400}%
\special{pa 2800 400}%
\special{fp}%
\special{sh 1}%
\special{pa 2800 400}%
\special{pa 2868 420}%
\special{pa 2854 400}%
\special{pa 2868 380}%
\special{pa 2800 400}%
\special{fp}%
\special{pa 2400 400}%
\special{pa 2000 400}%
\special{fp}%
\special{sh 1}%
\special{pa 2000 400}%
\special{pa 2068 420}%
\special{pa 2054 400}%
\special{pa 2068 380}%
\special{pa 2000 400}%
\special{fp}%
\special{pa 1600 400}%
\special{pa 1200 400}%
\special{fp}%
\special{sh 1}%
\special{pa 1200 400}%
\special{pa 1268 420}%
\special{pa 1254 400}%
\special{pa 1268 380}%
\special{pa 1200 400}%
\special{fp}%
\special{pa 800 400}%
\special{pa 400 400}%
\special{fp}%
\special{sh 1}%
\special{pa 400 400}%
\special{pa 468 420}%
\special{pa 454 400}%
\special{pa 468 380}%
\special{pa 400 400}%
\special{fp}%
\special{pa 4000 400}%
\special{pa 3600 400}%
\special{fp}%
\special{sh 1}%
\special{pa 3600 400}%
\special{pa 3668 420}%
\special{pa 3654 400}%
\special{pa 3668 380}%
\special{pa 3600 400}%
\special{fp}%
\special{pa 4000 400}%
\special{pa 4000 800}%
\special{fp}%
\special{sh 1}%
\special{pa 4000 800}%
\special{pa 4020 734}%
\special{pa 4000 748}%
\special{pa 3980 734}%
\special{pa 4000 800}%
\special{fp}%
\special{pa 3200 400}%
\special{pa 3200 800}%
\special{fp}%
\special{sh 1}%
\special{pa 3200 800}%
\special{pa 3220 734}%
\special{pa 3200 748}%
\special{pa 3180 734}%
\special{pa 3200 800}%
\special{fp}%
\special{pa 2400 400}%
\special{pa 2400 800}%
\special{fp}%
\special{sh 1}%
\special{pa 2400 800}%
\special{pa 2420 734}%
\special{pa 2400 748}%
\special{pa 2380 734}%
\special{pa 2400 800}%
\special{fp}%
\special{pa 1600 400}%
\special{pa 1600 800}%
\special{fp}%
\special{sh 1}%
\special{pa 1600 800}%
\special{pa 1620 734}%
\special{pa 1600 748}%
\special{pa 1580 734}%
\special{pa 1600 800}%
\special{fp}%
\special{pa 800 400}%
\special{pa 800 800}%
\special{fp}%
\special{sh 1}%
\special{pa 800 800}%
\special{pa 820 734}%
\special{pa 800 748}%
\special{pa 780 734}%
\special{pa 800 800}%
\special{fp}%
\special{pa 4000 1200}%
\special{pa 3600 1200}%
\special{fp}%
\special{sh 1}%
\special{pa 3600 1200}%
\special{pa 3668 1220}%
\special{pa 3654 1200}%
\special{pa 3668 1180}%
\special{pa 3600 1200}%
\special{fp}%
\special{pa 3200 1200}%
\special{pa 2800 1200}%
\special{fp}%
\special{sh 1}%
\special{pa 2800 1200}%
\special{pa 2868 1220}%
\special{pa 2854 1200}%
\special{pa 2868 1180}%
\special{pa 2800 1200}%
\special{fp}%
\special{pa 2400 1200}%
\special{pa 2000 1200}%
\special{fp}%
\special{sh 1}%
\special{pa 2000 1200}%
\special{pa 2068 1220}%
\special{pa 2054 1200}%
\special{pa 2068 1180}%
\special{pa 2000 1200}%
\special{fp}%
\special{pa 1600 1200}%
\special{pa 1200 1200}%
\special{fp}%
\special{sh 1}%
\special{pa 1200 1200}%
\special{pa 1268 1220}%
\special{pa 1254 1200}%
\special{pa 1268 1180}%
\special{pa 1200 1200}%
\special{fp}%
\special{pa 800 1200}%
\special{pa 400 1200}%
\special{fp}%
\special{sh 1}%
\special{pa 400 1200}%
\special{pa 468 1220}%
\special{pa 454 1200}%
\special{pa 468 1180}%
\special{pa 400 1200}%
\special{fp}%
\special{pa 4000 2000}%
\special{pa 3600 2000}%
\special{fp}%
\special{sh 1}%
\special{pa 3600 2000}%
\special{pa 3668 2020}%
\special{pa 3654 2000}%
\special{pa 3668 1980}%
\special{pa 3600 2000}%
\special{fp}%
\special{pa 3200 2000}%
\special{pa 2800 2000}%
\special{fp}%
\special{sh 1}%
\special{pa 2800 2000}%
\special{pa 2868 2020}%
\special{pa 2854 2000}%
\special{pa 2868 1980}%
\special{pa 2800 2000}%
\special{fp}%
\special{pa 2400 2000}%
\special{pa 2000 2000}%
\special{fp}%
\special{sh 1}%
\special{pa 2000 2000}%
\special{pa 2068 2020}%
\special{pa 2054 2000}%
\special{pa 2068 1980}%
\special{pa 2000 2000}%
\special{fp}%
\special{pa 1600 2000}%
\special{pa 1200 2000}%
\special{fp}%
\special{sh 1}%
\special{pa 1200 2000}%
\special{pa 1268 2020}%
\special{pa 1254 2000}%
\special{pa 1268 1980}%
\special{pa 1200 2000}%
\special{fp}%
\special{pa 800 2000}%
\special{pa 400 2000}%
\special{fp}%
\special{sh 1}%
\special{pa 400 2000}%
\special{pa 468 2020}%
\special{pa 454 2000}%
\special{pa 468 1980}%
\special{pa 400 2000}%
\special{fp}%
\special{pa 800 2000}%
\special{pa 800 1600}%
\special{fp}%
\special{sh 1}%
\special{pa 800 1600}%
\special{pa 780 1668}%
\special{pa 800 1654}%
\special{pa 820 1668}%
\special{pa 800 1600}%
\special{fp}%
\special{pa 1600 2000}%
\special{pa 1600 1600}%
\special{fp}%
\special{sh 1}%
\special{pa 1600 1600}%
\special{pa 1580 1668}%
\special{pa 1600 1654}%
\special{pa 1620 1668}%
\special{pa 1600 1600}%
\special{fp}%
\special{pa 2400 2000}%
\special{pa 2400 1600}%
\special{fp}%
\special{sh 1}%
\special{pa 2400 1600}%
\special{pa 2380 1668}%
\special{pa 2400 1654}%
\special{pa 2420 1668}%
\special{pa 2400 1600}%
\special{fp}%
\special{pa 3200 2000}%
\special{pa 3200 1600}%
\special{fp}%
\special{sh 1}%
\special{pa 3200 1600}%
\special{pa 3180 1668}%
\special{pa 3200 1654}%
\special{pa 3220 1668}%
\special{pa 3200 1600}%
\special{fp}%
\special{pa 4000 2000}%
\special{pa 4000 1600}%
\special{fp}%
\special{sh 1}%
\special{pa 4000 1600}%
\special{pa 3980 1668}%
\special{pa 4000 1654}%
\special{pa 4020 1668}%
\special{pa 4000 1600}%
\special{fp}%
\special{pa 4800 400}%
\special{pa 4400 800}%
\special{fp}%
\special{sh 1}%
\special{pa 4400 800}%
\special{pa 4462 768}%
\special{pa 4438 762}%
\special{pa 4434 740}%
\special{pa 4400 800}%
\special{fp}%
\special{pa 4800 2000}%
\special{pa 4400 1600}%
\special{fp}%
\special{sh 1}%
\special{pa 4400 1600}%
\special{pa 4434 1662}%
\special{pa 4438 1638}%
\special{pa 4462 1634}%
\special{pa 4400 1600}%
\special{fp}%
% LINE 2 0 3 0
% 2 4000 1200 200 1200
% 
\special{pn 8}%
\special{pa 4000 1200}%
\special{pa 200 1200}%
\special{fp}%
% LINE 2 0 3 0
% 2 4000 2000 200 2000
% 
\special{pn 8}%
\special{pa 4000 2000}%
\special{pa 200 2000}%
\special{fp}%
% LINE 2 0 3 0
% 2 4000 400 200 400
% 
\special{pn 8}%
\special{pa 4000 400}%
\special{pa 200 400}%
\special{fp}%
% LINE 2 0 3 0
% 6 4400 800 4000 1200 4000 1200 4400 1600 4800 2000 4800 400
% 
\special{pn 8}%
\special{pa 4400 800}%
\special{pa 4000 1200}%
\special{fp}%
\special{pa 4000 1200}%
\special{pa 4400 1600}%
\special{fp}%
\special{pa 4800 2000}%
\special{pa 4800 400}%
\special{fp}%
% LINE 2 0 3 0
% 10 4000 800 4000 1600 3200 800 3200 1600 2400 800 2400 1600 1600 800 1600 1600 800 800 800 1600
% 
\special{pn 8}%
\special{pa 4000 800}%
\special{pa 4000 1600}%
\special{fp}%
\special{pa 3200 800}%
\special{pa 3200 1600}%
\special{fp}%
\special{pa 2400 800}%
\special{pa 2400 1600}%
\special{fp}%
\special{pa 1600 800}%
\special{pa 1600 1600}%
\special{fp}%
\special{pa 800 800}%
\special{pa 800 1600}%
\special{fp}%
% STR 2 0 3 0
% 3 4000 100 4000 200 5 0
% $a$
\put(40.0000,-2.0000){\makebox(0,0){$a$}}%
% STR 2 0 3 0
% 3 4800 100 4800 200 5 0
% $a$
\put(48.0000,-2.0000){\makebox(0,0){$a$}}%
% STR 2 0 3 0
% 3 4000 2100 4000 2200 5 0
% $a$
\put(40.0000,-22.0000){\makebox(0,0){$a$}}%
% STR 2 0 3 0
% 3 4800 2100 4800 2200 5 0
% $a$
\put(48.0000,-22.0000){\makebox(0,0){$a$}}%
% STR 2 0 3 0
% 3 3200 2100 3200 2200 5 0
% $a+\omega_1$
\put(32.0000,-22.0000){\makebox(0,0){$a+\omega_1$}}%
% STR 2 0 3 0
% 3 2400 2100 2400 2200 5 0
% $a+\omega_2$
\put(24.0000,-22.0000){\makebox(0,0){$a+\omega_2$}}%
% STR 2 0 3 0
% 3 1600 2100 1600 2200 5 0
% $a+\omega_3$
\put(16.0000,-22.0000){\makebox(0,0){$a+\omega_3$}}%
% STR 2 0 3 0
% 3 800 2100 800 2200 5 0
% $a+\omega_4$
\put(8.0000,-22.0000){\makebox(0,0){$a+\omega_4$}}%
% STR 2 0 3 0
% 3 3200 100 3200 200 5 0
% $a+\omega_1$
\put(32.0000,-2.0000){\makebox(0,0){$a+\omega_1$}}%
% STR 2 0 3 0
% 3 2400 100 2400 200 5 0
% $a+\omega_2$
\put(24.0000,-2.0000){\makebox(0,0){$a+\omega_2$}}%
% STR 2 0 3 0
% 3 1600 100 1600 200 5 0
% $a+\omega_3$
\put(16.0000,-2.0000){\makebox(0,0){$a+\omega_3$}}%
% STR 2 0 3 0
% 3 800 100 800 200 5 0
% $a+\omega_4$
\put(8.0000,-2.0000){\makebox(0,0){$a+\omega_4$}}%
% STR 2 0 3 0
% 3 4200 1100 4200 1200 0 0
% 
\put(42.0000,-12.0000){\makebox(0,0)[lb]{}}%
% STR 2 0 3 0
% 3 4400 1100 4400 1200 5 0
% $a+\omega_1$
\put(44.0000,-12.0000){\makebox(0,0){$a+\omega_1$}}%
% STR 2 0 3 0
% 3 3600 1300 3600 1400 5 0
% $a+\omega_2$
\put(36.0000,-14.0000){\makebox(0,0){$a+\omega_2$}}%
% STR 2 0 3 0
% 3 2800 1300 2800 1400 5 0
% $a+\omega_3$
\put(28.0000,-14.0000){\makebox(0,0){$a+\omega_3$}}%
% STR 2 0 3 0
% 3 2000 1300 2000 1400 5 0
% $a+\omega_4$
\put(20.0000,-14.0000){\makebox(0,0){$a+\omega_4$}}%
% STR 2 0 3 0
% 3 1200 1300 1200 1400 5 0
% $a+\omega_5$
\put(12.0000,-14.0000){\makebox(0,0){$a+\omega_5$}}%
\end{picture}%
~\\
~\\
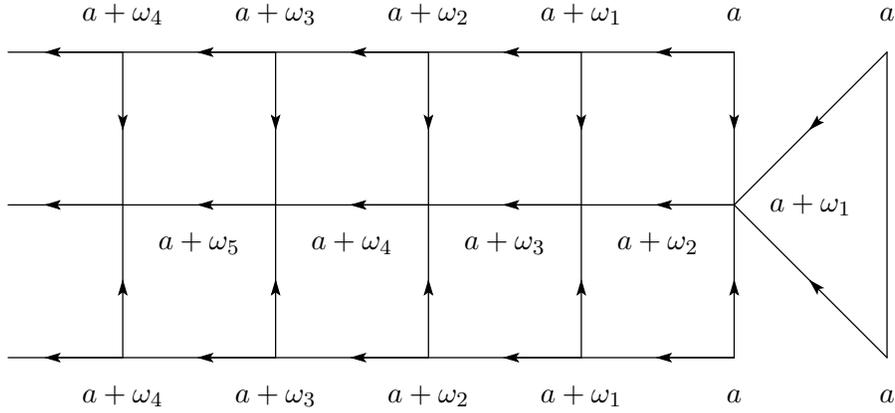
FIG.9.~Ground state

~\\
\end{center}
Hence, in what follows, 
we consider the space ${\cal H}_{l,k}$ for
$l=k=a+\rho$.
$${\cal H}_{k,k}.$$
Let us set the boundary state $|k \rangle_B$
by the half infinite plane figure : FIG.10.
\begin{center}

%WinTpicVersion3.08
\unitlength 0.1in
\begin{picture}( 49.4000, 32.0000)(  0.6000,-34.0000)
% LINE 2 0 3 0
% 28 810 600 3610 600 3610 1400 3610 600 3610 1400 3610 3000 3610 3000 810 3000 810 2200 3610 2200 3610 1400 810 1400 2810 600 2810 3000 2010 600 2010 3000 1210 3000 1210 600 4410 1400 4410 3000 4410 3000 3610 2200 3610 2200 4410 1400 4410 600 4410 1400 4410 1400 3610 600
% 
\special{pn 8}%
\special{pa 810 600}%
\special{pa 3610 600}%
\special{fp}%
\special{pa 3610 1400}%
\special{pa 3610 600}%
\special{fp}%
\special{pa 3610 1400}%
\special{pa 3610 3000}%
\special{fp}%
\special{pa 3610 3000}%
\special{pa 810 3000}%
\special{fp}%
\special{pa 810 2200}%
\special{pa 3610 2200}%
\special{fp}%
\special{pa 3610 1400}%
\special{pa 810 1400}%
\special{fp}%
\special{pa 2810 600}%
\special{pa 2810 3000}%
\special{fp}%
\special{pa 2010 600}%
\special{pa 2010 3000}%
\special{fp}%
\special{pa 1210 3000}%
\special{pa 1210 600}%
\special{fp}%
\special{pa 4410 1400}%
\special{pa 4410 3000}%
\special{fp}%
\special{pa 4410 3000}%
\special{pa 3610 2200}%
\special{fp}%
\special{pa 3610 2200}%
\special{pa 4410 1400}%
\special{fp}%
\special{pa 4410 600}%
\special{pa 4410 1400}%
\special{fp}%
\special{pa 4410 1400}%
\special{pa 3610 600}%
\special{fp}%
% STR 2 0 3 0
% 3 4010 300 4010 400 5 0
% $\vdots$
\put(40.1000,-4.0000){\makebox(0,0){$\vdots$}}%
% STR 2 0 3 0
% 3 2010 300 2010 400 5 0
% $\vdots$
\put(20.1000,-4.0000){\makebox(0,0){$\vdots$}}%
% STR 2 0 3 0
% 3 1210 300 1210 400 5 0
% $\vdots$
\put(12.1000,-4.0000){\makebox(0,0){$\vdots$}}%
% STR 2 0 3 0
% 3 4410 3100 4410 3200 5 0
% $a$
\put(44.1000,-32.0000){\makebox(0,0){$a$}}%
% STR 2 0 3 0
% 3 4610 1300 4610 1400 5 0
% $a$
\put(46.1000,-14.0000){\makebox(0,0){$a$}}%
% STR 2 0 3 0
% 3 4210 2100 4210 2200 5 0
% $u$
\put(42.1000,-22.0000){\makebox(0,0){$u$}}%
% STR 2 0 3 0
% 3 4210 700 4210 800 5 0
% $u$
\put(42.1000,-8.0000){\makebox(0,0){$u$}}%
% STR 2 0 3 0
% 3 3210 2500 3210 2600 5 0
% $u$
\put(32.1000,-26.0000){\makebox(0,0){$u$}}%
% STR 2 0 3 0
% 3 2410 2500 2410 2600 5 0
% $u$
\put(24.1000,-26.0000){\makebox(0,0){$u$}}%
% STR 2 0 3 0
% 3 1610 2500 1610 2600 5 0
% $u$
\put(16.1000,-26.0000){\makebox(0,0){$u$}}%
% STR 2 0 3 0
% 3 3210 1700 3210 1800 5 0
% $u$
\put(32.1000,-18.0000){\makebox(0,0){$u$}}%
% STR 2 0 3 0
% 3 2410 1700 2410 1800 5 0
% $u$
\put(24.1000,-18.0000){\makebox(0,0){$u$}}%
% STR 2 0 3 0
% 3 1610 1700 1610 1800 5 0
% $u$
\put(16.1000,-18.0000){\makebox(0,0){$u$}}%
% STR 2 0 3 0
% 3 3210 900 3210 1000 5 0
% $u$
\put(32.1000,-10.0000){\makebox(0,0){$u$}}%
% STR 2 0 3 0
% 3 2410 900 2410 1000 5 0
% $u$
\put(24.1000,-10.0000){\makebox(0,0){$u$}}%
% STR 2 0 3 0
% 3 1610 900 1610 1000 5 0
% $u$
\put(16.1000,-10.0000){\makebox(0,0){$u$}}%
% VECTOR 2 0 3 0
% 2 4410 1400 4010 1800
% 
\special{pn 8}%
\special{pa 4410 1400}%
\special{pa 4010 1800}%
\special{fp}%
\special{sh 1}%
\special{pa 4010 1800}%
\special{pa 4072 1768}%
\special{pa 4048 1762}%
\special{pa 4044 1740}%
\special{pa 4010 1800}%
\special{fp}%
% VECTOR 2 0 3 0
% 2 4410 3000 4010 2600
% 
\special{pn 8}%
\special{pa 4410 3000}%
\special{pa 4010 2600}%
\special{fp}%
\special{sh 1}%
\special{pa 4010 2600}%
\special{pa 4044 2662}%
\special{pa 4048 2638}%
\special{pa 4072 2634}%
\special{pa 4010 2600}%
\special{fp}%
% VECTOR 2 0 3 0
% 2 4410 1400 4010 1000
% 
\special{pn 8}%
\special{pa 4410 1400}%
\special{pa 4010 1000}%
\special{fp}%
\special{sh 1}%
\special{pa 4010 1000}%
\special{pa 4044 1062}%
\special{pa 4048 1038}%
\special{pa 4072 1034}%
\special{pa 4010 1000}%
\special{fp}%
% STR 2 0 3 0
% 3 2810 300 2810 400 5 0
% $\vdots$
\put(28.1000,-4.0000){\makebox(0,0){$\vdots$}}%
% STR 2 0 3 0
% 3 810 900 810 1000 5 0
% $\cdots$
\put(8.1000,-10.0000){\makebox(0,0){$\cdots$}}%
% STR 2 0 3 0
% 3 810 1700 810 1800 5 0
% $\cdots$
\put(8.1000,-18.0000){\makebox(0,0){$\cdots$}}%
% STR 2 0 3 0
% 3 810 2500 810 2600 5 0
% $\cdots$
\put(8.1000,-26.0000){\makebox(0,0){$\cdots$}}%
% VECTOR 2 0 3 0
% 16 3610 3000 3610 2600 3610 3000 3210 3000 2810 3000 2410 3000 2010 3000 1610 3000 1210 3000 810 3000 1210 3000 1210 2600 2010 3000 2010 2600 2810 3000 2810 2600
% 
\special{pn 8}%
\special{pa 3610 3000}%
\special{pa 3610 2600}%
\special{fp}%
\special{sh 1}%
\special{pa 3610 2600}%
\special{pa 3590 2668}%
\special{pa 3610 2654}%
\special{pa 3630 2668}%
\special{pa 3610 2600}%
\special{fp}%
\special{pa 3610 3000}%
\special{pa 3210 3000}%
\special{fp}%
\special{sh 1}%
\special{pa 3210 3000}%
\special{pa 3278 3020}%
\special{pa 3264 3000}%
\special{pa 3278 2980}%
\special{pa 3210 3000}%
\special{fp}%
\special{pa 2810 3000}%
\special{pa 2410 3000}%
\special{fp}%
\special{sh 1}%
\special{pa 2410 3000}%
\special{pa 2478 3020}%
\special{pa 2464 3000}%
\special{pa 2478 2980}%
\special{pa 2410 3000}%
\special{fp}%
\special{pa 2010 3000}%
\special{pa 1610 3000}%
\special{fp}%
\special{sh 1}%
\special{pa 1610 3000}%
\special{pa 1678 3020}%
\special{pa 1664 3000}%
\special{pa 1678 2980}%
\special{pa 1610 3000}%
\special{fp}%
\special{pa 1210 3000}%
\special{pa 810 3000}%
\special{fp}%
\special{sh 1}%
\special{pa 810 3000}%
\special{pa 878 3020}%
\special{pa 864 3000}%
\special{pa 878 2980}%
\special{pa 810 3000}%
\special{fp}%
\special{pa 1210 3000}%
\special{pa 1210 2600}%
\special{fp}%
\special{sh 1}%
\special{pa 1210 2600}%
\special{pa 1190 2668}%
\special{pa 1210 2654}%
\special{pa 1230 2668}%
\special{pa 1210 2600}%
\special{fp}%
\special{pa 2010 3000}%
\special{pa 2010 2600}%
\special{fp}%
\special{sh 1}%
\special{pa 2010 2600}%
\special{pa 1990 2668}%
\special{pa 2010 2654}%
\special{pa 2030 2668}%
\special{pa 2010 2600}%
\special{fp}%
\special{pa 2810 3000}%
\special{pa 2810 2600}%
\special{fp}%
\special{sh 1}%
\special{pa 2810 2600}%
\special{pa 2790 2668}%
\special{pa 2810 2654}%
\special{pa 2830 2668}%
\special{pa 2810 2600}%
\special{fp}%
% VECTOR 2 0 3 0
% 16 3610 1400 3610 1800 3610 1400 3210 1400 2810 1400 2410 1400 2010 1400 1610 1400 1210 1400 810 1400 1210 1400 1210 1800 2010 1400 2010 1800 2810 1400 2810 1800
% 
\special{pn 8}%
\special{pa 3610 1400}%
\special{pa 3610 1800}%
\special{fp}%
\special{sh 1}%
\special{pa 3610 1800}%
\special{pa 3630 1734}%
\special{pa 3610 1748}%
\special{pa 3590 1734}%
\special{pa 3610 1800}%
\special{fp}%
\special{pa 3610 1400}%
\special{pa 3210 1400}%
\special{fp}%
\special{sh 1}%
\special{pa 3210 1400}%
\special{pa 3278 1420}%
\special{pa 3264 1400}%
\special{pa 3278 1380}%
\special{pa 3210 1400}%
\special{fp}%
\special{pa 2810 1400}%
\special{pa 2410 1400}%
\special{fp}%
\special{sh 1}%
\special{pa 2410 1400}%
\special{pa 2478 1420}%
\special{pa 2464 1400}%
\special{pa 2478 1380}%
\special{pa 2410 1400}%
\special{fp}%
\special{pa 2010 1400}%
\special{pa 1610 1400}%
\special{fp}%
\special{sh 1}%
\special{pa 1610 1400}%
\special{pa 1678 1420}%
\special{pa 1664 1400}%
\special{pa 1678 1380}%
\special{pa 1610 1400}%
\special{fp}%
\special{pa 1210 1400}%
\special{pa 810 1400}%
\special{fp}%
\special{sh 1}%
\special{pa 810 1400}%
\special{pa 878 1420}%
\special{pa 864 1400}%
\special{pa 878 1380}%
\special{pa 810 1400}%
\special{fp}%
\special{pa 1210 1400}%
\special{pa 1210 1800}%
\special{fp}%
\special{sh 1}%
\special{pa 1210 1800}%
\special{pa 1230 1734}%
\special{pa 1210 1748}%
\special{pa 1190 1734}%
\special{pa 1210 1800}%
\special{fp}%
\special{pa 2010 1400}%
\special{pa 2010 1800}%
\special{fp}%
\special{sh 1}%
\special{pa 2010 1800}%
\special{pa 2030 1734}%
\special{pa 2010 1748}%
\special{pa 1990 1734}%
\special{pa 2010 1800}%
\special{fp}%
\special{pa 2810 1400}%
\special{pa 2810 1800}%
\special{fp}%
\special{sh 1}%
\special{pa 2810 1800}%
\special{pa 2830 1734}%
\special{pa 2810 1748}%
\special{pa 2790 1734}%
\special{pa 2810 1800}%
\special{fp}%
% VECTOR 2 0 3 0
% 8 3610 2200 3210 2200 2810 2200 2410 2200 2010 2200 1610 2200 1210 2200 810 2200
% 
\special{pn 8}%
\special{pa 3610 2200}%
\special{pa 3210 2200}%
\special{fp}%
\special{sh 1}%
\special{pa 3210 2200}%
\special{pa 3278 2220}%
\special{pa 3264 2200}%
\special{pa 3278 2180}%
\special{pa 3210 2200}%
\special{fp}%
\special{pa 2810 2200}%
\special{pa 2410 2200}%
\special{fp}%
\special{sh 1}%
\special{pa 2410 2200}%
\special{pa 2478 2220}%
\special{pa 2464 2200}%
\special{pa 2478 2180}%
\special{pa 2410 2200}%
\special{fp}%
\special{pa 2010 2200}%
\special{pa 1610 2200}%
\special{fp}%
\special{sh 1}%
\special{pa 1610 2200}%
\special{pa 1678 2220}%
\special{pa 1664 2200}%
\special{pa 1678 2180}%
\special{pa 1610 2200}%
\special{fp}%
\special{pa 1210 2200}%
\special{pa 810 2200}%
\special{fp}%
\special{sh 1}%
\special{pa 810 2200}%
\special{pa 878 2220}%
\special{pa 864 2200}%
\special{pa 878 2180}%
\special{pa 810 2200}%
\special{fp}%
% VECTOR 2 0 3 0
% 8 3610 1400 3610 1000 2810 1400 2810 1000 2010 1400 2010 1000 1210 1400 1210 1000
% 
\special{pn 8}%
\special{pa 3610 1400}%
\special{pa 3610 1000}%
\special{fp}%
\special{sh 1}%
\special{pa 3610 1000}%
\special{pa 3590 1068}%
\special{pa 3610 1054}%
\special{pa 3630 1068}%
\special{pa 3610 1000}%
\special{fp}%
\special{pa 2810 1400}%
\special{pa 2810 1000}%
\special{fp}%
\special{sh 1}%
\special{pa 2810 1000}%
\special{pa 2790 1068}%
\special{pa 2810 1054}%
\special{pa 2830 1068}%
\special{pa 2810 1000}%
\special{fp}%
\special{pa 2010 1400}%
\special{pa 2010 1000}%
\special{fp}%
\special{sh 1}%
\special{pa 2010 1000}%
\special{pa 1990 1068}%
\special{pa 2010 1054}%
\special{pa 2030 1068}%
\special{pa 2010 1000}%
\special{fp}%
\special{pa 1210 1400}%
\special{pa 1210 1000}%
\special{fp}%
\special{sh 1}%
\special{pa 1210 1000}%
\special{pa 1190 1068}%
\special{pa 1210 1054}%
\special{pa 1230 1068}%
\special{pa 1210 1000}%
\special{fp}%
% VECTOR 2 0 3 0
% 8 3610 600 3210 600 2810 600 2410 600 2010 600 1610 600 1210 600 810 600
% 
\special{pn 8}%
\special{pa 3610 600}%
\special{pa 3210 600}%
\special{fp}%
\special{sh 1}%
\special{pa 3210 600}%
\special{pa 3278 620}%
\special{pa 3264 600}%
\special{pa 3278 580}%
\special{pa 3210 600}%
\special{fp}%
\special{pa 2810 600}%
\special{pa 2410 600}%
\special{fp}%
\special{sh 1}%
\special{pa 2410 600}%
\special{pa 2478 620}%
\special{pa 2464 600}%
\special{pa 2478 580}%
\special{pa 2410 600}%
\special{fp}%
\special{pa 2010 600}%
\special{pa 1610 600}%
\special{fp}%
\special{sh 1}%
\special{pa 1610 600}%
\special{pa 1678 620}%
\special{pa 1664 600}%
\special{pa 1678 580}%
\special{pa 1610 600}%
\special{fp}%
\special{pa 1210 600}%
\special{pa 810 600}%
\special{fp}%
\special{sh 1}%
\special{pa 810 600}%
\special{pa 878 620}%
\special{pa 864 600}%
\special{pa 878 580}%
\special{pa 810 600}%
\special{fp}%
% STR 2 0 3 0
% 3 3610 3100 3610 3200 5 0
% $a$
\put(36.1000,-32.0000){\makebox(0,0){$a$}}%
% STR 2 0 3 0
% 3 2800 3100 2800 3200 5 0
% $\cdots$
\put(28.0000,-32.0000){\makebox(0,0){$\cdots$}}%
% STR 2 0 3 0
% 3 2000 3100 2000 3200 5 0
% $a$
\put(20.0000,-32.0000){\makebox(0,0){$a$}}%
% STR 2 0 3 0
% 3 1200 3100 1200 3200 5 0
% $a+\omega_1$
\put(12.0000,-32.0000){\makebox(0,0){$a+\omega_1$}}%
% STR 2 0 3 0
% 3 600 3100 600 3200 5 0
% $a+\omega_2$
\put(6.0000,-32.0000){\makebox(0,0){$a+\omega_2$}}%
% LINE 2 1 3 0
% 6 5000 200 5000 3400 5000 3400 200 3400 200 3400 200 200
% 
\special{pn 8}%
\special{pa 5000 200}%
\special{pa 5000 3400}%
\special{da 0.070}%
\special{pa 5000 3400}%
\special{pa 200 3400}%
\special{da 0.070}%
\special{pa 200 3400}%
\special{pa 200 200}%
\special{da 0.070}%
% STR 2 0 3 0
% 3 2400 3100 2400 3200 5 0
% $\cdots$
\put(24.0000,-32.0000){\makebox(0,0){$\cdots$}}%
% STR 2 0 3 0
% 3 3200 3100 3200 3200 5 0
% $\cdots$
\put(32.0000,-32.0000){\makebox(0,0){$\cdots$}}%
% STR 2 0 3 0
% 3 3800 1300 3800 1400 5 0
% $a$
\put(38.0000,-14.0000){\makebox(0,0){$a$}}%
\end{picture}%
~\\
~\\
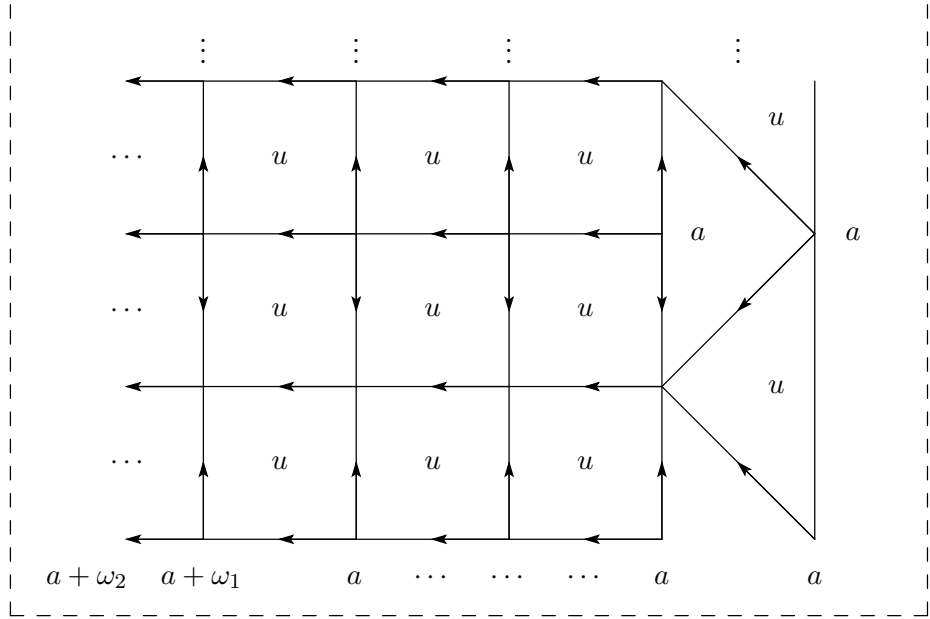
FIG.10.~Boundary state

~\\
\end{center}
The boundary state $|k\rangle_B$
and the infinite 
transfer matrix $T_B^{(a)}(u)$
satisfy the following relations.
\begin{eqnarray}
T_B^{(a)}(u)|k\rangle_B=|k\rangle_B.\nonumber
\end{eqnarray}
Let us set the dual boundary state $~_B\langle k|$
by the half infinite plane figure : FIG.11.

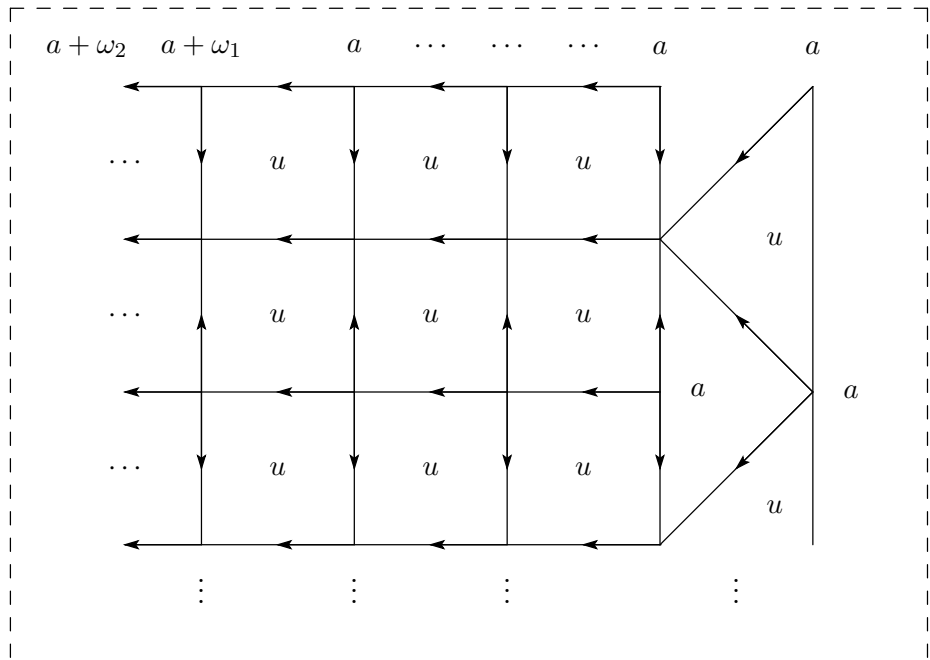
\begin{center}

%WinTpicVersion3.08
\unitlength 0.1in
\begin{picture}( 49.4000, 34.0000)(  2.7000,-35.9000)
% LINE 2 0 3 0
% 10 1010 600 3810 600 3810 600 3810 1400 3810 1400 3810 1400 3810 1400 3810 3000 3810 3000 1010 3000
% 
\special{pn 8}%
\special{pa 1010 600}%
\special{pa 3810 600}%
\special{fp}%
\special{pa 3810 600}%
\special{pa 3810 1400}%
\special{fp}%
\special{pa 3810 1400}%
\special{pa 3810 1400}%
\special{fp}%
\special{pa 3810 1400}%
\special{pa 3810 3000}%
\special{fp}%
\special{pa 3810 3000}%
\special{pa 1010 3000}%
\special{fp}%
% LINE 2 0 3 0
% 20 3810 2200 1010 2200 1010 1400 3810 1400 3010 3000 3010 600 2210 600 2210 3000 1410 3000 1410 600 3810 600 3810 600 4610 600 4610 3000 4610 600 3810 1400 3810 1400 4610 2200 4610 2200 3810 3000
% 
\special{pn 8}%
\special{pa 3810 2200}%
\special{pa 1010 2200}%
\special{fp}%
\special{pa 1010 1400}%
\special{pa 3810 1400}%
\special{fp}%
\special{pa 3010 3000}%
\special{pa 3010 600}%
\special{fp}%
\special{pa 2210 600}%
\special{pa 2210 3000}%
\special{fp}%
\special{pa 1410 3000}%
\special{pa 1410 600}%
\special{fp}%
\special{pa 3810 600}%
\special{pa 3810 600}%
\special{fp}%
\special{pa 4610 600}%
\special{pa 4610 3000}%
\special{fp}%
\special{pa 4610 600}%
\special{pa 3810 1400}%
\special{fp}%
\special{pa 3810 1400}%
\special{pa 4610 2200}%
\special{fp}%
\special{pa 4610 2200}%
\special{pa 3810 3000}%
\special{fp}%
% STR 2 0 3 0
% 3 4610 300 4610 400 5 0
% $a$
\put(46.1000,-4.0000){\makebox(0,0){$a$}}%
% STR 2 0 3 0
% 3 3810 300 3810 400 5 0
% $a$
\put(38.1000,-4.0000){\makebox(0,0){$a$}}%
% STR 2 0 3 0
% 3 4810 2100 4810 2200 5 0
% $a$
\put(48.1000,-22.0000){\makebox(0,0){$a$}}%
% VECTOR 2 0 3 0
% 6 4610 600 4210 1000 4610 2200 4210 1800 4610 2200 4210 2600
% 
\special{pn 8}%
\special{pa 4610 600}%
\special{pa 4210 1000}%
\special{fp}%
\special{sh 1}%
\special{pa 4210 1000}%
\special{pa 4272 968}%
\special{pa 4248 962}%
\special{pa 4244 940}%
\special{pa 4210 1000}%
\special{fp}%
\special{pa 4610 2200}%
\special{pa 4210 1800}%
\special{fp}%
\special{sh 1}%
\special{pa 4210 1800}%
\special{pa 4244 1862}%
\special{pa 4248 1838}%
\special{pa 4272 1834}%
\special{pa 4210 1800}%
\special{fp}%
\special{pa 4610 2200}%
\special{pa 4210 2600}%
\special{fp}%
\special{sh 1}%
\special{pa 4210 2600}%
\special{pa 4272 2568}%
\special{pa 4248 2562}%
\special{pa 4244 2540}%
\special{pa 4210 2600}%
\special{fp}%
% VECTOR 2 0 3 0
% 14 3810 600 3410 600 3010 600 2610 600 2210 600 1810 600 2210 600 2210 1000 1410 600 1410 1000 3010 600 3010 1000 3810 600 3810 1000
% 
\special{pn 8}%
\special{pa 3810 600}%
\special{pa 3410 600}%
\special{fp}%
\special{sh 1}%
\special{pa 3410 600}%
\special{pa 3478 620}%
\special{pa 3464 600}%
\special{pa 3478 580}%
\special{pa 3410 600}%
\special{fp}%
\special{pa 3010 600}%
\special{pa 2610 600}%
\special{fp}%
\special{sh 1}%
\special{pa 2610 600}%
\special{pa 2678 620}%
\special{pa 2664 600}%
\special{pa 2678 580}%
\special{pa 2610 600}%
\special{fp}%
\special{pa 2210 600}%
\special{pa 1810 600}%
\special{fp}%
\special{sh 1}%
\special{pa 1810 600}%
\special{pa 1878 620}%
\special{pa 1864 600}%
\special{pa 1878 580}%
\special{pa 1810 600}%
\special{fp}%
\special{pa 2210 600}%
\special{pa 2210 1000}%
\special{fp}%
\special{sh 1}%
\special{pa 2210 1000}%
\special{pa 2230 934}%
\special{pa 2210 948}%
\special{pa 2190 934}%
\special{pa 2210 1000}%
\special{fp}%
\special{pa 1410 600}%
\special{pa 1410 1000}%
\special{fp}%
\special{sh 1}%
\special{pa 1410 1000}%
\special{pa 1430 934}%
\special{pa 1410 948}%
\special{pa 1390 934}%
\special{pa 1410 1000}%
\special{fp}%
\special{pa 3010 600}%
\special{pa 3010 1000}%
\special{fp}%
\special{sh 1}%
\special{pa 3010 1000}%
\special{pa 3030 934}%
\special{pa 3010 948}%
\special{pa 2990 934}%
\special{pa 3010 1000}%
\special{fp}%
\special{pa 3810 600}%
\special{pa 3810 1000}%
\special{fp}%
\special{sh 1}%
\special{pa 3810 1000}%
\special{pa 3830 934}%
\special{pa 3810 948}%
\special{pa 3790 934}%
\special{pa 3810 1000}%
\special{fp}%
% VECTOR 2 0 3 0
% 10 3810 1400 3410 1400 3010 1400 2610 1400 2210 1400 1810 1400 1410 1400 1010 1400 1410 600 1010 600
% 
\special{pn 8}%
\special{pa 3810 1400}%
\special{pa 3410 1400}%
\special{fp}%
\special{sh 1}%
\special{pa 3410 1400}%
\special{pa 3478 1420}%
\special{pa 3464 1400}%
\special{pa 3478 1380}%
\special{pa 3410 1400}%
\special{fp}%
\special{pa 3010 1400}%
\special{pa 2610 1400}%
\special{fp}%
\special{sh 1}%
\special{pa 2610 1400}%
\special{pa 2678 1420}%
\special{pa 2664 1400}%
\special{pa 2678 1380}%
\special{pa 2610 1400}%
\special{fp}%
\special{pa 2210 1400}%
\special{pa 1810 1400}%
\special{fp}%
\special{sh 1}%
\special{pa 1810 1400}%
\special{pa 1878 1420}%
\special{pa 1864 1400}%
\special{pa 1878 1380}%
\special{pa 1810 1400}%
\special{fp}%
\special{pa 1410 1400}%
\special{pa 1010 1400}%
\special{fp}%
\special{sh 1}%
\special{pa 1010 1400}%
\special{pa 1078 1420}%
\special{pa 1064 1400}%
\special{pa 1078 1380}%
\special{pa 1010 1400}%
\special{fp}%
\special{pa 1410 600}%
\special{pa 1010 600}%
\special{fp}%
\special{sh 1}%
\special{pa 1010 600}%
\special{pa 1078 620}%
\special{pa 1064 600}%
\special{pa 1078 580}%
\special{pa 1010 600}%
\special{fp}%
% VECTOR 2 0 3 0
% 16 3810 2200 3410 2200 3010 2200 2610 2200 2210 2200 1810 2200 1410 2200 1010 2200 3810 3000 3410 3000 3010 3000 2610 3000 2210 3000 1810 3000 1410 3000 1010 3000
% 
\special{pn 8}%
\special{pa 3810 2200}%
\special{pa 3410 2200}%
\special{fp}%
\special{sh 1}%
\special{pa 3410 2200}%
\special{pa 3478 2220}%
\special{pa 3464 2200}%
\special{pa 3478 2180}%
\special{pa 3410 2200}%
\special{fp}%
\special{pa 3010 2200}%
\special{pa 2610 2200}%
\special{fp}%
\special{sh 1}%
\special{pa 2610 2200}%
\special{pa 2678 2220}%
\special{pa 2664 2200}%
\special{pa 2678 2180}%
\special{pa 2610 2200}%
\special{fp}%
\special{pa 2210 2200}%
\special{pa 1810 2200}%
\special{fp}%
\special{sh 1}%
\special{pa 1810 2200}%
\special{pa 1878 2220}%
\special{pa 1864 2200}%
\special{pa 1878 2180}%
\special{pa 1810 2200}%
\special{fp}%
\special{pa 1410 2200}%
\special{pa 1010 2200}%
\special{fp}%
\special{sh 1}%
\special{pa 1010 2200}%
\special{pa 1078 2220}%
\special{pa 1064 2200}%
\special{pa 1078 2180}%
\special{pa 1010 2200}%
\special{fp}%
\special{pa 3810 3000}%
\special{pa 3410 3000}%
\special{fp}%
\special{sh 1}%
\special{pa 3410 3000}%
\special{pa 3478 3020}%
\special{pa 3464 3000}%
\special{pa 3478 2980}%
\special{pa 3410 3000}%
\special{fp}%
\special{pa 3010 3000}%
\special{pa 2610 3000}%
\special{fp}%
\special{sh 1}%
\special{pa 2610 3000}%
\special{pa 2678 3020}%
\special{pa 2664 3000}%
\special{pa 2678 2980}%
\special{pa 2610 3000}%
\special{fp}%
\special{pa 2210 3000}%
\special{pa 1810 3000}%
\special{fp}%
\special{sh 1}%
\special{pa 1810 3000}%
\special{pa 1878 3020}%
\special{pa 1864 3000}%
\special{pa 1878 2980}%
\special{pa 1810 3000}%
\special{fp}%
\special{pa 1410 3000}%
\special{pa 1010 3000}%
\special{fp}%
\special{sh 1}%
\special{pa 1010 3000}%
\special{pa 1078 3020}%
\special{pa 1064 3000}%
\special{pa 1078 2980}%
\special{pa 1010 3000}%
\special{fp}%
% VECTOR 2 0 3 0
% 18 3810 3000 3810 3000 3810 2200 3810 1800 3010 2200 3010 1800 2210 2200 2210 1800 1410 2200 1410 1800 3810 2200 3810 2600 3010 2200 3010 2600 2210 2200 2210 2600 1410 2200 1410 2600
% 
\special{pn 8}%
\special{pa 3810 3000}%
\special{pa 3810 3000}%
\special{fp}%
\special{pa 3810 2200}%
\special{pa 3810 1800}%
\special{fp}%
\special{sh 1}%
\special{pa 3810 1800}%
\special{pa 3790 1868}%
\special{pa 3810 1854}%
\special{pa 3830 1868}%
\special{pa 3810 1800}%
\special{fp}%
\special{pa 3010 2200}%
\special{pa 3010 1800}%
\special{fp}%
\special{sh 1}%
\special{pa 3010 1800}%
\special{pa 2990 1868}%
\special{pa 3010 1854}%
\special{pa 3030 1868}%
\special{pa 3010 1800}%
\special{fp}%
\special{pa 2210 2200}%
\special{pa 2210 1800}%
\special{fp}%
\special{sh 1}%
\special{pa 2210 1800}%
\special{pa 2190 1868}%
\special{pa 2210 1854}%
\special{pa 2230 1868}%
\special{pa 2210 1800}%
\special{fp}%
\special{pa 1410 2200}%
\special{pa 1410 1800}%
\special{fp}%
\special{sh 1}%
\special{pa 1410 1800}%
\special{pa 1390 1868}%
\special{pa 1410 1854}%
\special{pa 1430 1868}%
\special{pa 1410 1800}%
\special{fp}%
\special{pa 3810 2200}%
\special{pa 3810 2600}%
\special{fp}%
\special{sh 1}%
\special{pa 3810 2600}%
\special{pa 3830 2534}%
\special{pa 3810 2548}%
\special{pa 3790 2534}%
\special{pa 3810 2600}%
\special{fp}%
\special{pa 3010 2200}%
\special{pa 3010 2600}%
\special{fp}%
\special{sh 1}%
\special{pa 3010 2600}%
\special{pa 3030 2534}%
\special{pa 3010 2548}%
\special{pa 2990 2534}%
\special{pa 3010 2600}%
\special{fp}%
\special{pa 2210 2200}%
\special{pa 2210 2600}%
\special{fp}%
\special{sh 1}%
\special{pa 2210 2600}%
\special{pa 2230 2534}%
\special{pa 2210 2548}%
\special{pa 2190 2534}%
\special{pa 2210 2600}%
\special{fp}%
\special{pa 1410 2200}%
\special{pa 1410 2600}%
\special{fp}%
\special{sh 1}%
\special{pa 1410 2600}%
\special{pa 1430 2534}%
\special{pa 1410 2548}%
\special{pa 1390 2534}%
\special{pa 1410 2600}%
\special{fp}%
% STR 2 0 3 0
% 3 1010 900 1010 1000 5 0
% $\cdots$
\put(10.1000,-10.0000){\makebox(0,0){$\cdots$}}%
% STR 2 0 3 0
% 3 1010 1700 1010 1800 5 0
% $\cdots$
\put(10.1000,-18.0000){\makebox(0,0){$\cdots$}}%
% STR 2 0 3 0
% 3 1010 2500 1010 2600 5 0
% $\cdots$
\put(10.1000,-26.0000){\makebox(0,0){$\cdots$}}%
% STR 2 0 3 0
% 3 1410 3100 1410 3200 5 0
% $\vdots$
\put(14.1000,-32.0000){\makebox(0,0){$\vdots$}}%
% STR 2 0 3 0
% 3 2210 3100 2210 3200 5 0
% $\vdots$
\put(22.1000,-32.0000){\makebox(0,0){$\vdots$}}%
% STR 2 0 3 0
% 3 3010 3100 3010 3200 5 0
% $\vdots$
\put(30.1000,-32.0000){\makebox(0,0){$\vdots$}}%
% STR 2 0 3 0
% 3 4210 3100 4210 3200 5 0
% $\vdots$
\put(42.1000,-32.0000){\makebox(0,0){$\vdots$}}%
% STR 2 0 3 0
% 3 3410 900 3410 1000 5 0
% $u$
\put(34.1000,-10.0000){\makebox(0,0){$u$}}%
% STR 2 0 3 0
% 3 2610 900 2610 1000 5 0
% $u$
\put(26.1000,-10.0000){\makebox(0,0){$u$}}%
% STR 2 0 3 0
% 3 1810 900 1810 1000 5 0
% $u$
\put(18.1000,-10.0000){\makebox(0,0){$u$}}%
% STR 2 0 3 0
% 3 4410 1300 4410 1400 5 0
% $u$
\put(44.1000,-14.0000){\makebox(0,0){$u$}}%
% STR 2 0 3 0
% 3 4410 2700 4410 2800 5 0
% $u$
\put(44.1000,-28.0000){\makebox(0,0){$u$}}%
% STR 2 0 3 0
% 3 3410 1700 3410 1800 5 0
% $u$
\put(34.1000,-18.0000){\makebox(0,0){$u$}}%
% STR 2 0 3 0
% 3 2610 1700 2610 1800 5 0
% $u$
\put(26.1000,-18.0000){\makebox(0,0){$u$}}%
% STR 2 0 3 0
% 3 1810 1700 1810 1800 5 0
% $u$
\put(18.1000,-18.0000){\makebox(0,0){$u$}}%
% STR 2 0 3 0
% 3 1810 2500 1810 2600 5 0
% $u$
\put(18.1000,-26.0000){\makebox(0,0){$u$}}%
% STR 2 0 3 0
% 3 2610 2500 2610 2600 5 0
% $u$
\put(26.1000,-26.0000){\makebox(0,0){$u$}}%
% STR 2 0 3 0
% 3 3410 2500 3410 2600 5 0
% $u$
\put(34.1000,-26.0000){\makebox(0,0){$u$}}%
% STR 2 0 3 0
% 3 3410 290 3410 390 5 0
% $\cdots$
\put(34.1000,-3.9000){\makebox(0,0){$\cdots$}}%
% STR 2 0 3 0
% 3 3010 290 3010 390 5 0
% $\cdots$
\put(30.1000,-3.9000){\makebox(0,0){$\cdots$}}%
% STR 2 0 3 0
% 3 2210 290 2210 390 5 0
% $a$
\put(22.1000,-3.9000){\makebox(0,0){$a$}}%
% STR 2 0 3 0
% 3 1410 290 1410 390 5 0
% $a+\omega_1$
\put(14.1000,-3.9000){\makebox(0,0){$a+\omega_1$}}%
% STR 2 0 3 0
% 3 810 290 810 390 5 0
% $a+\omega_2$
\put(8.1000,-3.9000){\makebox(0,0){$a+\omega_2$}}%
% LINE 2 1 3 0
% 6 410 3590 410 190 410 190 5210 190 5210 190 5210 3590
% 
\special{pn 8}%
\special{pa 410 3590}%
\special{pa 410 190}%
\special{da 0.070}%
\special{pa 410 190}%
\special{pa 5210 190}%
\special{da 0.070}%
\special{pa 5210 190}%
\special{pa 5210 3590}%
\special{da 0.070}%
% STR 2 0 3 0
% 3 2610 290 2610 390 5 0
% $\cdots$
\put(26.1000,-3.9000){\makebox(0,0){$\cdots$}}%
% STR 2 0 3 0
% 3 4010 2090 4010 2190 5 0
% $a$
\put(40.1000,-21.9000){\makebox(0,0){$a$}}%
\end{picture}%
~\\
~\\
FIG.11.~Boundary state

~\\
\end{center}
The dual boundary state $~_B\langle k|$ and the infinite
transfer matrix $T_B^{(a)}(u)$ 
satisfy the following relations.
\begin{eqnarray}
~_B\langle k|
T_B^{(a)}(u)=~_B\langle k|.\nonumber
\end{eqnarray}
For $k=a+\rho, l=b+\rho$,
if we set
\begin{eqnarray}
\Phi_{N}^{(a,a+\bar{\epsilon}_\mu)}(u)&=&
\Phi^{*(k,k+\bar{\epsilon}_\mu)}(z),\nonumber\\
\Phi_W^{(a+\bar{\epsilon}_\mu,a)}(u)&=&
\Phi^{(k+\bar{\epsilon}_\mu,k)}(z),\nonumber\\
T_B^{(a)}(u)&=&T_B(z),\nonumber
\end{eqnarray}
the above relations in the appendix are satisfied by
the free field realizations
on the Fock space ${\cal F}_{k,k}$.
However this is not perfect identification because
the space ${\cal H}_{k,k}$ and the space ${\cal F}_{k,k}$
have different characters.
We expect BRST cohomology of the certain complex consisting
of the space ${\cal F}_{l,k}$ provides the correct
identification of the space ${\cal H}_{l,k}$
\cite{FJMOP, MW, LP}.
Upon this assumption,
this appendix gives the physical interpretation
of the main text.

%%%%%%%%%%%%%%%%%%%%%%%%%%%%
%%%%%%%%%%%%%%%%%%%%%%%%%%%%

%\label{}

%\subsection{}

%\subsubsection{}

% If in two-column mode, this environment will change to single-column format so that long equations can be displayed. 

% Use only when necessary.

%\begin{widetext}

%$$\mbox{put long equation here}$$

%\end{widetext}

% Figures should be put into the text as floats. 

% Use the graphics or graphicx packages (distributed with LaTeX2e).

% See the LaTeX Graphics Companion by Michel Goosens, Sebastian Rahtz, and Frank Mittelbach for examples. 

%

% Here is an example of the general form of a figure:

% Fill in the caption in the braces of the \caption{} command. 

% Put the label that you will use with \ref{} command in the braces of the \label{} command.

%

% \begin{figure}

% \includegraphics{}%

% \caption{\label{}}%

% \end{figure}

% Tables may be be put in the text as floats.

% Here is an example of the general form of a table:

% Fill in the caption in the braces of the \caption{} command. Put the label

% that you will use with \ref{} command in the braces of the \label{} command.

% Insert the column specifiers (l, r, c, d, etc.) in the empty braces of the

% \begin{tabular}{} command.

%

% \begin{table}

% \caption{\label{} }

% \begin{tabular}{}

% \end{tabular}

% \end{table}

% If you have acknowledgments, this puts in the proper section head.

%\begin{acknowledgments}

% Put your acknowledgments here.

%\end{acknowledgments}

% Create the reference section using BibTeX:

%\bibliography{thebibliography}{99}

\end{document}